\documentclass[acmsmall]{acmart}

\usepackage{tikz}
\usepackage{textcomp}
\usepackage{listings}
\usepackage{xcolor}
\usepackage{filecontents}

\makeatletter

\newcommand\PrologPredicateStyle{}
\newcommand\PrologVarStyle{}
\newcommand\PrologAnonymVarStyle{}
\newcommand\PrologAtomStyle{}
\newcommand\PrologOtherStyle{}
\newcommand\PrologCommentStyle{}

\newif\ifpredicate@prolog@
\newif\ifwithinparens@prolog@

\lst@SaveOutputDef{`_}\underscore@prolog

\newcount\currentchar@prolog

\newcommand\@testChar@prolog%
{%
  \ifnum\lst@mode=\lst@Pmode%
    \detectTypeAndHighlight@prolog%
  \else
    \ifwithinparens@prolog@%
      \detectTypeAndHighlight@prolog%
    \fi
  \fi
  \global\predicate@prolog@false%
}

\newcommand\detectTypeAndHighlight@prolog
{%
  \def\lst@thestyle{\PrologAtomStyle}%
  \ifpredicate@prolog@%
    \def\lst@thestyle{\PrologPredicateStyle}%
  \else
    \expandafter\splitfirstchar@prolog\expandafter{\the\lst@token}%
    \expandafter\ifx\@testChar@prolog\underscore@prolog%
      \ifnum\lst@length=1%
        \let\lst@thestyle\PrologAnonymVarStyle%
      \else
        \let\lst@thestyle\PrologVarStyle%
      \fi
    \else
      \currentchar@prolog=65
      \loop
        \expandafter\ifnum\expandafter`\@testChar@prolog=\currentchar@prolog%
          \let\lst@thestyle\PrologVarStyle%
          \let\iterate\relax
        \fi
        \advance \currentchar@prolog by 1
        \unless\ifnum\currentchar@prolog>90
      \repeat
    \fi
  \fi
}
\newcommand\splitfirstchar@prolog{}
\def\splitfirstchar@prolog#1{\@splitfirstchar@prolog#1\relax}
\newcommand\@splitfirstchar@prolog{}
\def\@splitfirstchar@prolog#1#2\relax{\def\@testChar@prolog{#1}}

\def\beginlstdelim#1#2%
{%
  \def\endlstdelim{\PrologOtherStyle #2\egroup}%
  {\PrologOtherStyle #1}%
  \global\predicate@prolog@false%
  \withinparens@prolog@true%
  \bgroup\aftergroup\endlstdelim%
}

\newcommand\lang@prolog{Prolog-pretty}
\expandafter\lst@NormedDef\expandafter\normlang@prolog%
  \expandafter{\lang@prolog}

\expandafter\expandafter\expandafter\lstdefinelanguage\expandafter%
{\lang@prolog}
{%
  language            = Prolog,
  keywords            = {},      
  showstringspaces    = false,
  alsoletter          = (,
  alsoother           = @$,
  moredelim           = **[is][\beginlstdelim{(}{)}]{(}{)},
  MoreSelectCharTable =
    \lst@DefSaveDef{`(}\opparen@prolog{\global\predicate@prolog@true\opparen@prolog},
}

\newcommand\@ddedToOutput@prolog\relax
\lst@AddToHook{Output}{\@ddedToOutput@prolog}

\lst@AddToHook{PreInit}
{%
  \ifx\lst@language\normlang@prolog%
    \let\@ddedToOutput@prolog\@testChar@prolog%
  \fi
}

\lst@AddToHook{DeInit}{\renewcommand\@ddedToOutput@prolog{}}

\makeatother
\definecolor{PrologPredicate}{RGB}{000,031,255}
\definecolor{PrologVar}      {RGB}{024,021,125}
\definecolor{PrologAnonymVar}{RGB}{000,127,000}
\definecolor{PrologAtom}     {RGB}{186,032,032}
\definecolor{PrologComment}  {RGB}{063,128,127}
\definecolor{PrologOther}    {RGB}{000,000,000}

\renewcommand\PrologPredicateStyle{\color{PrologPredicate}}
\renewcommand\PrologVarStyle{\color{PrologVar}}
\renewcommand\PrologAnonymVarStyle{\color{PrologAnonymVar}}
\renewcommand\PrologAtomStyle{\color{PrologAtom}}
\renewcommand\PrologCommentStyle{\itshape\color{PrologComment}}
\renewcommand\PrologOtherStyle{\color{PrologOther}}

\lstdefinestyle{Prolog-pygsty}
{
  language     = Prolog-pretty,
  upquote      = true,
  stringstyle  = \PrologAtomStyle,
  commentstyle = \PrologCommentStyle,
  literate     =
    {:-}{{\PrologOtherStyle :-}}2
    {,}{{\PrologOtherStyle ,}}1
    {.}{{\PrologOtherStyle .}}1
}

\usepackage{booktabs}   
\usepackage{subcaption} 

\usepackage{amsmath} 
\usepackage{algorithm}
\usepackage{algorithmicx}
\usepackage[noend]{algpseudocode}
\usepackage{comment}
\RequirePackage{mathpartir}

\definecolor{mygreen}{rgb}{0,0.6,0}
\definecolor{mygray}{rgb}{0.5,0.5,0.5}
\definecolor{mymauve}{rgb}{0.58,0,0.82}

\definecolor{softred}{RGB}{239,154,154}

\colorlet{lightmygreen}{mygreen!40}

\colorlet{lightestorange}{orange!45}

\usepackage{listings}
\usepackage{xcolor}

\usepackage{mathtools, nccmath}
\usepackage{graphicx}
\usepackage{mathtools}
\usepackage{multirow}
\usepackage{xspace}

\usepackage{esvect}
\usepackage{macros}
\usepackage{comment}

\usepackage{tikz}
\usetikzlibrary{decorations.pathreplacing} 
\usetikzlibrary{positioning}
\usetikzlibrary{backgrounds}
\usetikzlibrary{calc}
\usetikzlibrary{fit}
\usetikzlibrary{positioning}
\usetikzlibrary{shadows}
\usetikzlibrary{shapes.geometric}
\usetikzlibrary{shapes.multipart}
\usetikzlibrary{backgrounds}
\usetikzlibrary{calc}
\usetikzlibrary{fit}
\usetikzlibrary{positioning}
\usetikzlibrary{shadows}
\usetikzlibrary{shapes.geometric}
\usetikzlibrary{shapes.multipart}
\definecolor{maincolor}{rgb}{0.1,0.5,1}
\colorlet{lightmain}{maincolor!75}
\colorlet{lightermain}{maincolor!50}
\colorlet{lightestmain}{maincolor!25}
\colorlet{darkmain}{maincolor!75!black}
\colorlet{darkermain}{maincolor!50!black}
\colorlet{darkestmain}{maincolor!25!black}
\definecolor{MistyRose}{rgb}{1.0, 0.89, 0.88}

\newcommand\jwnote[1]{\textcolor{purple}{{JW: #1}}}

\newcommand\todo[1]{{\colorbox{mygreen}{\textcolor{white}{\texttt{TODO}}}} \textcolor{darkerGreen}{#1}}

\newcommand\sh[1]{\textcolor{blue}{#1}}

\definecolor{darkerGreen}{RGB}{0,100,0} 

\usepackage[normalem]{ulem}

\usepackage{enumitem}

\usepackage{colonequals}

\usepackage{tcolorbox}
\usepackage{bm}

\usetikzlibrary{calc}

\usepackage{graphicx}
\usepackage{colortbl}
\usepackage{array}
\usepackage{pifont}

\usepackage{wrapfig,lipsum,booktabs}

\usepackage{soul}

\usepackage{caption}
\usepackage{soul}
\usepackage{multirow}

\AtBeginDocument{%
  }

\setcopyright{cc}
\setcctype{by}
\acmDOI{10.1145/3763058}
\acmYear{2025}
\acmJournal{PACMPL}
\acmVolume{9}
\acmNumber{OOPSLA2}
\acmArticle{280}
\acmMonth{10}
\received{2025-03-25}
\received[accepted]{2025-08-12}

\begin{document}

\title{Probabilistic Inference for Datalog  with Correlated Inputs}

\author{Jingbo Wang}
\orcid{0000-0001-5877-2677}
\affiliation{%
  \institution{Purdue University}
  \city{West Lafayette}
  \country{USA}
}
\email{wang6203@purdue.edu}

\author{Shashin Halalingaiah}
\orcid{0000-0002-1268-4345}
\affiliation{%
  \institution{University of Texas at Austin}
  \city{Austin}
  \country{USA}
}
\email{shashin@cs.utexas.edu}

\author{Weiyi Chen}
\orcid{0009-0009-6276-3525}
\affiliation{%
  \institution{Purdue University}
  \city{West Lafayette}
  \country{USA}
}
\email{chen5332@purdue.edu}

\author{Chao Wang}
\orcid{0009-0003-4684-3943}
\affiliation{%
  \institution{University of Southern California}
  \city{Los Angeles}
  \country{USA}
}
\email{wang626@usc.edu}

\author{Işıl Dillig}
\orcid{0000-0001-8006-1230}
\affiliation{%
  \institution{University of Texas at Austin}
  \city{Austin}
  \country{USA}
}
\email{isil@cs.utexas.edu}



\begin{abstract}

Probabilistic extensions of logic programming languages, such as ProbLog, integrate logical reasoning with probabilistic inference to evaluate probabilities of output relations; however, prior work does not account for potential statistical correlations among input facts. This paper introduces \toolname, a new extension to Datalog designed for precise probabilistic inference in the presence of (partially known) input correlations. We formulate the inference task as a constrained optimization problem, where the solution yields sound and precise probability bounds for output facts. However, due to the complexity of the resulting optimization problem, this approach alone often does not scale to large programs. To address scalability, we propose a more efficient $\delta$-exact inference algorithm that leverages constraint solving, static analysis, and iterative refinement. Our empirical evaluation on challenging real-world benchmarks, including side-channel analysis, demonstrates that our method not only scales effectively but also delivers tight probability bounds.

\end{abstract}





\maketitle

\section{Introduction}
Logic programming languages are powerful tools for modeling and reasoning about complex systems using well-defined rules, and they are also widely used in {graph analysis, bioinformatics, and} program analysis tasks~\cite{reps1995demand,zhang2014abstraction} such as detecting race conditions~\cite{naik2006effective,zhang2017effective} and side channels~\cite{wang2021data,wang2019mitigating}. However, real-world scenarios often involve uncertainty and incomplete information that traditional logic programming cannot handle effectively.  
{Probabilistic extensions to logic programming languages, such as ProbLog~\cite{de2007problog,dries2015problog2} and PPDL~\cite{barany2017declarative}, enhance traditional logic programming by incorporating probabilistic reasoning.
}
{For example, in medical diagnosis, symptom-disease associations are inherently probabilistic, and quantifying disease likelihood based on symptoms can enhance diagnostic accuracy. Such}
probabilistic reasoning is also useful for applications of logic programming in 
program analysis tasks~\cite{raghothaman2018user}: in the context of side channel detection,  understanding the probability of information leakage can significantly enhance security assessments.


However, these probabilistic extensions {to logic programming languages} typically assume the independence of input facts to simplify inference~\cite{de2007problog,barany2017declarative,li2023scallop}. 
In many real-world scenarios, this assumption does not hold: for example, in medical diagnostics, certain symptoms are often correlated, and treating them as independent can lead to incorrect diagnoses.
Similarly, in side-channel analysis, input facts such as memory accesses and branch outcomes are frequently correlated, and ignoring these correlations can produce unsound assessments.



This work aims to develop a novel probabilistic Datalog framework that can accommodate arbitrary statistical correlations between inputs. However, the removal of the independence assumption introduces several challenges: First, it significantly increases the complexity of probabilistic inference, requiring consideration of joint probabilities of inputs rather than treating them in isolation. Second,  since the exact conditional dependencies between input facts may not be known a priori, there can be \emph{implicit} dependencies that are not explicitly specified by users.  Finally, due to \emph{partial} knowledge about input dependencies, it may not always be possible to compute the \emph{exact} probability of an output fact, necessitating the computation of lower and upper probability bounds.  Our proposed framework addresses these challenges, providing an effective approach for accurate and reliable probabilistic reasoning in complex scenarios that involve correlated inputs.


\begin{wrapfigure}{r}{0.35\textwidth} 
    \vspace{-0.1in}
    \centering
    \includegraphics[width=\linewidth]{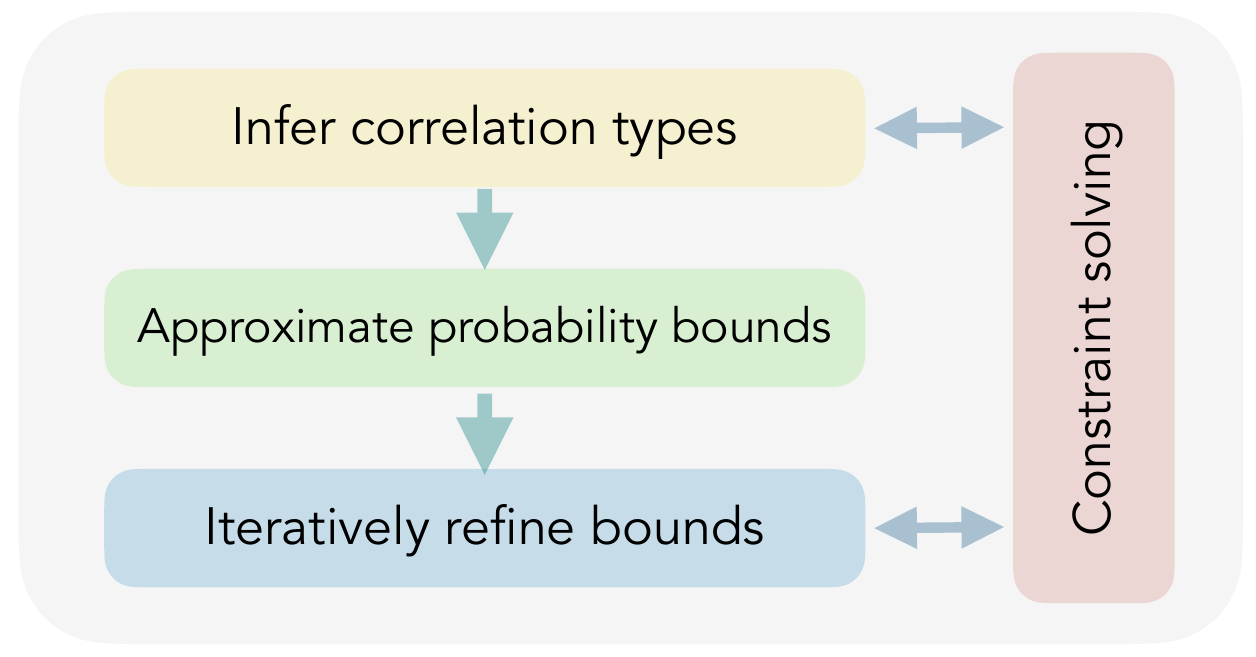}
    \vspace{-0.3in}
    \caption{Overview of our approach}
    \label{fig:sys}
    \vspace{-0.15in}
\end{wrapfigure}

At the core of our approach is a  formulation of probabilistic inference as a constrained optimization problem over \emph{joint probability variables}, which capture the joint probability distribution of correlated input facts. Probabilities specified in the Datalog program are translated into a system of constraints, and the probability of each output relation is expressed symbolically in terms of these variables. Our method then solves a constrained optimization problem to compute upper and lower bounds on the probability of the desired output fact.
However, a key challenge with this approach is that the generated constrained optimization problems can be very challenging to solve, particularly due to the non-linear nature of the objective function. 
To address this difficulty, we also present a more scalable $\delta$-exact algorithm~\cite{gao2012delta} that uses the optimization approach in a targeted way to solve simpler sub-problems.  Our  proposed method first computes \emph{approximate} probability bounds and then iteratively refines them until they are within a user-specified distance $\delta$ of the true probability bounds. As illustrated in Figure~\ref{fig:sys}, our technique comprises of three steps. First, it performs a combination of static analysis and constraint solving to infer whether a pair of facts are positively or negatively correlated, or independent. In the second step, it leverages the results of this ``correlation type analysis'' to compute approximate bounds on the probabilities of each predicate. Finally, it iteratively refines these approximate bounds to a user-specified tightness $\delta$ --- i.e., upon termination, the computed lower and upper bounds on the probabilities are guaranteed to be within $\delta$ of the ground truth.

%
%

We have implemented the proposed idea in a new Datalog variant called \toolname\footnote{Stands for PRobAbilistic Logical InfereNce Engine} and evaluate it on two domains, namely side channel vulnerability detection and inference for discrete Bayesian networks.   Our experiments show that \toolname can successfully infer accurate probability bounds for output facts, including for large benchmarks with hundreds of thousands of facts. We also conduct ablation studies to evaluate the impact of the three key ideas illustrated in Figure~\ref{fig:sys} and show that they have a significant impact on the scalability and precision of our approach.

To summarize, this paper makes the following key contributions:

\begin{itemize}[leftmargin=*]
    \item We introduce the first probabilistic Datalog framework that allows the user to specify arbitrary statistical dependencies between input facts.
    \item We propose a constrained optimization approach to compute exact probability bounds for outputs. 
    \item We present a $\delta$-exact algorithm that combines our basic constrained optimization formulation with  static analysis, approximation, and iterative refinement to improve scalability. 
    \item We perform an empirical evaluation  on 30 real-world probabilistic Datalog programs consisting of large-scale program analysis tasks and Bayesian inference benchmarks. Our results show that  \toolname can produce precise probability bounds, while scaling to large benchmarks.
\end{itemize}

\section{Motivating Examples}\label{sec:motivating}

In this section, we provide two simple examples to motivate the probabilistic inference capabilities of \toolname,  highlighting scenarios where input facts are correlated but the full joint probability distribution is unknown.

\begin{figure}[t]
\vspace{-0.2in}

    \begin{minipage}[t]{0.01\textwidth}
    \end{minipage}
       \begin{minipage}[t]{0.49\textwidth}
	\centering
\begin{lstlisting}[
    basicstyle=\scriptsize\ttfamily,  % Custom small font size
    style=Prolog-pygsty,
    escapechar=\%, 
    frame=none,
    numbers=left, 
    numberstyle=\tiny\color{gray}, 
    stepnumber=1,
    numbersep=4pt
]
%\textcolor{blue}{\% Input facts.} %
%\textcolor{blue}{\% Known input facts about patient} %
1.00 :: chest_pain.
%\textcolor{blue}{\% Unknown input facts about patient}%
%\textcolor{blue}{\% Probabilities estimated from average population.}%
0.03 :: abnormal_ecg.
0.02 :: troponin_high.
...
%\textcolor{blue}{\% Input correlations based on the literature}%
%\colorbox{lightmygreen}{0.90 :: abnormal\_ecg | chest\_pain, troponin\_high. }%
...
%\textcolor{blue}{\% Inference rules for heart disease assessment} %
0.85 :: heart_disease %$\colonminus$% abnormal_ecg, troponin_high, chest_pain.
0.60 :: heart_disease %$\colonminus$% abnormal_ecg, high_bnp, chest_pain, \+troponin_high.
%\textcolor{blue}{\% Query} % 
query%(heart\_disease)%.



\end{lstlisting}
    \end{minipage}
    \begin{minipage}[t]{0.49\textwidth}
    \centering
\begin{lstlisting}[
    basicstyle=\scriptsize\ttfamily,  % Custom small font size
    style=Prolog-pygsty,
    escapechar=\%, 
    frame=none,
    numbers=left, 
    numberstyle=\tiny\color{gray}, 
    stepnumber=1,
    numbersep=4pt
]
%\textcolor{blue}{\% Input Facts about input program from Fig.~\ref{fig:taintProg}} %
1.0 :: rand%(r3)%. % \textcolor{blue}{\% r3 is a random variable} %
%\textcolor{blue}{\% \texttt{XOR} operation highlighted in Fig.~\ref{fig:taintProg}} %
1.0 :: xor%(n5, n7, n8)%.
...
0.7 :: share%(n5, n8)%. %\textcolor{blue}{\% register sharing} 
%
0.8 :: dep%(n5, r3)%. %\textcolor{blue}{\% data dependency} %
0.8 :: dep%(n8, r3)%. 
...
%\textcolor{blue}{\% Input correlations} %
%\colorbox{lightmygreen}{0.9 :: share(n5,n8) | xor(n5,n7,n8), dep(n5,r3).
}%
...
%\textcolor{blue}{ \%  Rules for inferring leaks} %
0.7 :: t%(V1,V2)% %$\colonminus$% dep%(V1, R)%,rand%(R)%,dep%(V2, R)%.
0.8 :: leak%(V1, V2)% %$\colonminus$% share%(V1, V2)%, t%(V1,V2)%.
...
%\textcolor{blue}{\% Query} %
query%(%leak%(\_,\_)%)%.
\end{lstlisting}
    \end{minipage}
 \vspace{-0.1in}
    \caption{Two simple \toolname programs. Statistical correlations  between inputs (in \emph{green}) are specified using the syntax $p:: I \ | \  I_1, I_2$, indicating that probability of  $I$ given $I_1, I_2$ is $p$, and \texttt{\textbackslash+I} indicates the negation of~\texttt{I}.}\label{fig:motivating}
\vspace{-0.2in}
\end{figure}

\paragraph{Medical diagnosis.} The left side of Figure~\ref{fig:motivating} illustrates how a \toolname program can perform probabilistic inference for medical diagnosis. Specifically, it models the probability of a patient having heart disease based on observed symptoms (e.g., chest pain) and test results (e.g., ECG, blood biomarkers).  Lines 12--14 of the \toolname program encode how heart disease can be inferred from various diagnostic factors based on well-established medical literature. These factors include an abnormal ECG (\texttt{abnormal\_ecg}), elevated cardiac biomarkers such as troponin and B-type Natriuretic Peptide (BNP) (\texttt{troponin\_high}, \texttt{high\_bnp}), and the presence of chest pain (\texttt{chest\_pain}).

For example, the rule at line 13 states that the presence of \texttt{heart\_disease} (an output fact) can be deduced based on the input facts  \texttt{abnormal\_ecg} (test result), \texttt{troponin\_high}   (blood work result) and \texttt{chest\_pain} (symptom). The input fact in line 3 represents the patient’s known symptom (chest pain, with probability 1.0), while lines 6-7 encode test results that are currently unknown because the patient has not yet undergone an ECG or blood work. In clinical settings, when test results are unavailable, probabilities for these factors (e.g., \texttt{abnormal\_ecg}) are estimated based on population statistics. However, chest pain is often correlated with other markers of heart disease, such as abnormal ECG findings and elevated troponin levels. This dependence is captured in line 10, where the probability of an abnormal ECG increases to 0.9 given chest pain and high troponin levels. 
However, the full joint probability distribution of these input facts (e.g., symptoms and test results) is typically unknown, as it depends on numerous latent factors, such as medical history or environmental exposures.
The probabilistic reasoning capabilities of \toolname enable accurate estimation of heart disease risk even under such 
{\emph{partially-known} information.}
For example, assuming independence of input facts would lead us to estimate the probability of heart disease given chest pain as 
$0.07\%$ whereas the true probability is in the range $1.5-2.1\%$.\footnote{These probabilities are computed based on the simple \toolname program and do not reflect actual probability of heart disease given chest pain. }

\begin{wrapfigure}{r}{0.25\textwidth} 
    \vspace{-0.06in}
    \centering
\begin{lstlisting}[
    basicstyle=\scriptsize\ttfamily,  % Custom small font size
    style=Prolog-pygsty,
    escapechar=\%, 
    frame=none,
    numbers=left, 
    numberstyle=\tiny\color{blue}, 
    stepnumber=1,
    numbersep=4pt
]
bool mChi%(bool i1 ... i3, bool r1 ... r3)%   {  
bool b1, b2, b3... ; 
b1 = i1 %$\oplus$% r1; b2 = i2 %$\oplus$% r2; 
b3 = i3 %$\oplus$% r3;  n8 = r3 %$\wedge$% r2;
n7 = r3 %$\vee$%  b2;
%\colorbox{mygreen}{\textcolor{white}{n5 = n7 $\oplus$ n8;}}%
... }
\end{lstlisting}
    \vspace{-0.2in}
    \caption{Masked $\chi$ function from MAC-Keccak~\cite{eldib2014synthesis}}
    \label{fig:taintProg}
    \vspace{-0.2in}
\end{wrapfigure}

\paragraph{Quantitative program analysis.} Another motivating scenario for \toolname is quantitative Datalog-based program analysis, which has emerged as a powerful approach for static reasoning about program behavior~\cite{naik2006effective,raghothaman2018user,zhang2017effective, wang2021data, wang2019mitigating}. In this example, we consider a (drastically) simplified version of the Datalog-based side channel detection method described in~\cite{wang2019mitigating}. The \toolname program shown on the right side of Figure~\ref{fig:motivating} encodes an analysis for detecting information leaks caused by power side channels. In particular, the rules in lines 14-15 define how leakage is inferred based on data dependencies in the input program (\texttt{dep(X,Y)}), randomness (\texttt{rand(X)}), and register sharing between variables (\texttt{share(X,Y)}).

The input facts in lines 1--5 of Figure~\ref{fig:motivating} (right) correspond to properties of the analyzed program (see Figure~\ref{fig:taintProg}). Some of these facts, such as those in lines 2 and 4, have probability 1.0 because they correspond to known characteristics of the program. For instance, \texttt{r3} is annotated by the user to be a random variable, and the \texttt{xor} operation in line 4 corresponds to a specific highlighted statement in the source program from Figure~\ref{fig:taintProg}. However, not all input facts are deterministic. Whether two variables share a register, for example, depends on hardware constraints and register allocation policies, which introduce uncertainty. This uncertainty is reflected in line 6, where the probability of register sharing is estimated using empirical data from profiling a code corpus. Similarly, the input facts in lines 7-8 are probabilistic because they result from a pre-analysis~\cite{wang2019mitigating, zhang2018scinfer} that infers semantic data dependencies from syntactic ones. The probability associated with such dependencies is derived from prior empirical studies that measure how often syntactic dependencies lead to actual data dependencies in compiled programs~\cite{wang2019mitigating}.

Since register allocation and data dependencies are influenced by compiler optimizations and architectural constraints, certain input facts are naturally correlated. For example, as shown in line 11, the likelihood of two variables sharing a register is not independent of how frequently the compiler assigns dependent variables {(within the same instruction)} to the same register across different executions~\cite{alfred2007compilers}.  However, as in the medical diagnosis example, assuming access to the full joint probability distribution is impractical, as it would require exhaustively modeling all interactions between hardware configurations, compiler optimizations, etc. Instead, by combining Datalog inference with partially known probabilities (e.g., derived from empirical measurements and power consumption models), we enable \emph{quantitative} static analysis that estimates the \emph{severity} of an information leak rather than merely providing a binary vulnerability classification.

\section{Overview}
\label{motivation}

Before formalizing our technique in detail, we illustrate how \toolname performs probabilistic inference using the synthetic example in Figure~\ref{fig:probCond} that is crafted to give an overview of our approach without being overly complicated. The first six lines declare \emph{input facts}, specifying graph edges, with associated probabilities (e.g., line 1 states that an edge between nodes 5 and 7 exists with probability 0.7). Lines 10-11 are rules that define the notion of paths in the graph. Line 7 states that {\tt edge(1,4)}, {\tt edge(2,5)}, and {\tt edge(2,6)} are statistically correlated, while lines 8-9 specify the known conditional dependencies. For instance, line 8 states that, given {\tt edge(1,4)}, {\tt edge(2,5)} exists with probability 0.8. Finally, line 12 queries the probability of {\tt path(1,7)}. {{We refer to the input facts that are correlated (as declared on line 7) as a \emph{correlation class}. }}

\begin{figure}[t]
    \begin{minipage}[b]{0.05\textwidth}
    \end{minipage}
    \begin{minipage}[b]{0.35\textwidth}
    \centering
    \setcounter{lstlisting}{1}
\begin{lstlisting}[
  	basicstyle=  \footnotesize, %\tiny,or \small or \footnotesize
  	style      = Prolog-pygsty,
        %backgroundcolor=\color{white},
        escapechar=\%,
        frame=none,
        numbers=left, numberstyle=\tiny\color{gray}, stepnumber=1, numbersep= 4pt,
	basicstyle=\footnotesize\ttfamily
    ]
%\colorbox{lightestmain}{0.7::edge(5,7).}% 
%\colorbox{lightestmain}{0.6::edge(1,2).}% 
%\colorbox{lightestmain}{0.8::edge(6,7).}%
%\colorbox{lightestmain}{0.6::edge(2,5).}%
%\colorbox{lightestmain}{0.6::edge(1,4).}%
%\colorbox{lightestmain}{0.6::edge(2,6).}%
%\colorbox{lightmygreen}{corr(edge(1,4),edge(2,5),edge(2,6)).}% 
%\colorbox{lightmygreen}{0.80::edge(2,5) |   edge(1,4).}%
%\colorbox{lightmygreen}{0.83::edge(2,6)  | edge(1,4).}%
%\colorbox{lightestorange}{path(X,Y) $\colonminus$ edge(X,Y).}%
%\colorbox{lightestorange}{path(X,Y) $\colonminus$ path(X,Z), edge(Z,Y).}%
query%(%path%(1,7)%)%.
\end{lstlisting}
    \end{minipage}
    \begin{minipage}[b]{0.59\textwidth}
	\centering
	\scalebox{0.85}{\begin{tikzpicture}[font=\small] 
 \tikzstyle{arrow1}=[->,>=stealth, orange, thick]
 \tikzstyle{arrow2}=[thick,->,>=stealth,darkmain]

 \tikzstyle{medRec}=[%
rectangle, draw, minimum width=1.5cm, minimum height=0.9cm, inner sep = 0cm, outer sep = 0cm, align=center, text width=1.3cm]
\node [rectangle,draw] (P17) at (0,0) { \texttt{path(1,7)} };
\node [rectangle,draw] (P15) at (-2,-1.4) {  \texttt{path(1,5)} };
\node [rectangle,draw,fill=lightestmain] (E57) at (-4,-1.4)  { \texttt{edge(5,7)} };
\node [] (e1) at (-2,-0.7) { $e_1$ };
\node [rectangle,draw] (P16) at (2,-1.4) {  \texttt{path(1,6)} };
\node [rectangle,draw,fill=lightestmain] (E67) at (4,-1.4)  { \texttt{edge(6,7)} };
\node [] (e2) at (2,-0.7) { $e_2$ };
\node [rectangle,draw] (P12) at (0,-2.8) {  \texttt{path(1,2)} };
\node [rectangle,draw,fill=lightestmain] (E25) at (-2,-2.8) {  \texttt{edge(2,5)} };
\node [] (e3) at (-2,-2.1) { $e_3$ }; 
\node [rectangle,draw,fill=lightestmain] (E12) at (0,-4.2) {  \texttt{edge(1,2)} };
\node [] (e4) at (0,-3.5) { $e_4$ }; 

\node [rectangle,draw,fill=lightestmain] (E14) at (0,-4.9) {  \texttt{edge(1,4)} };

\node [rectangle,draw,fill=lightestmain] (E26) at (2,-2.8) {  \texttt{edge(2,6)} };
\node [] (e5) at (2,-2.1) { $e_5$ }; 
\node [] (e6) at (-2,-3.8) { \textcolor{darkerGreen}{\textbf{$e_6$}} }; 
\node [below=-0.2cm of e6] (probE6) {\textcolor{darkerGreen}{\textbf{0.8}}};
\node [] (e7) at (2,-3.8) { \textcolor{darkerGreen}{\textbf{$e_7$}} }; 
\node [below=-0.2cm of e7] (probE7) {\textcolor{darkerGreen}{\textbf{0.83}}};

\node [right=0.1cm of E12] (probE12) {\textcolor{blue}{0.6}};
\node [left=0.1cm of E25] (probE25) {\textcolor{blue}{0.6}};
\node [right=0.1cm of E26] (probE26) {\textcolor{blue}{0.6}};
\node [above=0.1cm of E57] (probE57) {\textcolor{blue}{0.7}};
\node [above=0.1cm of E67] (probE67) {\textcolor{blue}{0.8}};
\node [above right=0.1cm and -0.6 cm of P15] (probP15) {\textcolor{black}{\textbf{[0.36, 0.36]}}};
\node [above left=0.1cm and -0.6 cm of P16] (probP16) {\textcolor{black}{\textbf{[0.36, 0.36]}}};
\node [right=0.1cm of P17] (probP17) {\textcolor{black}{\textbf{[0.34, 0.42]}}};

  \path
        (P17) [-]  edge node { } (e1)
        (e1) [arrow1] edge node { } (E57)
        (e1) [arrow1] edge node { } (P15)
 	 (P17) [-]  edge node { } (e2)
	 (e2) [arrow1] edge node { } (E67)
	  (e2) [arrow1] edge node { } (P16)
 	(P15) [-] edge node {} (e3)	  
	(e3) [arrow1] edge node { } (P12)
	(e3) [arrow1] edge node { } (E25)
 	(P12) [-] edge node {} (e4)	  
	(e4) [arrow1] edge node { } (E12)	
 	(P16) [-] edge node {} (e5)	  
	(e5) [arrow1] edge node { } (P12)
	(e5) [arrow1] edge node { } (E26)	
	;
	 \draw [arrow1, darkerGreen, thick] (-2, -4.9) -- (E14);
	 \draw [-, darkerGreen, thick] (-2, -4.9) -- (probE6);
     \draw [-, darkerGreen, thick] (e6) -- ++(0, 0.7);
	\draw [-, darkerGreen, thick] (e7) -- ($(e7) + (0, .7)$);
        \draw [-, darkerGreen, thick] (2, -4.9) -- (probE7);
        \draw [arrow1, darkerGreen, thick] (+2, -4.9) -- (E14) ;
\

 \end{tikzpicture}}
    \end{minipage}
    \hspace{-2em}    
        \begin{minipage}[t]{.39\textwidth}  
       \vspace{-0.15in}
  	\caption{A \toolname program}  
  	\label{fig:probCond}  
	\end{minipage} 
 	\begin{minipage}[t]{.6\textwidth}
        \vspace{-0.15in}
  	\caption{Derivation graph}  
  	\label{fig:treeOurs}  
        \end{minipage}%
\end{figure}

%
%
%


This example exhibits a salient feature we wish to highlight: Given the information specified by the user, the probability of the predicate \texttt{path(1,7)} cannot be determined \emph{exactly}; instead, we can \emph{only} derive upper and lower bounds. In particular, to compute this probability exactly, the user would  need to give the full {conditional probability table}~\cite{darwiche2009modeling} between the predicates \texttt{edge(1,4),edge(2,5)}, and \texttt{edge(2,6)};  but, in the absence of such information, the only sound conclusion that can be reached is that the probability of \texttt{path(1,7)} being true is in the range $[0.34, 0.42]$. 
%
%
To the best of our knowledge, there is no existing technique that can perform precise and accurate probabilistic inference for this type of scenario.
In the rest of this section, we illustrate how our proposed inference technique addresses this challenge.

\subsection{Basic Approach: Inference via Constrained Optimization} 
\label{ss:overview-basic}

Our starting point  is a formulation of this probabilistic inference task as a {constrained optimization} problem. {{To formulate this optimization problem, our approach introduces a set of \emph{joint probability variables}, each representing the unknown value of an entry in the conditional probability table for a given correlation class.}} {In our example, the predicates \texttt{edge(2,5)},
\texttt{edge(1,4)}, and \texttt{edge(2,6)} form a correlation class, while \texttt{edge(5,7)}, \texttt{edge(6,7)}, and \texttt{edge(1,2)} are mutually independent. For the singleton correlation class containing \texttt{edge(5,7)}, we introduce two joint probability variables, $V_1[1]$ and $V_1[0]$, representing the probabilities of the predicate \texttt{edge(5,7)} being true and false, respectively. Since its probability is given as 0.7, the first constraint in Figure~\ref{fig:opt-example} sets $V_1[1] = 0.7$ and, by the \texttt{SumToOne} constraint, sets $V_1[0] = 0.3$. 
The constraints for the correlation class with predicates \texttt{edge(2,5)},
\texttt{edge(1,4)}, and \texttt{edge(2,6)} are more involved. Since this class contains 3 predicates,
%
it requires 8 joint probability variables, representing all boolean combinations of these predicates. 
For example, the variable $V_4[100]$ in Figure~\ref{fig:opt-example} represents the probability of \texttt{edge(2,5)} being true and \texttt{edge(1,4)} and \texttt{edge(2,6)} being false.
Line 4 of Figure~\ref{fig:probCond} enforces the constraint $V_4[100] + V_4[101] + V_4[110] + V_4[111] = 0.6$, ensuring that the sum of joint probability variables where \texttt{edge(2,5)} holds is 0.6. Similarly, the conditional probability from line 8 induces the constraint $V_4[110] + V_4[111] = 0.48$, capturing $P(\texttt{edge(2,5)} \wedge \texttt{edge(1,4)}) = P(\texttt{edge(2,5)} | \texttt{edge(1,4)}) \times P(\texttt{edge(1,4)})$. Figure~\ref{fig:opt-example} presents all such constraints for the example in Figure~\ref{fig:probCond}.}


\begin{figure}[t]
\footnotesize     
\[
\begin{array}{ll}
\textbf{\texttt{Constraints}} & \\
\colorbox{lightestmain}{\texttt{edge(5,7) = 0.7} } &\mapsto V_1[1] = 0.7 \\
\colorbox{lightestmain}{\texttt{edge(1,2) = 0.6} } &\mapsto  V_2[1] = 0.6 \\
\colorbox{lightestmain}{\texttt{edge(6,7) = 0.8} } &\mapsto  V_3[1] = 0.8 \\
\colorbox{lightestmain}{\texttt{edge(2,5) = 0.6} } &\mapsto  V_4[100]+V_4[101]+V_4[110]+V_4[111] = 0.6 \\
\colorbox{lightestmain}{\texttt{edge(1,4) = 0.6} } &\mapsto  V_4[010]+V_4[011]+V_4[110]+V_4[111] = 0.6 \\
\colorbox{lightestmain}{\texttt{edge(2,6) = 0.6} } &\mapsto  V_4[001]+V_4[011]+V_4[101]+V_4[111] = 0.6 \\
\colorbox{lightmygreen}{0.80 :: \texttt{edge}(2,5) $|$ \texttt{edge}(1,4).} &\mapsto  V_4[110] + V_4[111] = 0.8*0.6 \\
\colorbox{lightmygreen}{0.83 :: \texttt{edge}(2,6) $|$ \texttt{edge}(1,4).}&\mapsto  V_4[011] + V_4[111] = 0.83*0.6 \\
\colorbox{red}{\textcolor{white}{ \texttt{SumToOne}} } &\mapsto  \sum_{b\in\mathbb{B}}V_1[b] = 1 \quad \quad \sum_{b\in\mathbb{B}}V_2[b] = 1 \quad \sum_{b\in\mathbb{B}}V_3[b] = 1 \quad \sum_{b\in\mathbb{B}^3}V_4[b] = 1 \\
\colorbox{red}{\textcolor{white}{ \texttt{Input}} } &\mapsto  
\forall b \in \mathbb{B}, V_1[b] \in [0,1],~V_2[b] \in [0,1],~ V_3[b] \in [0,1], \forall b \in \mathbb{B}^3, V_4[b] \in [0,1] \\
\textbf{\texttt{Objective}} & \\
\colorbox{lightestmain}{\texttt{path(1,7)}} &\mapsto V_1[1]V_2[1]V_3[1](V_4[111]+V_4[101]) + V_1[1]V_2[1]V_3[0](V_4[110]+V_4[100]) + \\
& V_1[1]V_2[1]V_3[1](V_4[110]+V_4[100]) + V_1[1]V_2[1]V_3[0](V_4[111]+V_4[101]) + \\
& V_1[0]V_2[1]V_3[1](V_4[011]+V_4[001]) +V_1[0]V_2[1]V_3[1](V_4[101]+V_4[111]) +\\
&V_1[1]V_2[1]V_3[1](V_4[011]+V_4[001])
\end{array}
\]
\vspace{-0.17in}
\caption{Optimization problem for running example}\label{fig:opt-example}
\vspace{-0.25in}
\end{figure}

 In addition to generating these constraints, our method also expresses the unknown probability of output facts in terms of the joint probability variables using the logical derivation graph of the input program. In particular, Figure~\ref{fig:treeOurs} shows the logical derivation of the output predicate \texttt{path(1,7)} as a graph where nodes correspond to predicates and edges correspond to Datalog rule applications. For instance, according to Figure~\ref{fig:treeOurs}, there are two ways to derive the predicate \texttt{path(1,7)}: one using \texttt{edge(5,7)} and  \texttt{path(1,5)}  and another using \texttt{path(1,6)} and  \texttt{edge(6,7)}. As shown in Figure~\ref{fig:opt-example} under  \texttt{Objective}, our method uses this information to express the probability of predicate \texttt{path(1,7)} being true in terms of the joint probability variables.
 Finally, our method computes the probability bounds for \texttt{path(1,7)} being true by minimizing and maximizing the objective function subject to the constraints shown in Figure~\ref{fig:opt-example}. The bold annotations on the derivation graph in Figure~\ref{fig:treeOurs} show the probability bounds obtained for each relation using our  method. It is worth re-iterating that these probabilities are ranges rather than absolute values \emph{not} because of some imprecision in this solution but rather because of unknown conditional dependencies between some of the input facts.  The details of this optimization approach are presented in Section~\ref{sec:basic-algo}.

\subsection{Scalable $\delta$-exact Inference}  
\label{ss:overview-delta}
\begin{wrapfigure}{r}{0.5\textwidth}
    \vspace{-0.2in}
    \centering
    \includegraphics[width=\linewidth]{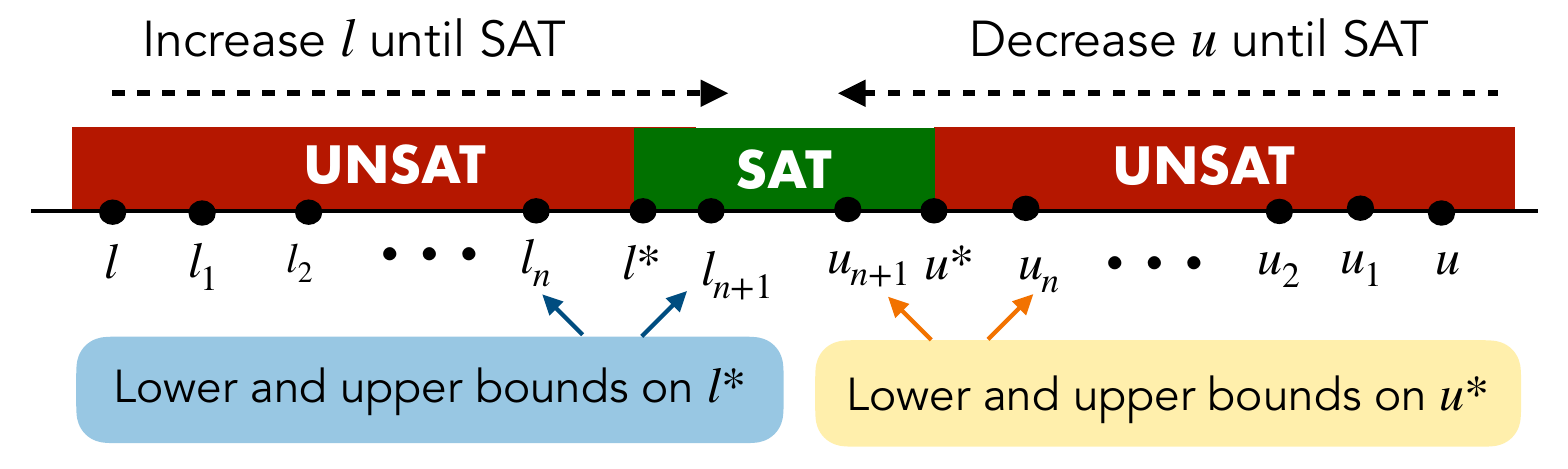}
    \vspace{-0.35in}
    \caption{Iterative strengthening. $l^*, u^*$ are the exact  probability bounds; $l, u$ are \emph{approximate}  bounds from Step 2, and $l_n, u_n$ are bounds computed by our  method.}
    \label{fig:refine}
    \vspace{-0.19in}
\end{wrapfigure}
The  approach introduced above guarantees precise lower and upper bounds, but it can be quite expensive due to the complexity of the resulting constrained optimization problem. Thus, to scale our approach to large Datalog programs with complex statistical correlations, we propose a $\delta$-exact algorithm that first computes (loose) approximate probabilities for each output fact and then iteratively tightens these probabilities until they are within $\delta$ of the true lower and upper bounds. 
Our novel $\delta$-exact algorithm can be reduced to three steps, described next.

\vspace{0.05in}
\noindent
{\bf Step 1: Correlation type inference.} 
 Starting with tighter initial approximations accelerates the overall refinement process, and just knowing the \emph{polarity} of the statistical correlation between predicates can help us compute more precise approximations. 
For example, if two predicates are positively correlated (i.e., one is more likely true if the other is), we can derive tighter bounds than if the correlation type were unknown. Thus, our method first infers \emph{correlation types} between predicates, classifying them as \emph{positive}, \emph{negative}, \emph{independent}, or \emph{unknown}.
To do so, we adapt a simpler version of our constrained optimization method to deduce \emph{how} input facts are correlated. For instance, in  our  example, while the program specifies that  \texttt{edge(2,5)}, \texttt{edge(2,6)}, and \texttt{edge(1,4)} are correlated,  it does not  specify \emph{how}. Using our analysis, we first deduce that  \texttt{edge(2,5)}, \texttt{edge(2,6)}  are positively correlated, which helps establish that  \texttt{path(1,5)}, \texttt{path(1,6)} are \emph{also} positively correlated. We present the details of our correlation analysis method in Section~\ref{sec:corr}.

\vspace{0.05in}
\noindent
{\bf Step 2: Deriving approximate probability bounds.}
Next, our algorithm uses the correlation analysis results to compute sound but approximate bounds on output probabilities. This involves bottom-up propagation of probabilities from the leaves to the roots of the derivation graph, applying pre-defined approximation rules at each internal node.
In our example, to approximate the probability of \texttt{path(1,7)}, the algorithm first computes that \texttt{path(1,5)} and \texttt{path(1,6)} each have a probability of $0.36$ as they consist of two independent events, each with a probability of 0.6. It then moves to \texttt{path(1,7)}, which is true if either \texttt{path(1,5)} $\land$ \texttt{edge(5,7)} (event $e_1$) or \texttt{path(1,6)} $\land$ \texttt{edge(6,7)} (event $e_2$) is true. To compute the exact probability of \texttt{path(1,7)}, the conditional probabilities $P(e_1 | e_2)$ and $P(e_2 | e_1)$ are needed, but are not provided.
In the absence of such information, the Fréchet inequality~\cite{ruschendorf1991frechet} gives the upper bound $P(e_1 \lor e_2) = \textsf{min}(1, 0.54)$. However, by utilizing the information  that \texttt{path(1,5)} and \texttt{path(1,6)} are positively correlated, we can  compute a tighter upper bound of 0.467. Details of this method are in Section~\ref{sec:approxBound}.

\vspace{0.05in}
\noindent
{\bf Step 3: Iterative refinement.}
While sound, the probability bounds from the previous step may be far from the true values. Hence, our method iteratively tightens the bounds until they are within a user-specified distance $\delta$ of the true bounds. The intuition is to gradually adjust the bounds while checking whether the resultant bound is consistent with the constraint system induced by the program. 

As an illustration, consider the output fact \texttt{path(1,7)} in our  example.
Starting from the bounds $[l, u] = [0.288, 0.467]$ computed by the previous step, the iterative refinement step repeatedly attempts to increase $l$ and decrease $u$ by some amount $\epsilon$ until the constraint system becomes satisfiable, as shown in Figure~\ref{fig:refine}. When this step terminates, we know that the the true lower (resp. upper) bound is between $l_n$ and $l_{n+1}$ (resp. between $u_n$ and $u_{n+1}$) in Figure~\ref{fig:refine}. In a second   tightening step, the algorithm performs binary search between $l_n$ and $l_{n+1}$ (and between $u_n$ and $u_{n+1}$)  until the two values are within distance $\delta$ of each other. Thus, upon termination, it can guarantee that the computed lower and upper bounds are always within $\delta$ of the ground truth. 
{Assuming a user-specified \( \delta = 0.05 \) in our running example, this approach refines the initial approximate bounds for \texttt{path(1,7)} from \([0.288, 0.467]\) to \([0.338, 0.417]\), ensuring they lie within \( \delta \) of the true bounds $[0.34, 0.42]$. The details of our iterative refinement procedure are provided in Section~\ref{sec:iterative}.}

\section{Preliminaries}
\label{sec:prelim}


A Datalog program consists of a set of rules $R$, where each rule  is a Horn clause of the form 
$R(\vec{x}) \  \colonminus  \ \odot R_1(\vec{y}_1), \ldots, \odot R_n(\vec{y}_1)$ where $\odot$ denotes an optional negation operator. We refer to $R(\vec{x})$ as the \emph{head} of the rule and the right hand side  as its \emph{body}. Given a rule $r$, we write $\emph{Body}^+$ to denote all predicates without a negation symbol in the front and $\emph{Body}^-$ to denote all predicates that are negated.  Finally, we refer to rules without a body as \emph{input facts}.  As standard, we can obtain a solution to a Datalog program by first grounding all predicates over the Herbrand universe and then applying the ground rules to a fixed-point.
In the rest of this paper, we represent the \emph{solution} to a Datalog program in terms of a \emph{derivation graph}, which shows how each ground output predicate can be derived using ground input predicates.  

\begin{definition}[{\bf Derivation graph}]\label{def:deriv-graph}
A \emph{derivation graph} for a Datalog program  is a  hypergraph $(V, E)$ where $V$ is a set of nodes representing ground predicates, and $E$ is a set of directed hyperedges $(h, B^+, B^-, r)$ representing a ground rule $r$ with head $h$, body $B^+ = \emph{Body}^+(r)$ and $B^- = \emph{Body}^-(r)$. The source vertex set of this hyper-edge is $\{h\}$ and the target vertices are $B^+ \cup B^-$. Given a node $n$, we write $\mathsf{Pred}(n)$ to denote the ground predicate represented by $n$.
\end{definition}

\begin{example}
{In the derivation graph of Figure~\ref{fig:treeOurs}, the edge \( e_3 \) has source node \texttt{path(1,5)} and target nodes \texttt{edge(2,5)} and \texttt{path(1,2)}. The quadruple \( (\texttt{path(1,5)}, \{ \texttt{edge(2,5)}, \texttt{path(1,2)} \}, \emptyset, r_2) \) is used to represent \( e_3 \), where $r_2$ denotes the rule 1 :: \texttt{path(1,5)} $\colonminus$ 
 \texttt{path(1,2)}, \texttt{edge(2,5)}. 
}
\end{example}

Note that different ways of deriving the same ground predicate correspond to multiple hyperedges that share the same source. Given a Datalog program $\dprog$, we write $\dgraph(\dprog)$ to denote its derivation graph. Intuitively, the derivation graph encodes all possible ways of deriving a ground predicate given the input facts.  In the rest of the paper, we assume that the derivation graph is acyclic, as it corresponds to the result of the solver's internal fixed-point computation.\footnote{The derivation graph may contain auxiliary relations that are used for breaking cycles. We refer the interested reader to prior work for details on this semantics-preserving transformation~\cite{fierens2015inference}.}\textit{}

We also define a \emph{flattened} version of the derivation graph that represents each hyperedge with a \emph{set} of regular edges. As we will see later, this flattened representation is useful for understanding basic relationships between output and input predicates.

\begin{definition}[{\bf Flattened derivation graph}]
Given a derivation graph $G = (V,E)$,  $\mathsf{Flatten}(G)$ yields a graph $G' = (V', E')$ where $V' = V$ and there exists an edge $(v, v', \sigma) \in E'$ iff there exists a hyperedge $(v, B^+, B^-, r) \in E$ where $v' \in B^+ \cup B^-$ and $\sigma = 1$ if $v' \in B^+$ and $\sigma= -1$ otherwise. 
\end{definition}

\begin{example}
{Consider hyperedge $e_3$ from Figure~\ref{fig:treeOurs}. This corresponds to two separate edges $e_3^1$ and $e_3^2$ in the flattened derivation graph, where $e_3^1$ is from \texttt{path(1,5)} to \texttt{edge(2,5)} and $e_3^2$ is from \texttt{path(1,5)} to \texttt{path(1,2)}. Since there are no negated predicates, both edges have a $\sigma$ value 1.  }
\end{example}

The flattened derivation graph is useful for determining whether some output facts depend only positively or negatively on an input fact. Given a path $\pi = (v, v_1, \sigma_1), \ldots, (v_n, v', \sigma_n)$ from node $v$ to $v'$ in the flattened derivation graph, we define the polarity of the path as $\Pi_{i=1}^n \sigma_i$, denoted $\polarity(\pi)$. We can then  classify dependencies between different output and input facts as follows:

\vspace{-.25em}
\begin{definition}[{\bf Dependence}]
\label{def:inDependence}
Given a Datalog program $\dprog$ with flattened derivation graph $G$, we say that an output predicate $O$ \emph{depends on} input predicate $I$, denoted $I \dep O$ if there exists a path from $O$ to $I$ in $G$. 
We say that $O$ depends positively (resp. negatively) on $I$ iff there exists a path from $O$ to $I$ with a positive (resp. negative) polarity.
We write $I \pdep O$ to denote (logical) positive dependence and $I \ndep O$ for (logical) negative dependence.  
\end{definition}

The notion of dependence defined above refers to \emph{logical} rather than \emph{statistical} dependence, and it is possible for an output relation to depend both positively \emph{and} negatively on an input fact.

\section{PRALINE: Probabilistic Datalog with Correlated Inputs}
\label{sec:syntax}

In this section, we introduce the syntax of \toolname, our probabilistic Datalog variant that allows correlated input facts. 
A pair of  input facts $I_1, I_2$ are correlated iff $P(I_1 | I_2) \neq P(I_1)$, and a 
\emph{correlation class} represents a set of input facts that may be correlated with each other. 

\vspace{-.25em}
\begin{definition}{\bf (Praline program)} A \toolname program is a tuple $(\classes, \rules, \ruleprob, \inprob)$ where $\classes$ is a set of correlation classes, $R$ is a set of rules whose heads are  output relations, $\ruleprob$ is a mapping from rules in $R$ to their probabilities, and $\inprob$ is a set of  (conditional) probabilities about input facts.
\end{definition}

\vspace{-.5em}
\paragraph{Syntax.} Borrowing notation from prior work~\cite{dries2015problog2,raghothaman2018user}, we express  rule probabilities in \toolname using the notation
$
p:: R_i$ meaning that the probability of rule $R_i \in \rules$ is $p$, {and we assume that rules are statistically independent of each other}. {However, unlike rules, input facts can be statistically dependent.}
We express such conditional dependencies  using the notation $p::  I\ | \ S$, meaning that the probability of input fact $I$ given input fact \emph{set} $S$ is $p$.
When $S$ is the empty set,  such a rule simply declares the probability of an input fact.  
Finally, we  use the notation $\mathsf{Class}(I)$ to denote the correlation class $C_i \in \classes$ that input fact $I$ belongs to, 
and we assume that each input fact $I \in C_i$ has an index, denoted $\mathsf{Index}(I, C_i) \in [1, |C_i|]$, that can be used to uniquely identify $I$ within $C_i$. 

\vspace{-.25em}
\paragraph{Semantics.} We define the semantics of \toolname as an extension of the least-fixed-point semantics of Datalog, incorporating probabilities  through a \emph{possible-worlds interpretation}. In particular, each Herbrand model of a \toolname program represents a deterministic instantiation of probabilistic rules and facts, forming a possible world $\omega$. Given a Praline program $D$, we define an interpretation of $D$ as a pair $(\omega, \mu)$ where $\omega$ is a possible world and $\mu$ is a full joint probability distribution over instantiated rules and input facts in $D$. We say that $(\omega, \mu)$ is a model of $D$, denoted $(\omega, \mu) \models D$ if $\omega$ is a possible world of $D$ (under the standard Datalog semantics) and $\mu$ is consistent with both $\omega$ and the probabilities in $D$.  Then, given such a $\mu$, each possible world $\omega$ has an exact probability associated with it, denoted as $P_\mu(\omega)$. Given an output fact $O$, we can now define $P_\mu(O)$ as follows:
\[
P_\mu(O) = \sum_{\omega \in \Omega } P_\mu(\omega) \ \textrm{where} \ \Omega = \{ \omega \ | \  (\omega, \mu) \models D, {\omega \models O} \}  
\] 
Finally, given an output fact $O$, the \emph{set} of possible probabilities of $O$ is given by:
\[
P(O) =  \{  P_\mu(O) \ | \ \exists \omega. (\omega, \mu) \models D \} 
\]

We refer interested readers to Appendix~\ref{appendix:semantics} in the Supplementary Material for a more formal treatment of the semantics.

\vspace{-.5em}
\paragraph{Problem definition.}  We conclude this section by defining the probabilistic inference problem addressed in the remainder of this paper.

\begin{definition}{\bf (Exact probabilistic inference)} Given a \toolname program $D = (\classes, \rules, \ruleprob, \inprob)$, the goal of exact probabilistic inference is to produce a mapping from each derived output fact $O$ to a probability interval $[l, u]$ such that  $P(O) = \{ x \ | \  l \leq x \leq u\}$. 
\end{definition}

In Section~\ref{sec:basic-algo}, we provide a method for solving the exact probabilistic inference problem defined above. However, since exact inference is often computationally intractable, we also introduce the notion of $\delta$-exact inference, which we address in Section~\ref{sec:delta}:

\begin{definition}{\bf ($\delta$-exact probabilistic inference)} Given a \toolname program $D = (\classes, \rules, \ruleprob, \inprob)$, the goal of $\delta$-exact probabilistic inference is to produce a mapping from each derived output fact $O$ to a probability interval $[l, u]$ such that $l \geq l^* - \delta$ and $u \leq u^*+ \delta$ where $P(O) = \{ x \ | \ l^*  \leq x \leq u^* \}$.
\end{definition}

\label{sec:language}

\section{Exact Probabilistic Inference via Constrained Optimization}
\label{sec:basic-algo}

\begin{algorithm}[t]
\caption{\textsc{Solve}($D$)}
\label{alg:baseline:opt}
{\footnotesize
\begin{algorithmic}[1]
\Input  Datalog program $D = (\classes, R, P_R, P_I)$ 
\Output Mapping $M$ from output relations to their probability intervals


\State $S_o, G ~\leftarrow~ \textsc{SolveStandard}(D)$
\Comment{\textcolor{mygray}{\textsc{SolveStandard} produces  output relations $S_o$ and  derivation graph $G$}}
\State $ V \leftarrow$ \textsc{JointProbabilityVars}($D$) 
\Comment{\textcolor{mygray}{Introduce joint probability variables,  as discussed in Section~\ref{sec:vars}}}
\State $ \exptemplate \leftarrow$ \textsc{GenExprTemplate}($V$) 
\Comment{\textcolor{mygray}{Generate probability expression templates as discussed in Section~\ref{sec:template}}}

\State $\phi~\leftarrow~$\textsc{GenConstraint}($V, D, \exptemplate$)
\Comment{\textcolor{mygray}{Generate constraints $\phi$ as discussed in Section~\ref{sec:constraint}}}

\For {$R_o \in S_o$}
\State $\Phi ~\leftarrow~\textsc{GenObjective}(R_o, V, G, \exptemplate)$
\Comment{\textcolor{mygray}{Generate objective $\Phi$ as discussed in Section~\ref{sec:obj}}}
	
\State $M[R_o] \leftarrow$ \textsc{Optimize}($\Phi, \phi$)
\Comment{\textcolor{mygray}{Use  MIP solver to minimize/maximize $\Phi$ subject to $\phi$}}
\EndFor
\State \textbf{return} $M$

\end{algorithmic}

}
\end{algorithm}

In this section, we address the \emph{exact} probabilistic inference problem using  constrained optimization. As summarized in  Algorithm~\ref{alg:baseline:opt}, our method consists of five steps. First, for each correlation class $C_i$, it introduces a set of variables that express the joint probabilities of input facts in that class $C_i$ (line 2). 
Second, at line 3, it generates a \emph{probability expression template} $\exptemplate$ over these joint probability variables --- the key idea is to express the probabilities of all relations symbolically as instantiations of the same shared template. Then, at line 4, it invokes {\sc GenConstraint} to express requirements on the joint probability variables based on the facts in the Datalog program. Finally, for each fact $R_o$, the algorithm  first invokes
{\sc GenObjective} (line 6) to express the probability of $R_o$ symbolically as an instantiation $\Phi$ of $\exptemplate$ and then optimizes $\Phi$ (line 7) subject to the constraints generated earlier.


\subsection{Joint Probability Variables}\label{sec:vars}

The variables  in our encoding represent \emph{joint probabilities} over input facts; hence, we refer to them as \emph{joint probability variables}. For each correlation class $C$, we introduce a map $V_C$ of joint probability variables, denoted $\repvar(C)$ (for ``representative''). Given a correlation class $C$ of size $n$, $V_C$ contains $2^n$ variables, one for each possible boolean assignment  to  input facts in $C$. We use bitvectors to represent boolean assignments and write $V_C[b]$ to denote the corresponding joint probability variable. For example, for $C = \{I_1, I_2, I_3 \} $, $V_C[001]$ denotes the joint probability of $I_1, I_2$ being false and $I_3$ being true. We also write $\mathbb{B}^n$ to denote the set of all bitvectors of size $n$. Given an input predicate $I$ and a joint probability variable $v$, $v\models I$ (resp. $v \not \models I$) indicates that $v$ represents an event in which $I$ is true (resp. false). In our example, we have $ V_C[001] \models I_3$ and $ V_C[001] \not \models I_1$.

\subsection{Probability Expressions and  Templates}\label{sec:template}
As explained earlier, a key idea underlying our algorithm is to express the probability of each output relation as a symbolic expression over the joint probability variables. Because all of these expressions are instantiations of the same template, we first explain what these templates look like.

\begin{definition}{\bf (Product term)}\label{def:pterm} Let $C_1 \ldots C_n$ be the set of all correlation classes, i.e., $C_i \in \mathcal{C}$. A product term $\pterm$ is a product of joint probability variables $v_1 \ldots v_n$ where each $v_i$ is a joint probability variable associated with class $C_i$, i.e., $v_i = \mathsf{Rep}(C_i)[b]$ for some bitvector $b$.
\end{definition}
Intuitively, a product term represents the probability of a particular truth assignment to \emph{all} input relations in the program. For example, if we have two correlation classes $C_1 = \{I_1, I_2\}$ and $C_2 = \{ I_3, I_4\}$, then the product term $\psi = V_1[10] \times V_2[11]$ represents the probability of  $I_2$ being false and $I_1, I_3, I_4$ being true. Extending our previous notation, given a product term $\pterm$ and input relation $I$, we write $\pterm \models I $ (resp. $\pterm \not \models I$) to denote that $\pterm$ represents the probability of an event in which $I$ is true (resp. false). For instance, in our previous example, we have $\pterm \models I_1$ and $\pterm \not \models I_2$. 
%

\begin{definition}{\bf (Probability expression template)} 
\label{def:exprTemplate}
Let $\Psi$ be the set of all possible product terms. A \emph{probability expression template} $\exptemplate$ is a sum-of-product expression of the form
$
\sum_{\pterm_i \in \Psi}  \square_i \times \pterm_i
$.
\end{definition} 

For instance, if we have two correlation classes each with a single input fact, the probability expression template would be of the form:
\[ \hole_1 \times V_1[0]V_2[0] + \hole_2 \times V_1[0]V_2[1] + \hole_3 \times  V_1[1] V_2[0] + \hole_4 \times V_1[1] V_2[1] \] 
Intuitively, the probability of every relation in the program can be expressed symbolically as an instantiation of a probability expression template, where holes $\square$ are filled by \emph{coefficient terms} $\lambda$:

\begin{definition}{\bf (Coefficient term)}\label{def:coef-term} A coefficient term $\cterm$ is an expression of the form
$
\sum_i  \lambda_i^+ \times \lambda_i^-
$
where $\cterm_i^+ = \ruleprobvar_1 \times \ldots \times \ruleprobvar_n$ and $\cterm_i^- = (1-\ruleprobvar'_1) \times \ldots \times (1-\ruleprobvar'_k)$  and each $\ruleprobvar_j, \ruleprobvar'_j$ is a \emph{variable} representing the probability of some rules in the Datalog program.
\end{definition}

A coefficient term represents the probability of an output fact being derived by applying a specific sequence of rules. However, since rule bodies  can include negations, the derivation of some facts may depend on certain rules \emph{not} being applied. As a result, coefficient terms also incorporate factors of the form $(1 - r)$, which represent the probability of a rule \emph{not} being applied. 



\begin{definition}{\bf (Template instantiation)}\label{def:temp-inst} An \emph{instantiation} of template $\exptemplate$ either replaces each hole with a $c \in \{0,1\}$ or with a coefficient term $\cterm$. Given a vector of expressions $\sigma$, we use the notation $\exptemplate[\sigma]$ to denote a probability  expression that is obtained by filling hole $\hole_i$ in $\exptemplate$ with expression $\sigma_i$.
\end{definition}

Intuitively, template instantiations that fill holes with coefficient terms represent probabilities of output facts whereas those that use constants represent probabilities of input relations. Given a relation $R$, \textsf{Expr}($R$) denotes the symbolic expression representing the probability of $R$ and is obtained through some instantiation $\sigma$ of the holes in $\exptemplate$ --- i.e.,  $\mathsf{Expr}(R) = \exptemplate[\sigma]$ for some $\sigma$.

\begin{definition}{\bf (Input expression)}\label{def:input-expr}
Let $\exptemplate$ be  a template with $\pterm_i$ as its $i$'th  term. The \emph{probability expression} for input fact $I$, denoted $\mathsf{Expr}(I)$, is given by ${\sigma_i} = 1$ if $ \psi_i \models I$ and ${\sigma_i} = 0$ otherwise.

\end{definition}
In other words, the symbolic expression for an input relation is obtained by choosing $1$ as the coefficient of a product term $\pterm_i$ if $\pterm_i$ represents the probability of an event in which $I$ is true and $0$ otherwise. It is easy to see that $\mathsf{Expr}(I)$ symbolically represents the probability of an input relation in terms of the joint probability variables.

\begin{figure*}
{\small
$
\begin{array}{llll}
     \llbracket D \rrbracket &= &\bigwedge_{C_i \in \mathcal{C}} \llbracket {C_i} \rrbracket 
    ~\land~ \bigwedge_{I_r \in P_I}  \llbracket I_r \rrbracket 
    ~\land~ \bigwedge_{r \in R} \textsf{ProbVar}(r) = P_R(r) & 
    \\
    \llbracket {C} \rrbracket &= & \forall  b \in \mathbb{B}^{|C|}. ~~ \textsf{Rep}(C)[b] \in [0, 1]  ~\land~ \sum_{{b \in \mathbb{B}^{|C|}}} \textsf{Rep}(C)[b] = 1   & \\
    \llbracket I_r \rrbracket &= &\textsf{Expr}(I)  = p  &\textsf{if} \quad  I_r = (p:: I | \varnothing) \\
     \llbracket I_r \rrbracket &= & \otimes_{i=0}^n \textsf{Expr}(I_i) = p \times \otimes_{i=1}^n \textsf{Expr}(I_i) &\textsf{if} \quad I_r = (p :: I_0 \ | \ I_1, \ldots, I_n)
\end{array}
$
}
\caption{Rules for producing constraints from a Datalog program $D = (\mathcal{C}, R, P_R, P_I)$. \textsf{ProbVar}(r) is a fresh variable representing the probability of rule $r$.
}
\label{fig:constraint-gen}
\end{figure*}

\subsection{Constraint Generation}\label{sec:constraint}

We now turn to the \texttt{{\sc GenConstraint}} procedure used in Algorithm~\ref{alg:baseline:opt}, with its implementation summarized in Figure~\ref{fig:constraint-gen}. Given a program \( D = (\mathcal{C}, R, P_R, P_I) \), this procedure generates three types of constraints. First, for each correlation class \( C_i \) with representative \( V_i \), it introduces a constraint ensuring that all variables in \( V_i \) are valid probabilities in the range \([0, 1]\) and that they sum to 1 (line 2 of Figure~\ref{fig:constraint-gen}). Second, the procedure introduces constraints to encode the (conditional) input probabilities in the Datalog program. Specifically, for each rule of the form \( p :: I \), a constraint is introduced stating that \( \mathsf{Expr}(I) = p \), where \( \mathsf{Expr}(I) \) is defined in Definition~\ref{def:input-expr}. Additionally, for conditional probability declarations of the form \( p :: I_0 | I_1 \ldots I_n \) (line 4 of Figure~\ref{fig:constraint-gen}), a constraint is introduced stating that \( P(I_0 \land \ldots \land I_n) = p \times P(I_1 \land \ldots \land I_n) \). Note that the probability \( P(I_0 \land \ldots \land I_n) \) is computed using a special multiplication operator \( \otimes \), which is explained in the next section. Finally, the procedure generates a third type of constraint that relates variables in the coefficient terms to the actual rule probabilities given by \( P_R \) (the last conjunct in the first line). 

\subsection{Generating Optimization Objective}\label{sec:obj}

In this final subsection, we focus on the \texttt{{\sc GenObjective}} procedure. As a reminder, the purpose of this procedure is to express the probability of each output relation as a symbolic expression over the joint probability variables—i.e., as an instantiation of the probability expression template defined in Definition~\ref{def:temp-inst}. The key idea behind this procedure is to use the logical derivation graph $G$ to find a coefficient term that can be used to fill each hole in the expression template. In particular, the algorithm performs bottom-up traversal of the derivation graph to construct a symbolic probability expression of each node, utilizing the probability expressions of its children. 

\begin{figure*}
{\footnotesize

\begin{mathpar}
\inferrule*[right=(\textsc{Leaf})]{
v \in \mathsf{Leaves}(G) \quad \mathsf{Pred}(v) = I
}
{ G \vdash v \hookrightarrow \mathsf{Expr}(I) }
\end{mathpar}
\vspace{-0.5em}
\begin{mathpar}
{\footnotesize
\inferrule*[right=(\textsc{Node})]{
\mathsf{OutgoingEdges}(G, v) = \{e_1, \ldots, e_n \}
\quad \quad 
G \vdash e_i \hookrightarrow E_i \ \mathsf{for} \ i \in [1, n] \quad \quad
E_1 \oplus  \dots \oplus E_{n}
 \rightsquigarrow E
}
{G \vdash v  \hookrightarrow E} 
}
\end{mathpar}
\vspace{-0.5em}
\begin{mathpar}
{\footnotesize
\inferrule*[right=(\textsc{Edge})]{
 B^+ = \{v_1^+, \dots v_m^+ \}  \quad
  G \vdash v_i^+  \hookrightarrow E_i^+ \ ~\mathsf{for}~  \ i \in [1, m] \quad \quad 
 B^- = \{v_1^-,  \dots v_n^-\} \quad
\quad   G \vdash v_i^- \hookrightarrow E_i^- \ ~\mathsf{for}~  \ i \in [1, n] ~\\\\
 (E^+_1 \otimes  \dots \otimes E^+_m) \rightsquigarrow E^+  \quad \quad (\ominus(E^-_1)) \otimes \dots \otimes ~(\ominus(E^-_n))
 \rightsquigarrow E^- \quad E^+ \otimes E^- \rightsquigarrow E
}
{ {G \vdash (\_, B^+, B^-, r)  \hookrightarrow \mathsf{ProbVar}(r) 
\times E}}
}
\end{mathpar}
}
\caption{Rules for inferring symbolic probability expression for output relations. 
%
}
\label{fig:objectivePre}
\end{figure*}

Our bottom-up traversal algorithm is presented in Figure~\ref{fig:objectivePre}. The base case is the rule labeled \texttt{{\sc Leaf}}, which corresponds to input relations. Since we already know how to compute the probability expression for an input relation (see Definition~\ref{def:input-expr}), the \texttt{{\sc Leaf}} rule simply produces \( \mathsf{Expr}(I) \), where \( I \) is the input relation represented by the node \( v \).
Next, the rule labeled \texttt{{\sc Node}} describes how to compute the probability expression for an internal node \( v \) representing an output relation. If a node labeled \( O \) has \( n \) outgoing edges, this means \( O \) can be derived in \( n \) different ways. Therefore, the rule first computes a probability expression \( E_i \) for each edge \( e_i \), representing the probability of a possible derivation of \( O \). It then computes the probability of \( P(e_1 \lor \ldots \lor e_n) \) using a special \( \oplus \) operator, which we explain later.
Finally, the last rule, labeled \texttt{{\sc Edge}}, computes the probability expression for a derivation. Recall that a hyperedge with source \( v \) and target nodes \( B^+ \) and \( B^- \) represents the application of a rule \( r \), whose body consists of positive facts \( B^+ \) and negative facts \( B^- \). This rule first computes the probability expressions \( E_i^+ \) for each \( v_i^+ \in B^+ \) and \( E_i^- \) for each \( v_i^- \in B^- \), respectively. It then combines these to obtain a probability expression for the edge. 

To combine these expressions, we first note that the probability expression for \( \neg R \) is given by \( \ominus E \), where \( E \) denotes the probability expression for \( R \), i.e., \( E = \mathsf{Expr}(R) \). Next, we define an operator \( \otimes \) (explained later) to compute the probability of \( P(v_1 \land v_2) \) as \( P(v_1) \otimes P(v_2) \). Therefore, the \texttt{{\sc Edge}} rule first computes the probability of all positive facts being true as \( E^+ = E_1^+ \otimes \ldots \otimes E_m^+ \), and the probability of all negative facts being true as \( E^- = \ominus(E_1^-) \otimes \ldots \otimes \ominus(E_n^-) \). The probability of both positive and negative facts being satisfied is then given by \( E = E^+ \otimes E^- \). Finally, since the rule itself has a probability, the final probability expression is obtained as \( \mathsf{ProbVar}(r) \times E \), where \( \mathsf{ProbVar}(r) \) is the variable representing the probability of rule \( r \).

\begin{figure*}

\begin{mathpar}
{
\hspace{-4ex}

\footnotesize
\inferrule*[right=(\textsc{Neg})]
 {
E = \sum \cterm_i \pterm_i
 }
 {
\ominus(E) \rightsquigarrow \sum (1-\cterm_i)\pterm_i
 } 
}

\hspace{-7ex}

{\footnotesize
\inferrule*[right=(\textsc{Mul})]{
E_1 = \sum \cterm_{1i}\pterm_i \quad 
E_2 = \sum \cterm_{2i}\pterm_i  ~\\\\
{\mathsf{JointProb}(\cterm_{1i}, \cterm_{2i})} \rightsquigarrow  \cterm_i
}
{
\vdash E_1 \otimes E_2  \rightsquigarrow \sum  \cterm_i \pterm_i
}

\hspace{-11ex} 

\inferrule*[right=(\textsc{Add})]{
E_1 = \sum \cterm_{1i}\pterm_i \quad 
E_2 = \sum \cterm_{2i}\pterm_i ~\\\\
{\mathsf{JointProb}(\cterm_{1i}, \cterm_{2i})} \rightsquigarrow \cterm_i
}
{
 E_1 \oplus E_2  \rightsquigarrow  \sum (\cterm_{1i} + \cterm_{2i} - \cterm_i)\pterm_i
}  

}
\end{mathpar}

\vspace{-.5em}

\begin{mathpar}
{\footnotesize
\hspace{-20ex}

%

\footnotesize
\inferrule*[right=(\textsc{Joint})]
 {
 \cterm_1 = \sum\nolimits_{i=1}^n e_i \quad
 \cterm_2 = \sum\nolimits_{j=1}^m e_j \quad
{\forall (i,j).~e_i \diamond e_j \rightsquigarrow e_{ij}}
 }
 {
{
 \textsf{JointProb}(\cterm_1, \cterm_2) }\rightsquigarrow \sum\nolimits_{(i, j) = 1}^{(n,m)}  e_{ij}
 } 
}

\hspace{-8ex}

\footnotesize
\inferrule*[right=($\diamond$)]
 {
e_i = X_i^+  \times  X_i^- \quad     
e_j =  Y_j^+ \times Y_j^- ~\\\\
X_i^+ \blacklozenge Y_j^+ \rightsquigarrow Z^+   \quad
X_i^- \blacklozenge Y_j^- \rightsquigarrow Z^-
 }
 {
e_i \diamond e_j \rightsquigarrow  \textsf{Disjoint}(X_i, Y_j) ~?~ Z^+ Z^- ~:~ 0 
 } 
\end{mathpar}
\vfill
\vspace{-0.2in}
\begin{mathpar}
{
\footnotesize

\footnotesize
 \inferrule*[right=($\blacklozenge$)]
 {
 \vspace{0.05in}
 }
 {
 \prod\nolimits_{i=1}^p  \bigcirc (x_i) ~\blacklozenge~
 \prod\nolimits_{j=1}^q  \bigcirc (y_j) \rightsquigarrow  \prod\nolimits_{z_k \in (\bigcup_{i=1}^p x_i ~\cup~ \bigcup_{j=1}^q y_j)} \bigcirc(z_k) 
 } 
}
\end{mathpar}
\vfill
\caption{Rules  defining $\ominus, \otimes$ and $\oplus$ operations.  \textsf{Disjoint}$(X_i, Y_j)$ is true iff $\mathsf{Vars}(X_i^+) \cap \mathsf{Vars}(Y_j^-) = \emptyset ~\wedge~ \mathsf{Vars}(X_i^-) \cap \mathsf{Vars}(Y_j^+) = \emptyset$. 
$\mathsf{Vars}(X)$ denotes the set of rule probability variables present in $X$.
In the $\blacklozenge$ rule, $\bigcirc(x)$ represents a term of the form $x$ or $1-x$. }
\label{fig:objectiveRule}
\end{figure*}

Finally, we turn our attention to the definitions of the $ \ominus $, $ \otimes $, and $ \oplus $ operators from Figure~\ref{fig:objectiveRule}, which are used to compute probability expressions for conjunctions and disjunctions of events. All of these rules rely on the fact that probability expressions are in a normalized form, consisting of sums of terms of the form $ \cterm \times \pterm $, where $ \cterm $ is a coefficient term (Def.~\ref{def:coef-term}) and $ \pterm $ is a product term (Def.~\ref{def:pterm}).
%
%
First, given a symbolic expression $ E $ representing the probability of some relation $ R $, the \texttt{{\sc Neg}} rule computes the probability of $ \neg R $ by simply replacing all coefficient terms $ \cterm $ with $ 1 - \cterm $. Second, the \texttt{{\sc Mul}} rule computes the probability of a conjunction of events by combining the coefficients of each product term using the \textsf{JointProb} function (explained later). Similarly, the \texttt{{\sc Add}} rule computes the probability of a disjunction of events by updating the coefficient of each term $ \pterm_i $ as $ \cterm_{1i} + \cterm_{2i} - \textsf{JointProb}(\cterm_{1i}, \cterm_{2i}) $. Intuitively, this corresponds to an application of the inclusion-exclusion principle.

Next, we focus on the last three rules for computing the joint probability of two coefficient terms $ \cterm_1 $ and $ \cterm_2 $. The rule labeled \textsf{{\sc Joint}} essentially distributes multiplication over addition, as we require each coefficient term to be a sum of products. The second rule, labeled $\diamond$, considers multiplication expressions of the form $ (X_i^+ \times X_i^-) $ and $ (Y_i^+ \times Y_i^-) $. In this rule, $ Z^+ $ denotes a product of variables $ v $, and $ Z^- $ represents a product of terms of the form $ 1 - v $, where $ v $ represents a rule probability. Intuitively, $ Z^+ $ indicates the rules that must be applied to derive a fact, while $ Z^- $ indicates the rules that must not be applied. 
Thus, if a variable $ v $ appears in $ X_i^+ $ (or $ Y_i^+ $) and $ 1 - v $ appears in $ Y_j^- $ (or $ X_i^- $), this results in a contradiction, and the probability of the term is zero. Otherwise, if the disjointness condition is satisfied, the probability is computed using the rule labeled $ \blacklozenge $. The intuition behind the $ \blacklozenge $ rule is as follows: If a variable $ v $ appears in both $ X_i $ and $ Y_i $, it indicates that the same fact is derived using the same rule. To avoid overcounting, we do not multiply probabilities repeatedly, as that rule only needs to be applied once. Therefore, the $ \blacklozenge $ rule multiplies probabilities after ensuring that variables associated with the same rule are treated uniquely.

\vspace{-.5em}
\begin{example}
\label{example:baseline}
Consider the \textsc{Praline} program on the left side of Figure~\ref{fig:baseline}, with the corresponding probability expressions for both input  and output facts displayed on the right. In this example,  $I_1$ and $I_2$ are correlated, so they belong to the same correlation class $V$ --- e.g., $V[01]$ represents the joint probability of $I_1$ being false and $I_2$ being true. The right side of Figure~\ref{fig:baseline} shows the symbolic probability expressions for each  fact.  The parts  highlighted in red show the need for the disjointness check in the $\diamond$ rule; and the parts highlighted in blue illustrate the need for the $\blacklozenge$ operator. 

\end{example}

\begin{figure}[t]
\centering
{
\footnotesize
$
\begin{array}{l|ll}
    p_{1}::I_1 & Pr(I_1) &= 0*V[00] + 0 * V[01] + 1*V[10] + 1*V[11]  
    \\
    p_{2}::I_2 & Pr(I_2) &= 0*V[00] + 1 * V[01] + 0*V[10] + 1*V[11]  
    \\
    p_3:: I_2 | I_1 & &
    \\
    r_1:: A \colonminus I_1 & Pr(A) &= 0*V[00] + 0 * V[01] + r_1*V[10] + r_1*V[11] 
    \\
    r_2:: B \colonminus A & Pr(B) &= 0*V[00] +  0*V[01] + r_1r_2*V[10] + r_1r_2 V[11]
    \\
    r_3:: C \colonminus \text{\textbackslash+} A, I_2 & Pr(\text{\textbackslash+} A) &= 1* V[00] + 1*V[01] + (1-r_1) V[10] + (1-r_1) V[11]
    \\
    & Pr(C) &=r_3 \times (Pr(\text{\textbackslash+} A) \otimes Pr(I_2)) = 0*V[00] + r_3* V[01] + 0* V[10] + r_3(1-r_1)V[11] 
    \\
    r_4:: D \colonminus B, A & Pr(D) &= r_4 \times (Pr(B) \otimes Pr(A)) = 0*V[00] + 0*V[01] + \textcolor{blue}{r_1r_2r_4} V[10] + \textcolor{blue}{r_1r_2r_4}V[11] 
    \\
    r_5::E \colonminus C & Pr(E_1) &= 0*V[00] + r_3r_5*V[01] + 0*V[10] + r_3r_5(1-r_1) V[11]
    \\
    r_6:: E \colonminus D & Pr(E_2) & = 0*V[00] + 0*V[01] + r_1r_2r_4r_6 V[10] + r_1r_2r_4r_6 V[11]
    \\
    & Pr(E) &= Pr(E_1) \oplus Pr(E_2) 
    \\
    & &= 0*V[00] + r_3r_5 V[01] + r_1r_2r_4r_6 V[10] + (r_3r_5\textcolor{red}{(1-r_1)} + \textcolor{red}{r_1}r_2r_4r_6 - \textcolor{red}{0}) V[11]
\end{array}
$
}
\vspace{-.5em}
    \caption{Left: \textsc{Praline} program. Right: Probability expressions for both input and output facts.}
    \label{fig:baseline}
\end{figure}

\vspace{-1em}
\begin{theorem}
\label{theorem:AndOr}
Let $E_1$ and $E_2$  denote the probability expressions of events $A$ and $B$ respectively. Then, we have: (1) If  $\ominus E_1 \rightsquigarrow E$, then $E$ represents the probability of $\neg A$; (1) if \ $ E_1 \oplus E_2 \rightsquigarrow E$, then $E$ represents the probability of event $A \lor B$
(3) iff \  $ E_1 \otimes E_2 \rightsquigarrow E$,then $E$ represents the probability of event $A \land B$

\end{theorem}

\section{$\delta$-Exact Probabilistic Inference}
\label{sec:delta}


Building on our approach from Section~\ref{sec:basic-algo}, we now present a more practical $\delta$-exact probabilistic inference algorithm,  summarized in Algorithm~\ref{alg:SolveDelta}.
Similar to the previous algorithm, it first uses the underlying Datalog solver to obtain a derivation graph $G$. Next, it invokes the {\sc InferCorrelationType} function to compute the \emph{type} of the statistical correlation between 
facts, where a correlation type is either positive, negative,  independent, or unknown. 
In the third step, the algorithm uses this correlation type environment $\Sigma$ to derive \emph{approximate} probability bounds on output facts. 
Finally, the call to {\sc MakeDeltaPrecise} at line 5 keeps refining the inferred bounds until the derived bounds are within $\delta$ of the ground truth. 

\begin{algorithm}[t]
\caption{\textsc{Solve$^{\pm\delta}$}($D, \delta$)}
\label{alg:SolveDelta}
{\footnotesize
\begin{algorithmic}[1]
\Input  Datalog program $D$, precision bound $\delta$
\Output Mapping  $M$ that maps output facts to probabilities
\vspace{0.03in}
\State $S_o,G ~\leftarrow~ \texttt{SolveStandard}(D)$ \Comment{\textcolor{mygray}{\texttt{SolveStandard} produces  output facts $S_o$ and  derivation graph $G$}}
\State $\phi \leftarrow$  \textsc{GenConstraints}($D, G$)
 \Comment{\textcolor{mygray}{Use technique from Section~\ref{sec:constraint} to generate constraints}}
\State $\Sigma \leftarrow $ \textsc{InferCorrelationType}($G, \phi$) \Comment{\textcolor{mygray}{Infer type of statistical correlation between 
facts}}
\State $M \leftarrow $ \textsc{DeriveApproximateBounds}($G, D, \Sigma$) \Comment{\textcolor{mygray}{Approximate bounds using static analysis and constraint solving}}
\State $M' \leftarrow \textsc{MakeDeltaPrecise}(M, S_o, \phi, \delta)$ \Comment{\textcolor{mygray}{Iterative refinement of bounds}}

\vspace{0.02in}
\State \Return $M'$

\end{algorithmic}

}
\end{algorithm}

\subsection{Inference of Correlation Types}
\label{sec:corr}

In this section, we present the {\sc InferCorrelationType} algorithm that can be used to infer whether a pair of relations are positively/negatively correlated or whether they are independent.

\begin{definition}{\bf (Statistical correlation)}
Two events $X$ and $Y$ are positively  (resp. negatively) correlated if $P(X|Y) > P(X)$ (resp. $P(X | Y) < P(X)$). If $P(X | Y) = P(X)$, then $X$ and $Y$ are independent.
\end{definition}

Note that the notion of statistical correlation is symmetric. That is, if $X$ is positively correlated with $Y$, then $Y$ is also positively correlated with $X$. 
A \emph{correlation type} for a pair of Datalog facts is one of $\mathsf{Pos}~(+), \mathsf{Neg}~(-), \bot, \top $, where $\bot, \top$ indicate independence and  unknown correlation respectively. 
As stated earlier, identifying  correlation types allows us to derive tighter approximate bounds than would otherwise be possible. 
To infer these correlation types, our method proceeds in two phases: First, it infers statistical correlations between input facts. In the second phase, it uses this information to infer statistical correlations between outputs. \\

\begin{figure*}
\vspace{-0.1in}
{\small
\begin{mathpar}
{\footnotesize
\inferrule*[right=(\textsc{Id})]
{
p::(I | \varnothing) \in {P_I}
\quad p < 1
}
{D, \phi \vdash I \pdepin I} 
\hspace{2ex}
\inferrule*[right=(\textsc{Indep})]
{
\depClass(I_1) \neq \depClass(I_2)
}
{D, \phi \vdash I_1 \indepin I_2} 
\hspace{2ex}
\inferrule*[right=(\textsc{Symm})]
{
I_1 \blacktriangleright^\star I_2 \ \ ( \star \in \{ +, -, \bot \})
}
{D, \phi \vdash I_2 \blacktriangleright^\star I_1} 
}
\end{mathpar}
}


\begin{mathpar}
{\small
\inferrule*[right=(\textsc{Semantic})]
{
 E_1 = \mathsf{Expr}(I_1) \quad E_2 = \mathsf{Expr}(I_2) \ \quad   E_1 \otimes E_2 \rightsquigarrow E_\land \quad 
 \models \phi \Rightarrow E_\land \boxdot E_1 \times E_2 \quad  \star = \mathsf{Sign}(\boxdot)
 }
{D, \phi \vdash I_1 \blacktriangleright^{\star} I_2}
}
\end{mathpar}
\caption{Inference of statistical correlation between input variables. Here, $\boxdot \in \{ <, >, =\}$, and 
$\mathsf{Sign}(\boxdot)$ yields $+$ for $>$, $-$ for $<$, and $\bot$ for $=$.  }
\label{fig:corr-input}
\vspace{-0.1in}
\end{figure*}

\noindent
{\bf Phase 1: Inferring correlation types between input facts.} Figure~\ref{fig:corr-input} presents our method for inferring correlation types for pairs of input facts using the judgment 
$\
D, \phi \vdash I_1 \blacktriangleright^\star I_2
$
where $\star \in \{ +, -, \bot \}$ indicates positive and negation correlation and statistical independence respectively. Here, $D$ is the Datalog program and $\phi$ is the set of constraints generated from the Datalog program as described in Section~\ref{sec:constraint}.   The first rule, labeled \textsc{Id} indicates that the input fact $I$ is positively correlated with itself.
The next rule, labeled {\textsc{Indep}} applies to two input facts that do not belong to the same correlation class. 
Finally, the last rule labeled {\sc Semantic} uses the constraint-based technique from Section~\ref{sec:basic-algo} to check for statistical correlation. 
The basic idea is to use the algorithm of Section~\ref{sec:basic-algo} to check  how the joint probability of $I_1 \land I_2$ relates to the product of the individual probabilities of $I_1$ and $I_2$. To do so, it first generates symbolic expressions $E_1, E_2, E_\land$ for $I_1, I_2$, and $I_1 \land I_2$ respectively. Then, it uses a solver to check whether the constraints $\phi$ (encoding the input probabilities) logically imply whether $E_\land \boxdot E_1 \times E_2$, where $\boxdot$ denotes one of $<, >, =$. Note that this semantic approach is not as susceptible to the scalability challenges discussed earlier 
because we  consider \emph{only} input facts and \emph{only} those that belong to the same correlation class. 

\begin{theorem}
\label{thm:input}
Suppose that we derive $I_1 \blacktriangleright^\star I_2$ using the rules from Figure~\ref{fig:corr-input}. Then, $p(I_1 | I_2) > p(I_1)$ if $\star = +$, $p(I_1 | I_2) < p(I_1)$ if $\star = -$, and $p(I_1 | I_2) = p(I_1)$ if $\star = \bot$.
\end{theorem}

\begin{figure*}
\footnotesize
\[
\begin{array}{llllll}
\textsf{Dep}(O) & = & \{I ~|~ I \dep O \} & \textsf{Dep}^\star(O) & = & \{I ~|~ I \dep^\star O  \} \\


\textsf{Dep}(E_1  ~\wedge~ E_2) & = & \textsf{Dep}(E_1) \cup \textsf{Dep}(E_2) \quad \quad &
\textsf{Dep}^\star(E_1  ~\wedge~ E_2) & = & \textsf{Dep}^\star(E_1) \cup \textsf{Dep}^\star(E_2) \\

\textsf{Dep}(E_1  ~\vee~ E_2) & = & \textsf{Dep}(E_1) \cup \textsf{Dep}(E_2) \quad \quad &
\textsf{Dep}^\star(E_1  ~\vee~ E_2) & = & \textsf{Dep}^\star(E_1) \cup \textsf{Dep}^\star(E_2) \\

\textsf{Dep}(\neg E) & = & \textsf{Dep}(E)  & \textsf{Dep}^\star(\neg E) & = & \textsf{Dep}^{\overline{\star}}(E)  \\

\end{array}
\]
\caption{Auxiliary $\deps$ function. $O$ denotes an output fact  and $\star \in \{+, -\}$. $\overline{\star}$ is $+$ if $\star$ is $-$ and vice versa. 
}
\label{fig:dep-def}
\end{figure*}

\begin{figure*}
{\tiny

\begin{mathpar}
{\scriptsize
\inferrule*[right=(\textsc{NonInterfere})]{
\forall (x,y) \in \textsf{Dep}(E).~x \neq y \rightarrow 
\textsf{Class}(x) \neq \textsf{Class}(y)
}
{ \vdash \chi(E)}
}
\end{mathpar}
\vspace{-.3em}
\begin{mathpar}
{\scriptsize
\inferrule*[right=(\textsc{May-1})]{
\exists (x,y) \in \textsf{Dep}^{\star}(E_1) \times \textsf{Dep}^{\star}(E_2). ~x \depin y
}
{ \vdash E_1 \rightharpoonup^{\natural} E_2}
\hspace{2ex}
\inferrule*[right=(\textsc{May-2})]{
\exists (x,y) \in \textsf{Dep}^{\star}(E_1) \times \textsf{Dep}^{\overline{\star}}(E_2). ~x \depin y
}
{ \vdash E_1 \rightharpoonup^{\overline{\natural}} E_2}
}
\end{mathpar}
%
\vspace{-.3em}
%
%
\vspace{-.3em}
\begin{mathpar}
{\scriptsize
\inferrule*[right=(\textsc{Pos})]
{
\vdash \chi(E_1) \quad  \vdash \chi(E_2)  \quad 
\vdash E_1 \rightharpoonup^{+} E_2 \quad
\not \vdash  E_1 \rightharpoonup^{-} E_2
}
{ \vdash E_1 \cpos E_2} 
\hspace{2ex}
\inferrule*[right=(\textsc{Neg})]
{
\vdash \chi(E_1)~ \quad \vdash \chi(E_2) \quad
\vdash E_1 \rightharpoonup^{-} E_2 \quad
\not \vdash  E_1 \rightharpoonup^{+} E_2
}
{ \vdash E_1 \cneg E_2} 
}
\end{mathpar}
%
\vspace{-.3em}
\begin{mathpar}
{\scriptsize
\inferrule*[right=(\textsc{Indep})]
{
\forall x, y \in \deps(E_1) \times \deps(E_2). ~ 
\textsf{Class}(x) \neq \textsf{Class}(y)
}
{ \vdash E_1 \cindep E_2} 
\hspace{2ex}
\inferrule*[right=(\textsc{Unknown})]
{
\not \vdash E_1 \cpos E_2 \quad 
\not \vdash E_1 \cneg E_2
\quad \not \vdash E_1 \cindep E_2
}
{ \vdash E_1 \cunknown E_2} 
}
\end{mathpar}
}
\caption{Inference rules for computing statistical correlation between expressions, where $\natural \in \{+, -\}$.  Note that predicates  $x \blacktriangleright^{\star} y$ are derived using Figure~\ref{fig:corr-input}, and $\overline{\star}$ (resp. $\overline{\natural}$)  is $+$ if $\star$ (resp. $\natural$) is $-$ and vice versa. }
\label{fig:corr}
\end{figure*}

\noindent
{\bf Phase 2: Inferring correlation types between outputs.} The second phase of our inference algorithm uses the results of the first phase to infer correlation types between expressions involving outputs. Note that we could, \emph{in principle}, use the same constraint-based approach 
presented  Figure~\ref{fig:corr-input} to infer correlation types between arbitrary expressions; however, such an approach does not scale well.
To overcome this scalability bottleneck, we instead utilize  lightweight static analysis.

The key idea underlying our method is to utilize the derivation graph, along with the known correlations between input facts, to infer correlation types between arbitrary expressions (i.e., boolean combinations of ground predicates). Given a pair of expressions $E_1, E_2$, our method first uses the derivation graph to identify the set $S_1, S_2$ of input facts, along with their polarity, that $E_1$ and $E_2$ \emph{logically} depend on; it then analyzes the \emph{statistical} correlations between elements in $S_1, S_2$ to decide whether we can determine the  correlation type between $E_1$ and $E_2$.

Our analysis is summarized in Figures~\ref{fig:dep-def} and ~\ref{fig:corr}, where the former defines  two auxiliary functions $\textsf{Dep}, \textsf{Dep}^\star (\star \in \{+,-\})$ used in Figure~\ref{fig:corr}. As shown in Figure~\ref{fig:dep-def},  $\textsf{Dep}(E)$ simply yields the set of all input facts that $E$ is \emph{logically} dependent on according to the derivation graph (recall Def~\ref{def:inDependence}). Similarly, $\textsf{Dep}^+(E)$  (resp. $\textsf{Dep}^-(E)$) yields the set of input facts that $E$ depends \emph{positively} (resp. \emph{negatively}) on. For example, consider an output fact $O$ that can be derived using $I_1 \land \neg I_2$ or using only $I_2$. In this case, both $\deps(O)$ and $\pdeps(O)$ include $I_1, I_2$ but $\ndeps(O)$ only includes $I_2$. 



Figure~\ref{fig:corr}  uses these auxiliary functions  to infer correlation types between arbitrary  expressions. The basic idea is to infer whether two expressions $E_1$ and $E_2$ \emph{may} be positively or negatively correlated (rules labeled {\sc May}), meaning that a pair of  shared input predicates in $E_1$ and $E_2$ have the potential to introduce a positive or negative correlation. Then, according to the rules labeled {\sc Pos} and {\sc Neg}, if we find that $E_1$ and $E_2$ may be positively (resp. negatively) correlated and there is nothing that introduces a potential negative (resp. positive) correlation, we can conclude that $E_1$ and $E_2$ are definitely positively (resp. negatively) correlated {as long as both expressions exhibit a certain non-interference property} shown in the {\sc NonInterfere} rule as $\chi(E)$. Intuitively, the non-interference property is necessary because, if an input fact $I$ is positively correlated with $I_1$ and $I_2$ individually, it does not necessarily mean that it is positively correlated with $I_1 \land I_2$. 
{At the end of the correlation type analysis, the inferred dependencies are stored in $\Sigma$ and used in Algorithm~\ref{alg:SolveDelta}.}

\begin{theorem}
\label{theorem:corr}
If $E_1 \cpos E_2 $ is derivable using the rules in Figure~\ref{fig:corr}, then  $p(E_1 | E_2) > p(E_1)$.  Similarly, $\vdash E_1 \cneg E_2 $ implies $p(E_1 | E_2) < p(E_1)$ and $\vdash E_1 \cindep E_2$ implies $p(E_1 | E_2) = p(E_1)$.
\end{theorem}

\subsection{Computing Approximate Probability Bounds}
\label{sec:approxBound}

\begin{figure}[t]
\centering
\hspace{-5ex}
\begin{minipage}[t]{0.42\textwidth} 
    \vspace{0pt} 
    \begin{algorithm}[H]
        \caption{\textsc{DeriveApproximateBounds}($G, D, \Sigma$)}
        \label{alg:approxInterval}
        {\footnotesize
        \input{figs/approxInterval}
        }
    \end{algorithm}
\end{minipage}
\hspace{-.5ex}
\begin{minipage}[t]{0.5\textwidth} 
    \vspace{0pt} 
    \[
    \footnotesize
    \begin{array}{llll}
    & & & \\
     \llbracket n  \rrbracket{(G)} &  =  & \mathsf{Pred}(n)  & \text{if $n$ is a leaf node} \\
     \llbracket n  \rrbracket{(G)} &  =  & \bigvee_{e \in E} (\llbracket e \rrbracket{(G)}, \mathbb{P}(e)) & \text{where \textsf{OutEdges($G,n$) $= E$}} \\
     \llbracket e \rrbracket{(G)} & =  & \bigwedge_{n \in B^+} \llbracket n \rrbracket{(G)} & \text{where } e = (n_s, B^+, B^-, r) \\
     & &\wedge \bigwedge_{n \in B^-} \neg \llbracket n \rrbracket{(G)}  & 
    \end{array}
    \]
    \vspace{-1em} 
    \caption{Computation of derivation expressions. $\mathbb{P}(e)$ yields the probability of the rule labeling edge~$e$.}
    \label{fig:deriv-expr}
\end{minipage}
\end{figure}


\begin{table}[t]
\scriptsize
\centering
\setlength{\tabcolsep}{4pt} 
\footnotesize
\begin{tabular}{ccccc}
\toprule
Operation &  $\star=+$  & $\star=-$ & $\star=\bot$ & $\star=\top$
\\
\midrule
\textsf{CL}$(e_1,e_2,\star)$ & $e_1 \times e_2$ & \textsf{max}($e_1+e_2-1,0$) & $e_1 \times e_2$ & \textsf{max}($e_1+e_2-1,0$) \\
\textsf{CU}$(e_1,e_2,\star)$ & \textsf{min}($e_1, e_2$) & $e_1 \times e_2$ & $e_1 \times e_2$ & \textsf{min}($e_1, e_2$) \\
\textsf{DL}$(e_1,e_2,\star)$ & \textsf{max}($e_1, e_2$) & $1 - (1-e_1)(1-e_2)$ & $1 - (1-e_1)(1-e_2)$ & \textsf{max}($e_1, e_2$) \\
\textsf{DU}$(e_1,e_2,\star)$ & $1 - (1-e_1)(1-e_2)$ & \textsf{min}($1, e_1+e_2$) & $1 - (1-e_1)(1-e_2)$ & \textsf{min}($1, e_1+e_2$) \\
\bottomrule
\end{tabular}
\vspace{.5em}
\caption{\textsf{CL/CU/DL/DU computation rules.} \textsf{CL} and \textsf{CU} denote the lower and upper bounds of the conjunction operation, respectively, while \textsf{DL} and \textsf{DU} represent the lower and upper bounds of the disjunction operation.
}
\label{table: CLCU}
\end{table}

\begin{figure*}
\vspace{-0.25in}
{\footnotesize
\begin{mathpar}
\inferrule*[right=(\textsc{In})]
{
I \in \textsf{InputFacts}(D) \quad
p :: (I \ | \ \varnothing) \in \textsf{InputProbs}(D)
}
{\Sigma,D  \vdash I \curvearrowright [p,p]} 
\hspace{2ex}
\inferrule*[right=(\textsc{Neg})]
{
\Sigma,D \vdash E \curvearrowright [l, u]
}
{\Sigma,D \vdash \neg E \curvearrowright [1-u, 1-l] } 
\end{mathpar}
%
\vspace{-.5em}
\begin{mathpar}
\inferrule*[right=(\textsc{Conjunct})]
{
\Sigma,D  \vdash E_1 \curvearrowright [l_1, u_1] \quad
\Sigma,D \vdash E_2 \curvearrowright [l_2, u_2] \quad
\Sigma(E_1, E_2) = \star
}
{\Sigma,D  \vdash E_1 \wedge E_2 \curvearrowright [\textsf{CL}(l_1, l_2, \star), \textsf{CU}(u_1, u_2, \star)]} 
\end{mathpar}
\vspace{-.5em}
\begin{mathpar}
\inferrule*[right=({\textsc{Disjunct}})]
{
\Sigma,D \vdash E_1 \curvearrowright [l_1, u_1] \quad
\Sigma,D \vdash E_2 \curvearrowright [l_2, u_2] \quad
\Sigma(E_1, E_2) = \star 
}
{\Sigma,D  \vdash (E_1, p_1) \vee (E_2, p_2) \curvearrowright [\textsf{DL}(l_1\times p_1, l_2\times p_2, \star), \textsf{DU}(u_1\times p_1, u_2\times p_2, \star)]} 
\end{mathpar}
}
\caption{Inference rules for computing approximate probability bounds.
}
\label{fig:approx}
\end{figure*}

In this section, we present a technique, summarized in Algorithm~\ref{alg:approxInterval},  for deriving \emph{approximate} probability bounds on output relations. 
For each node in the derivation graph, this algorithm computes a so-called \emph{derivation expression} $E$ that summarizes all ways in which a given relation can be derived ({line 2}). For example, if there are two rules $p_1:: R \colonminus A, B$ and $p_2:: R \colonminus C$, then the derivation expression is of the form $(A \land B) \lor C$. However, because we also need to keep track of the rule probabilities, expressions inside a disjunct also have a corresponding probability, represented as $(A \land B, p_1) \lor (C, p_2)$ for this example. Figure~\ref{fig:deriv-expr} presents the rules for generating derivation expressions for each node.
Then, given the derivation expression $E_n$ for node $n$, the {\sc ApproxExpr} procedure (called at line 3 in Algorithm~\ref{alg:approxInterval}) computes the approximate upper and lower bounds for $E_n$ using the rules presented in Figure~\ref{fig:approx}. Given a derivation expression $E$, these rules derive judgments of the form $\Sigma, D \vdash E \curvearrowright [l, u]$ indicating that $l, u$ are lower and upper bounds on the probability of expression $E$ evaluating to true. To compute these lower and upper bounds, we leverage the results of the correlation type analysis (stored in $\Sigma$) as well as known statistical inequalities provided in Table~\ref{table: CLCU} for different correlation types ~\cite{ruschendorf1991frechet}.

\vspace{-.5em}
\begin{theorem}
\label{theorem:approx}
    Let $D$ be a \toolname program with derivation graph $G$, and suppose $\Sigma$ is a sound correlation environment for $D$. Also, let $M = \textsc{Solve}(D)$ (Algorithm~\ref{alg:baseline:opt}) and let $M' = \textsc{DeriveApproxBounds}(G, D, \Sigma)$. For every output relation $O$ of $D$ such that $M(O) = (l^*, u^*)$ and $M'(O) = (l, u)$, we have $l~\leq~l^*~\leq~u^*~\leq~u$.
\end{theorem}

\subsection{Iterative Refinement of Probability Bounds}
\label{sec:iterative}

In this section, we describe the {\sc MakeDeltaPrecise} algorithm that iteratively tightens the computed probability bounds until it is within some $\delta$ of the ground-truth. The key idea is to combine the algorithm from Section~\ref{sec:basic-algo} with the approximate bounds as illustrated in Figure~\ref{fig:refine}. In this Figure~\ref{fig:refine}, $l^*$ and $u^*$ denote the ground truth (but unknown) probability bounds for relation $R$, and $l$ and $u$ denote the approximate probability bounds for $R$, computed as described in Section~\ref{sec:approxBound}. Thus,  it is always the case that $l \leq l^*$ and $u^* \leq u$. Our key observation is that the constraint $\phi$ generated in Section~\ref{sec:constraint} partitions this space into three regions:
\begin{itemize}[leftmargin=*]
\item {\bf Region 1:} This is the region $\psi_1 = (l \leq \mathsf{Expr}(R) < l^*)$, where $\mathsf{Expr}(R)$ denotes the symbolic expression generated for $R$, as described in Section~\ref{sec:obj}. Since the ground truth is $l^* \leq \mathsf{Expr}(R) \leq u^*$, $\psi_1 \land \phi$ must be unsatisfiable. 
\item {\bf Region 2:} This is the ``ground truth'' region $\psi_2 = l^* \leq \mathsf{Expr}(R) \leq u^*$; thus, $\phi \land \psi_2$ is satisfiable.
\item {\bf Region 3:} This is the region $\psi_3 = (u^* < \mathsf{Expr}(R) \leq u)$, so  $\psi_3 \land \phi$ is again unsatisfiable.  
\end{itemize}

\begin{figure}[t]
\vspace{-0.1in}
    \centering
    \begin{minipage}{0.48\textwidth}
        \centering
    \begin{algorithm}[H]
\caption{\textsc{MakeDeltaPrecise}($M, S_o, \phi, \delta$)}
\label{alg:MakeDeltaPrecise}
{\footnotesize
\input{figs/make-delta}
}
\end{algorithm}
    \end{minipage}\hfill
    \begin{minipage}{0.48\textwidth}
\begin{algorithm}[H]
\caption{\textsc{BoundBounds}($M, \phi, S_o$)}
\label{alg:epilson}
{\footnotesize
\input{figs/epsilon.tex}
}
\end{algorithm}
    \end{minipage}
\end{figure}

As illustrated in Figure~\ref{fig:refine}, the idea is to repeatedly increase $l$ (resp. $u$) by  $\epsilon$ until the formula $l_i \leq \mathsf{Expr}(R) \leq l_i+\epsilon \land \phi$ (resp. $u_i-\epsilon \leq \mathsf{Expr}(R) \leq u_i \land \phi$) becomes satisfiable. When this procedure terminates, we can obtain lower and upper  bounds $(l^-, l^+)$ for $l^*$ as well as bounds $(u^-, u^+)$ for $u^*$. We can then perform binary search  until the distance between the two becomes less than $\delta$.

This discussion is summarized in Algorithm~\ref{alg:MakeDeltaPrecise}. The {\sc MakeDeltaPrecise} procedure first calls {\sc BoundBounds} to compute upper and lower bounds for $l^*, u^*$, as depicted in Figure~\ref{fig:refine}. As shown in Algorithm~\ref{alg:epilson} (and its auxiliary procedure {\sc MakeSAT} in Algorithm~\ref{alg:make-sat}), {\sc BoundBounds} increments (resp. decrements) $l$ (resp. $u$) until we get into the SAT region in Figure~\ref{fig:refine}. Upon termination of {\sc MakeSAT}, $(l^-, l^+)$ (resp. $(u^-, u^+)$) provide lower and upper bounds for $l^*$ (resp. $u^*$). Then, for each relation $R$, {\sc MakeDeltaPrecise} calls {\sc BinarySearch} (Algorithm~\ref{alg:boundByDelta}) to find a $\delta$-optimal solution. 
When \textsc{BinarySearch} terminates, the returned interval $[l^-, l^+]$ is guaranteed to contain $l^*$, with $l^+ - l^- \leq \delta$. The same guarantee also applies to the upper bound.

\vspace{-.5em}
\begin{theorem}
\label{thm:delta}
Let $(l^*, u^*)$ be the ground truth probability bounds for relation $R$, and let $(l, u$) be the bounds computed by {\sc MakeDeltaPrecise}. Then, we have $l^* -\delta \leq l \leq l^*$ and $u^* \leq u \leq u^* + \delta$.
\end{theorem}
\vspace{-.5em}

\begin{figure}[t]

\vspace{-0.25in}
    \centering
    \begin{minipage}{0.44\textwidth}
        \centering
\begin{algorithm}[H]
\caption{\textsc{MakeSAT}($l, u, \phi, e,\mathsf{b}$)}
\label{alg:make-sat}
{\footnotesize
\input{figs/search.tex}
}
\end{algorithm}
    \end{minipage}\hfill
    \begin{minipage}{0.5\textwidth}
\begin{algorithm}[H]
\caption{\textsc{BinarySearch}($l, u, \phi, \delta, R, \mathsf{b}$)}
\label{alg:boundByDelta}
{\footnotesize
\input{figs/BoundDelta.tex}
}
\end{algorithm}
    \end{minipage}
    \vspace{-0.15in}
\end{figure}

\section{Implementation}
\label{sec:impl}

We have implemented our proposed approach in a tool called \toolname
written in C++.  
\toolname instruments the solving procedure of \textsc{Souffle}~\cite{jordan2016souffle} to generate the derivation graph and  utilizes the \textsc{Gurobi}~\cite{bixby2007gurobi} solver for optimization and the \textsc{Cvc5}~\cite{barbosa2022cvc5} SMT solver for  satisfiability. 


\vspace{0.05in}
\noindent
\textbf{\emph{Derivation graph generation.}}
Datalog solvers such as \textsc{Souffle}  avoid generating the same relation from different rules. For instance, if an output relation $O$ has already been derived using a rule $R$, the Datalog solver would avoid applying other rules to derive $O$  again. 
However, to compute the probability of $O$, we need all possible ways of deriving it; thus, our implementation modifies \textsc{Souffle} to generate the complete derivation graph. It also augments the derivation graph to keep track of rule probabilities.

\vspace{0.05in}
\noindent
\emph{\bf Inference of correlation classes.} While \toolname allows the user to explicitly specify correlation classes (e.g., via the \texttt{corr} declaration) , it  does not require them to do so. In particular, \toolname constructs a dependency graph between input facts based on the specified conditional probabilities and assumes that a pair of input facts are in the same correlation class iff they belong to the same connected component. This default behavior can be overridden by users by explicitly specifying correlation classes.  

%

\vspace{0.05in}
\noindent
\textbf{\emph{Optimized satisfiability checks.}} Recall that the iterative refinement technique from Section~\ref{sec:iterative} requires repeatedly checking satisfiability until the constraint becomes satisfiable. However, because the overwhelming majority of these calls return unsatisfiable, we simplify the problem by overapproximating the constraints until the overapproximation becomes satisfiable, in which case we switch to the exact encoding. The key idea underlying our over-approximation is as follows. While the exact encoding introduces joint probability variables over the input facts, we can instead introduce joint probability variables over \emph{intermediate relations} that are $k$ steps from the root node and encode the known correlations between them as part of the constraint. We provide an example of this encoding in Appendix~\ref{appendix:SAT}. 
\( k \) is not a fixed value; instead, \( k \) is selected dynamically. Details on how \( k \) is determined are provided in the Appendix.

\vspace{0.05in}
\noindent
\textbf{\emph{Handling very large correlation classes.}} In cases where correlation classes become prohibitively large, even computing joint probability distributions within a \emph{single} correlation class may be infeasible. For instance, some outliers in our experimental evaluation have hundreds of input facts in the same correlation class, making it infeasible to reason precisely about the joint probability for the entire class.  In such cases, our implementation retains soundness but  may compromise $\delta$-exactness. In particular, for correlation classes whose size exceeds a predefined threshold, 
we do not compute  correlation types as described in { Phase 1 of Section~\ref{sec:corr}}; however, we still perform the static analysis from Phase 2, assuming that 
input correlations
within the same class are unknown.  Second, when performing iterative refinement of the probability bounds, we approximate the satisfiability check as described in the previous paragraph and do \emph{not} switch to the fully precise encoding. However, we emphasize that even this approximate solution for handling excessively large correlation classes utilizes the exact same machinery described in the rest of the paper.

%
%
%
%

\section{Evaluation}
\label{sec:eval}

In this section, we now describe the results for the  evaluation that is designed to answer the following  questions:
\begin{itemize}[leftmargin=*]
\item {\bf RQ1:} How accurate are the probability bounds inferred by \toolname?
\item {\bf RQ2:} How efficient/scalable is our method in inferring the probability bounds? 
\item {\bf RQ3:} How impactful are the key technical ingredients underlying \toolname? 
\end{itemize}

\begin{wraptable}{R}{0.35\textwidth}
\vspace{-0.2in}
    \centering
    \scalebox{0.7}{
    \begin{tabular}{|c|ccccc|}
\hline
& & {\bf Min} & {\bf Max} & {\bf Avg} &  {\bf Med}\\ \hline
\multirow{3}{*}{\rotatebox[origin=c]{90}{\textbf{SC}}} 
    &  {\bf \# nodes} & 56 & 200,164 & 43,366 & 9,093 \\ \cline{2-6}
    & {\bf \# edges} & 3 & 173,617 & 33,284 &  5,919\\ \cline{2-6}
    & {\bf  CC size} & 8 & 731 & 133 & 31 \\ \hline \hline
\multirow{3}{*}{\rotatebox[origin=c]{90}{\textbf{Bayes}}} 
    & {\bf \# nodes} & 10 & 223 & 79 & 60 \\ \cline{2-6}
    & {\bf \# edges} & 11 & 328 & 128 & 77 \\ \cline{2-6}
    & {\bf CC size} & 2 & 13 & 5 & 4 \\ 
\hline
\end{tabular}
}
\vspace{0.1in}
    \caption{Benchmark statistics, CC denotes ``correlation class''.}
    \label{fig:stat}
    \vspace{-0.3in}
\end{wraptable}

\vspace{-.5em}
\subsection{Application Domains and Benchmarks} 
We evaluate \toolname on two different application domains spanning 30  benchmarks, summarized in Table~\ref{fig:stat}.
  The first category, labeled \textbf{SC} (for Side Channel) in Table~\ref{fig:stat}, corresponds to 19 Datalog-based program analyses for detecting power side-channel leaks. The second category, labeled \textbf{Bayes}, consists of 11 Bayesian networks sourced from the \textsf{bnlearn} repository. We provide more information about each of these application domains below.   

\vspace{-.5em}
\paragraph{Side channel benchmarks.}
Our first application domain is power side channel detection -- a problem that  has received significant attention in recent work~\cite{wang2021data,wang2019mitigating}. For this  domain,  input facts in the Datalog program are derived from the programs under analysis and include implementations of well-known cryptographic protocols such as AES, SHA3, and MAC-Keccak. The Datalog rules describing the side channel  analysis are taken from~\cite{wang2019mitigating}. Hence, each benchmark in the side channel category corresponds to the ``cross product'' of an existing side channel detector~\cite{wang2019mitigating} and a real-world implementation of a cryptographic protocol. However, since the original Datalog-based analyzer only outputs a yes/no answer (indicating a potential power side channel vulnerability), we extend the analysis to quantify leakage severity.
Our extension retains the existing Datalog rules and input facts as is, but augments them with probabilities  as well as conditional dependencies/probabilities between input facts.   


In more detail, we obtain the probabilities of the Datalog rules describing the program analysis using a similar methodology to what has been described in prior work on quantitative Datalog-based  race detection~\cite{raghothaman2018user}.    This involves instrumenting the program to count rule firings and using these counts to estimate the probability of each rule applying in practice.
To obtain the probabilities of input facts, we first note that while some input facts remain deterministic, others are probabilistic due to variations in register allocation algorithms and hardware architectures. 
For instance, the probability of register sharing is estimated using empirical data from profiling a code corpus. Similarly, certain input facts are probabilistic as they stem from pre-analysis~\cite{wang2019mitigating, zhang2018scinfer} that infers semantic data dependencies from syntactic ones. The probability of such dependencies is derived from prior empirical studies measuring how often syntactic dependencies translate into actual data dependencies in compiled programs~\cite{wang2019mitigating}.
To quantify conditional probabilities between input facts, we leverage empirical co-occurrence statistics from an existing code corpus~\cite{wang2021data}.  We analyze execution traces to measure how often certain conditions –such as register sharing information and key-related data dependencies – appear together. These conditional probabilities are essential for accurately assessing the severity of detected side channels, as some Datalog rules depend on both register-sharing behavior and secret-dependent data flows between variables. Consequently, the overall probability of data leakage must account for these correlations between (probabilistic) input facts rather than treating them as independent events.

\vspace{-.5em}
\paragraph{Bayesian network benchmarks.} Our second set of benchmarks, labeled \textbf{Bayes} in Table~\ref{fig:stat}, consists of 11 discrete Bayesian networks sourced from the \textsf{bnlearn} repository~\cite{scutari2009learning}, encompassing a range of network sizes, including \emph{small}, \emph{medium}, and \emph{large} examples. We derive these benchmarks by preserving the original network structure, designating nodes without incoming edges as input facts and all other nodes as output facts. To introduce partially known statistical correlations between input facts, we inject dependencies that reflect realistic co-occurrence patterns observed in empirical data. For example, in medical networks, we introduce correlations between demographic and lifestyle factors, such as smoking and being Asian, while in weather models, we correlate atmospheric conditions like humidity and precipitation, which often vary together.
%
{These correlations are not deterministically defined,}
meaning that while input facts are not mutually independent, their exact joint distribution remains unknown. This approach enables us to evaluate inference under conditions where  partial dependency information is available.



\vspace{-0.06in}
\paragraph{Benchmark statistics.} Table~\ref{fig:stat} summarizes key statistics for both benchmark categories. \textbf{\#nodes} and \textbf{\#edges} represent the number of nodes and hyperedges in the derivation graph, while \textbf{CC size} denotes the size of the largest correlation class. We report the minimum, maximum, average, and median values across all benchmarks in each category. As shown in Table~\ref{fig:stat}, the 19 side-channel benchmarks present a greater computational challenge than the 11 Bayesian network benchmarks from~\cite{scutari2009learning}, highlighting the complexity of 
probabilistic inference
in security applications. These two categories illustrate distinct but complementary use cases for Praline, both of which involve correlated inputs where exact dependencies are not fully known. 

\subsection{Experimental Methodology and Set-up} 
\label{sec:setup}

Existing probabilistic extensions of logic programming languages do not account for conditional dependencies between inputs. To assess whether \textsc{ProbLog}\cite{de2007problog} could serve as a baseline, we attempted to encode input correlations\footnote{For a fair comparison, we also constructed modified versions of our benchmarks where all input facts are treated as independent; in these cases, \textsc{Praline} and \textsc{ProbLog} produced identical results whenever \textsc{ProbLog} successfully terminated.} using its \emph{evidence} predicate (see Appendix~\ref{appendix:evidence} for details). However, \textsc{ProbLog} successfully terminated in only 17 of 30 benchmarks and failed to terminate on the remaining 13. Moreover, even when it did terminate, it produced unsound results due to its inability to faithfully represent \textsc{Praline} programs. This limitation stems from fundamental expressiveness gaps—accurately encoding a single \textsc{Praline} program in \textsc{ProbLog} would require generating an infinite number of distinct \textsc{ProbLog} programs (Appendix~\ref{appendix:evidence}).

Given these limitations, we evaluate \textsc{Praline} against the constrained optimization approach introduced in Section~\ref{sec:basic-algo}, 
which we use as the baseline to produce the ground truth.
%
%
Our evaluation compares this baseline with the proposed $\delta$-exact algorithm, which enhances scalability while preserving precise probability bounds. Throughout the remainder of this section,  \toolname refers to the $\delta$-exact method from Section~\ref{sec:delta}, and “{\sc Constrained Optimization}” refers to the baseline.

All experiments were conducted on macOS Sonoma 14.4.1 with a 3-hour time limit and a memory cap of 16GB. In our evaluation, we set the $\delta$ parameter to $0.01$ for the $\delta$-exact algorithm,
as it offered a practical balance between runtime and accuracy. In general, $\delta$ controls a clear trade-off: smaller values produce tighter probability bounds but incur longer runtimes due to additional refinement iterations, whereas larger values yield faster computations at the cost of looser bounds.


\subsection{Accuracy Evaluation}

In this section, we evaluate the accuracy of the probability bounds inferred by \toolname in two ways. First, for those benchmarks where exact inference (\textsc{Constrained Optimization}) terminates, we compare the bounds computed by \toolname against the ground truth. Second, because \toolname may compromise $\delta$-exactness for very large correlation classes (recall Section~\ref{sec:impl}), we evaluate the percentage of output facts for which \toolname guarantees $\delta$-exact inference.

\subsubsection{Comparison with ground truth.}
\begin{wraptable}{R}{0.3\textwidth}
\vspace{-0.1in}
    \centering
    \scalebox{0.8}{
    \begin{tabular}{|c|c|c|}
    \hline  
        & \textbf{LB Error} & \textbf{UB Error} \\ \hline
        \textbf{Average}  & 0.00168 &  0.00177 \\
        \textbf{Min} & 0.00000 & 0.00000 \\
        \textbf{Max} & 0.00904 & 0.00935 \\
        \hline
        \end{tabular}
        }
    \vspace{0.05in}
    \caption{Accuracy Comparison}
    \label{tab:acc}
    \vspace{-0.35in}
\end{wraptable}

For 12 of the 30 benchmarks used in our evaluation (covering 3,179 queried facts), exact inference terminates within the 3-hour time limit, allowing us to evaluate \toolname's results against the ground truth values. 
Table~\ref{tab:acc} presents the results of this evaluation, where the \textbf{LB Error} column reports the average, minimum, and maximum deviations in the lower bound, and the \textbf{UB Error} column provides the same for the upper bound. Specifically, \textbf{LB Error} corresponds to \( l^* - l \), while \textbf{UB Error} denotes \( u - u^* \), where \( (l, u) \) are the probability bounds computed by the $\delta$-exact procedure and \( (l^*, u^*) \) are the exact values obtained via constrained optimization. 
Despite the substantial efficiency gains of the $\delta$-exact method (evaluated more thoroughly in Section~\ref{sec:eval-ablation}), its computed probability bounds remain highly precise. Across all 12 benchmarks, the average lower and upper bound errors are just \textbf{0.00168} and \textbf{0.00177}, respectively—well within the specified $\delta$ threshold of 0.01 (i.e., $1\%)$. These results demonstrate that our approximate inference method provides an effective alternative to exact inference while maintaining near-optimal accuracy.

\subsubsection{Evaluation of $\delta$-exactness.}

\begin{wraptable}{R}{0.26\textwidth}
\vspace{-0.15in}
    \centering
    \scalebox{0.9}{
    \begin{tabular}{|c|cc|}
    \hline
        & \textbf{\#facts} & \textbf{$\delta$\%} \\ \hline 
       \textbf{SC} & 96,393 & 73\%  \\
      \textbf{Bayes} & 663 & 100\% \\ 
      \textbf{Overall} & 97,056 & 73\% \\ \hline
        \end{tabular}
        }
    \vspace{0.05in}
    \caption{$\delta$-exactness rate}
    \label{tab:deltaComplete}
    \vspace{-0.3in}
\end{wraptable}

As mentioned in Section~\ref{sec:impl}, the implementation of \toolname gives up on $\delta$-exactness in some cases to scale to programs with very large correlation classes. In this section, we evaluate for what percent of output facts \toolname can guarantee $\delta$-exactness of inference. In particular, the result of inference for an output fact is guaranteed to be $\delta$-exact if either (1) the length of the inferred interval is $\leq \delta$, or (2) the (final) satisfiability check in Section~\ref{sec:iterative} uses the exact encoding over input facts, theoretically guaranteeing $\delta$-exactness. Note that while these constitute sufficient conditions for $\delta$-exact inference, they are not necessary conditions, meaning that the numbers reported here form a \emph{lower bound} on the percentage of $\delta$-exact results.  The result of this evaluation is presented in Table~\ref{tab:deltaComplete}, where \#facts denotes the number of output facts queried by the program. \textit{Overall, at least 73\% of the queried output facts are guaranteed to be $\delta$-exact}. 
As stated earlier, this number is merely a lower bound on the percentage of $\delta$-exact results, owing to the simple reason that we do not have a scalable method of computing ground truths for the remaining facts.
\\

\vspace{-0.05in}
\idiotbox{RQ1}{For the  benchmarks where ground truth bounds are available, \toolname produces precise probability bounds, with  average lower/upper bound errors of $0.1\%$. Over all 30 benchmarks comprising 97,056 queried facts, \emph{at least} 73\% are guaranteed to be $\delta$-exact.}

\subsection{Inference Time Evaluation}

\begin{wraptable}{R}{0.16\textwidth}
\vspace{-0.4in}
    \centering
    \scalebox{0.8}{
    \begin{tabular}{c|c}
     \textbf{Time}   & \textbf{Rate} 
        \\ \hline 
    < 1s & 50\% \\
     1-10s & 10\% \\
     10-100s & 16\% \\
     100-1000s & 17\% \\
     >1000s & 7\% \\
     \end{tabular}
        }
    \vspace{0.05in}
    \caption{Runtime}
    \label{tab:inferenceTime}
    \vspace{-0.4in}
\end{wraptable}

To answer our second research question, we evaluate \toolname's inference efficiency and scalability. 

\subsubsection{Inference time}

Table~\ref{tab:inferenceTime} shows the percentage of benchmarks 
that can be solved within a given time limit. As shown in this table, \toolname is able to analyze $60\%$ of the benchmarks  in under  $10$ seconds and $76\%$  in under 100 seconds. All 30 benchmarks terminate within the specified time limit, with the largest benchmark (with $200164$ nodes, $173617$ edges, $116$ as the average correlation class size) taking 1414.23 seconds.  This result shows that \toolname is able to achieve practical inference times even for complex benchmarks. 

\subsubsection{Scalability Evaluation} 

\begin{wrapfigure}{R}{0.56\textwidth}
    \centering
    \begin{subfigure}[b]{0.27\textwidth}
        \centering
        \includegraphics[width=0.95\textwidth]{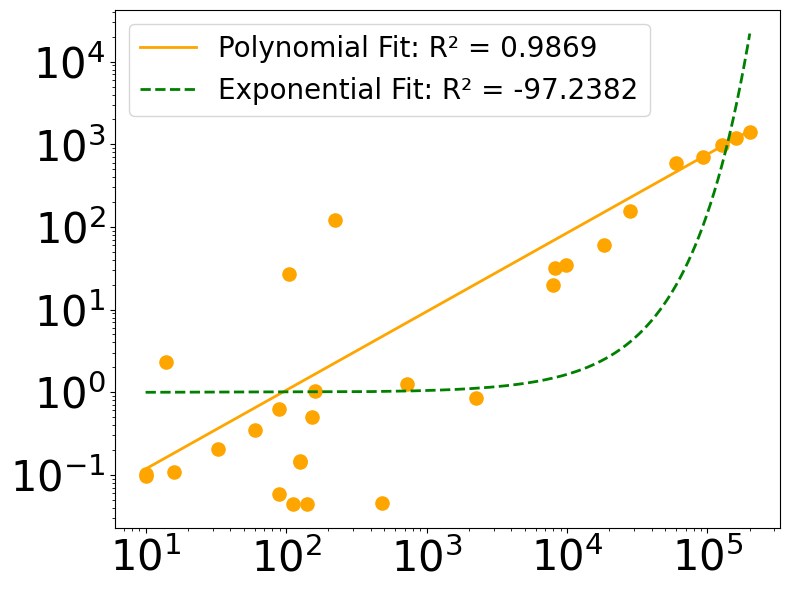}
        \caption{X = \#Node}
        \label{fig:nodeFitting}
    \end{subfigure}
    \vspace{-0.1in}  
    \begin{subfigure}[b]{0.27\textwidth}
        \centering
        \includegraphics[width=0.95\textwidth]{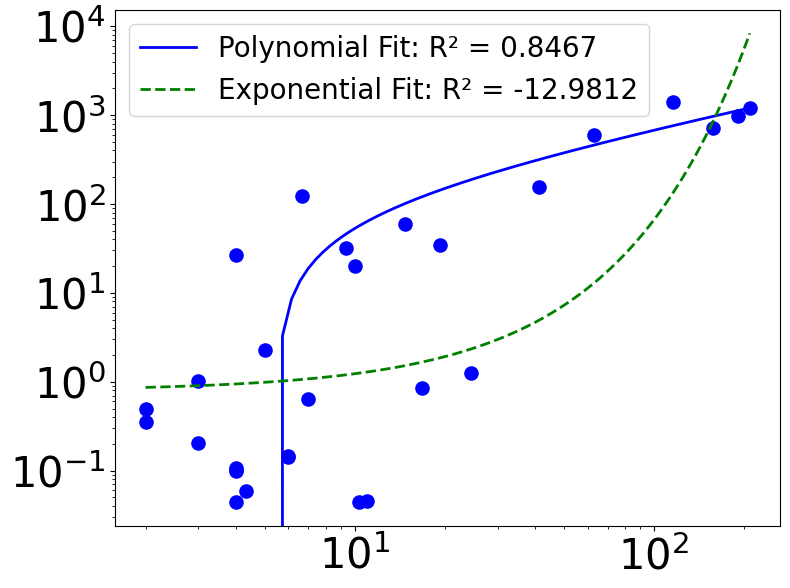}
        \caption{X = Average \#Corr}
        \label{fig:corrFitting}
    \end{subfigure}
    \vspace{-0.05in}
    \caption{Runtime (Y-axis) vs. benchmark complexity (X-axis).}
    \label{fig:timeNodeCorr}
    \vspace{-0.15in}
\end{wrapfigure}
To evaluate the scalability of \toolname with respect to benchmark complexity, Figure~\ref{fig:timeNodeCorr} plots runtime against two  complexity metrics. Figure~\ref{fig:nodeFitting} shows runtime (Y-axis,  in seconds) versus the number of nodes (X-axis), while Figure~\ref{fig:corrFitting} plots runtime against the average correlation class size. In Figure~\ref{fig:nodeFitting}, the orange curve {(a \emph{ polynomial fit})} aligns closely with the data, achieving a high correlation coefficient of 0.9869, whereas the exponential fit (green dashed line) does not capture the trend well. Similarly, in Figure~\ref{fig:corrFitting}, the blue curve {(an \emph{approximately polynomial fit})} provides a better fit than the exponential alternative.  \\


\idiotbox{RQ2}{\toolname is able to perform efficient probabilistic inference, with  76\% of benchmarks being solved in under 100 seconds. {Empirically, \toolname scales polynomially with both the number of nodes and the average correlation class size.}}




\subsection{Ablation Study}\label{sec:eval-ablation}

In this section, we describe a series of ablation studies designed to evaluate the impact of key ingredients of our approach. Specifically, we compare \toolname against the following ablations:

\begin{wrapfigure}{R}{0.4\textwidth}
        \vspace{-0.3in}
        \includegraphics[width=0.45\textwidth]{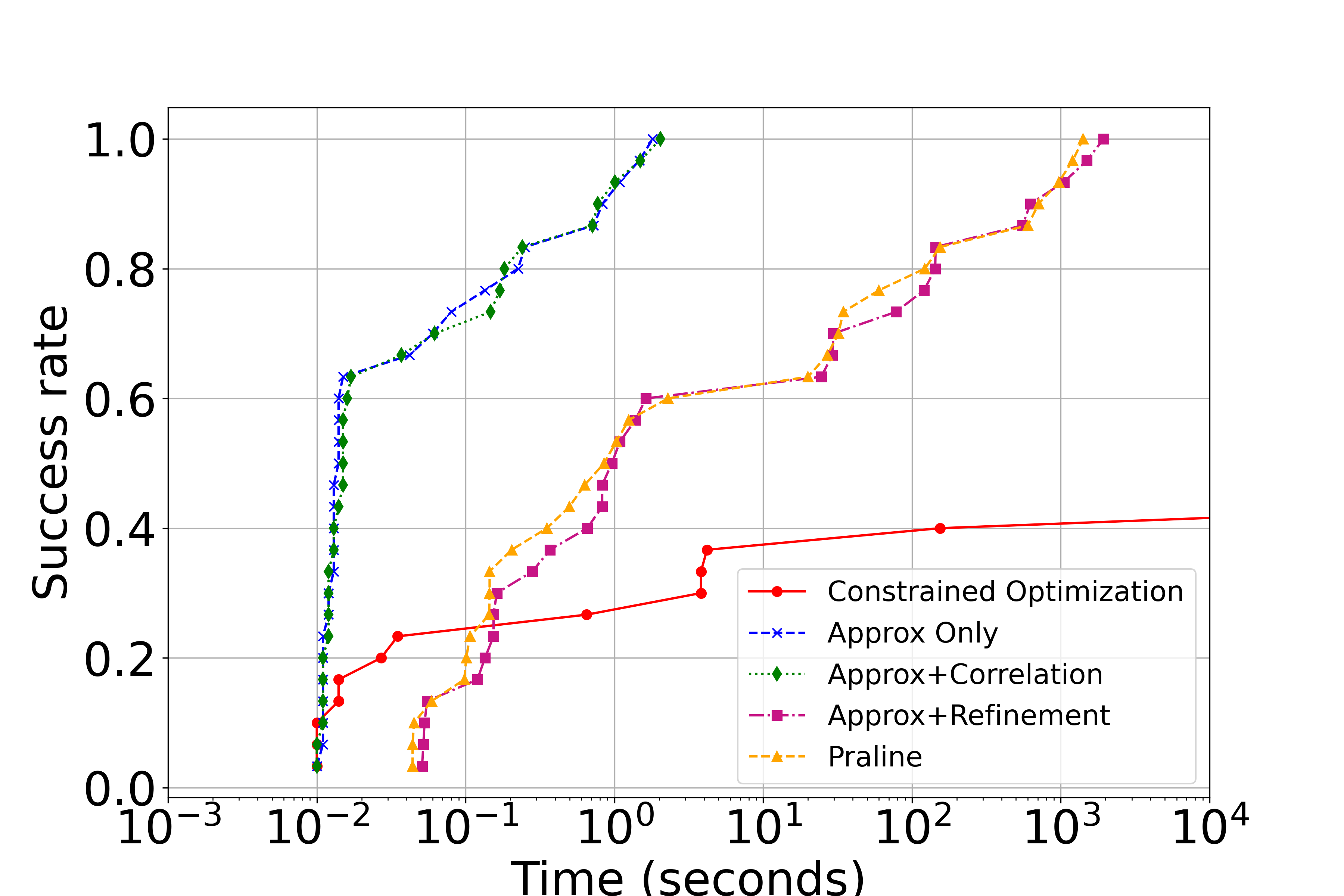}
        \vspace{-0.3in}
        \caption{Inference time.}
        \label{fig:cdf2}
        \vspace{-0.2in}
\end{wrapfigure}

\begin{itemize}[leftmargin=*]
\item  \textbf{\textsc{Constrained optimization}} This variant implements  Algorithm~\ref{alg:baseline:opt}.
\item \textbf{\textsc{Approx Only} (AO)} This is a variant of \textsc{Praline} that computes loose approximate bounds using the approximated bound computation technique from Section~\ref{sec:approxBound}. However, it does not utilize correlation types (correlation types are unknown), and it also does not perform refinement.  
\item \textbf{\textsc{Approx+Correlation} (AC)} This variant does not use  iterative refinement for tightening the bound. That is, it computes approximate probabilities while leveraging correlation types. 
\item \textbf{\textsc{Approx+Refinement} (AR)} This ablation  does not compute correlation types to assist the approximated bound computation. However, it does perform the refinement method of Section~\ref{sec:iterative}.
\end{itemize}
Among these ablations, we note that  \textbf{AO} and \textbf{AC} do not provide any precision guarantees. 
Next, we evaluate the impact of each key ingredient on both inference time as well as accuracy. 

\subsubsection{Evaluation of inference time.} Figure~\ref{fig:cdf2} explores the impact of various design choices on \emph{inference time}.
As we can see from Figure~\ref{fig:cdf2}, the variants of \textbf{\textsc{Praline}}  that do not have precision guarantees can perform inference more efficiently than all of the others. As expected, \textbf{\textsc{Constrained Optimization}} is the slowest and times out on the majority of the benchmarks.   In contrast, both \textbf{\textsc{Praline}} and \textbf{AR} solve all benchmarks within the three-hour time limit, exhibiting similar performance in inference time. To see why \textbf{\textsc{Praline}}  and \textbf{AR} have similar performance, note that computing correlation types adds some overhead, however, it also reduces time to perform iterative refinement, as the refinement algorithm starts with tighter bounds. Thus, overall, the computation of correlation types does not end up having a significant impact on inference time.  

\subsubsection{Accuracy evaluation.}

{We now consider how the different variants of \toolname perform in terms of accuracy. To quantify the impact of removing a specific feature in terms of accuracy, we consider a \emph{bound tightness ratio (BTR)} metric, defined as follows:
\[
\mathsf{BTR}({\textsc{Praline}}^\Delta) = \frac{u'-l'}{u-l}
\]

 \begin{wraptable}{R}{0.3\textwidth}
\vspace{-0.8in}
    \centering
     \scalebox{0.9}{
    \begin{tabular}{c|cccc}
    & & Med & Avg & Max 
    \\ \hline
   \multirow{3}{*}{\rotatebox[origin=c]{90}{BTR}} &  \textbf{AO}  &  22.50& 28.20 & 82.24  \\ 
      &  \textbf{AC} & 11.33 &  17.16 & 52.08  \\ 
      &  \textbf{AR} & 2.20 &  2.38 & 5.53  \\
        \end{tabular}
        }
        \vspace{0.05in}
    \caption{Ablation results of \textsf{BTR}}
    \label{fig:ablationAcc}
    \vspace{-0.4in}
\end{wraptable}

 Here, ${\textsc{Praline}}^\Delta$ refers to a specific ablation for \toolname and $[l, u]$ and $[l', u']$ are the bounds produced by \toolname and ${\textsc{Praline}}^\Delta$ respectively. Intuitively, the closer the BTR is to $1$, the more accurate the ablation. The results of this evaluation are shown in Table~\ref{fig:ablationAcc}.
For each ablation (e.g., \textbf{AO}), we compute the \textsf{BTR} value for all output facts, and report the median, average, and maximum values.
As is evident from the data, the intervals computed by \toolname are much tighter compared to the other ablations, with \textbf{AO} producing the least precise intervals, followed by \textbf{AC}, and then \textbf{AR}. 

Theoretically, \textbf{AR} and \textbf{\textsc{Praline}} should achieve similar accuracy, as both employ iterative refinement via binary search to tighten probability bounds. In practice, however, a key difference in how satisfiability checking is performed during refinement leads to \textbf{\textsc{Praline}} achieving at least twice the tightness of \textbf{AR}, as shown in Table~\ref{fig:ablationAcc}.

This accuracy gap stems from our implementation (Section~\ref{sec:impl}). When correlation classes are large, the exact SAT encoding becomes computationally expensive and often intractable. To mitigate this, we introduce an optimized satisfiability checking strategy that leverages over-approximation. The precision of this optimization depends on the availability of accurate correlation type information. \textbf{\textsc{Praline}} explicitly infers correlation types in earlier stages (Section~\ref{sec:corr}), enabling it to apply tighter over-approximations that closely match the exact encoding. In contrast, \textbf{AR} does not perform correlation type inference and conservatively treats all correlations as \textsc{Unknown}, resulting in looser encodings and reduced accuracy.

\vspace{1em}
\idiotbox{RQ3}{\toolname strikes an effective balance between precision and inference time,  delivering the most precise results across ablations that terminate on all benchmarks. {Variants of \toolname{} that do not perform refinement result in bounds that are $17-28\times$ worse on average.}}

\section{Related Work}
\label{sec:related}
\textbf{\emph{Probabilistic logic programming.}}
Probabilistic programming allows programmers to model distributions and perform probabilistic sampling and inference, with systems like Pyro~\cite{bingham2019pyro}, Turing~\cite{ge2018turing}, Hakaru~\cite{narayanan2016probabilistic}, SPPL~\cite{saad2021sppl}, Dice~\cite{holtzen2020scaling} and PPL~\cite{van2018introduction} leading the way. Recently, there has been significant progress in integrating logical reasoning into probabilistic programming to capture richer logical formalisms such as Horn clauses and first-order logic. Notable examples include probabilistic relational models~\cite{getoor2007probabilistic}, Markov logic networks~\cite{niu2011tuffy}, Bayesian logic programs~\cite{kersting2007bayesian}, 
 and probabilistic logic programming languages such as PRISM~\cite{sato1995statistical}, LPADs~\cite{vennekens2004logic}, Blog~\cite{milch2007blog}, CP-logic\cite{vennekens2009cp}, 
PPDL~\cite{barany2017declarative,grohe2022generative}, Datalogp~\cite{fuhr1995probabilistic}, Scallop~\cite{li2023scallop}, and ProbLog~\cite{de2007problog,dries2015problog2}.
These formalisms extend existing logic
programming languages like Prolog and Datalog by associating each  rule with 
 probabilities.
%
%
%
%

Among these languages,
ProbLog~\cite{de2007problog,dries2015problog2} and Scallop~\cite{li2023scallop} focus on discrete distributions, which are closely related to our work. 
These techniques reduce probabilistic inference to weighted model counting (WMC)~\cite{vlasselaer2016tp} and employ representations like binary decision diagrams (BDD)~\cite{bryant1986graph} to support efficient WMC.
However, both ProbLog~\cite{de2007problog,dries2015problog2} and Scallop~\cite{li2023scallop} largely assume that input facts are independent and do not allow expressing general forms of conditional dependencies (other than providing the ability to express mutual exclusion between predicates).
{In contrast, \textsc{Praline} offers syntactic support to declare general forms of conditional dependency between input facts, assign numerical probabilities to such dependencies, and assumes independence only in the absence of such declarations.}
To the best of our knowledge,  JudgeD~\cite{wanders2016judged} is the only current Datalog extension that allows expressing dependencies between  clauses by associating each input fact with a logical sentence. However,  it neither supports negations nor does it allow specifying numerical probabilities for conditional dependencies.

\vspace{0.05in}

\noindent
\textbf{\emph{Datalog for program analysis.}} 
Logic programming languages like Datalog have found numerous applications in program analysis, including for data-race detection~\cite{naik2006effective,raghothaman2018user,zhang2017effective}, thread-modular analysis~\cite{kusano2016flow,kusano2017thread}, side-channel detection~\cite{wang2021data,wang2019mitigating}, and points-to analysis~\cite{zhang2014abstraction,madsen2016datalog,bravenboer2009strictly,smaragdakis2014introspective,whaley2005using}. Traditionally, Datalog-based analyses have been qualitative, but recent work~\cite{raghothaman2018user,zhang2014abstraction} has investigated quantitative analysis methods for inferring the likelihood of data races by incorporating probabilistic reasoning. 
{These approaches, however, are constrained by their 
assumption about independence of input predicates, which our approach aims to address.
}

\vspace{0.05in} \noindent
\textbf{\emph{Exact probabilistic inference.}} Our method also relates to exact inference in graphical models like Bayesian networks and Markov networks. Exact inference techniques  include weighted model counting~\cite{holtzen2020scaling,vlasselaer2016tp,fierens2015inference}, symbolic analysis~\cite{gehr2016psi}, variable elimination~\cite{li2024compiling}, conjugacy~\cite{hoffman2018autoconj}, generating functions~\cite{klinkenberg2023exact,klinkenberg2024exact}, and 
optimization methods~\cite{daxberger2021mixed,bach2017hinge}. However, applying these techniques directly to our problem domain is challenging for several reasons. 
Aside from the obvious structural differences between a Bayesian network and a Datalog derivation graph, our work also distinguishes itself from the Bayesian network setting by allowing for the specification of \emph{incomplete} conditional dependencies for which it is not possible to compute a single probability value.
Because of these important differences, prior approaches~\cite{holtzen2020scaling, li2024compiling} for  speeding up probabilistic inference are unlikely to be effective in our setting. 

\vspace{0.05in} \noindent
\textbf{\emph{Approximate probabilistic inference.}} Approximate inference techniques  are  primarily based on sampling methods~\cite{koller2009probabilistic,darwiche2009modeling} such as Importance Sampling (IS), Markov Chain Monte Carlo (MCMC) and variational inference. 
However, these methods do not provide guarantees for  results produced within a finite time frame. Some approaches~\cite{beutner2022guaranteed,sankaranarayanan2013static,cousot2012probabilistic,albarghouthi2017fairsquare,feng2023lower} infer approximate posterior probabilities with guaranteed bounds;
however, these methods typically focus on continuous rather than discrete distributions. Additionally, their guarantees rely on a countable set of sampled interval traces, which scale exponentially with the model's dimension~\cite{beutner2022guaranteed}. 

\vspace{0.05in} \noindent
\textbf{\emph{Verifying probabilistic properties.}}
There is large body of work on   verifying probabilistic properties of programs,
 such as differential privacy and demographic fairness.
For example, differential privacy can be expressed as relational properties of probabilistic computations involving expected values.
Barthe et al.~\cite{barthe2015higher} propose a relational refinement type system and use approximate coupling to construct proofs. Albarghouthi and Hsu~\cite{albarghouthi2017synthesizing} and Wang et al.~\cite{wang2019proving} simplify approximate coupling proofs to make it more automated. 
%
FairSquare~\cite{albarghouthi2017fairsquare}, on the other hand, uses symbolic solving to verify if a program meets specified demographic fairness properties. While these approaches deal with probabilistic properties, they are largely orthogonal to our approach.




\section{Conclusion}
In this paper, we introduced a new probabilistic Datalog framework, \textsc{Praline}, which allows users to specify arbitrary statistical correlations between input facts, addressing a significant limitation in existing methods.  Importantly, \textsc{Praline} is designed to handle scenarios where the statistical correlations between inputs are not fully known, allowing accurate probabilistic inference even under partial information. To solve this problem, 
we first proposed a constrained optimization approach that can compute exact probability bounds. Then, to address the scalability limitations of exact inference, we  used this constrained optimization method in a more lightweight manner as the basis of $\delta$-exact algorithm that can approximate the true bounds, while guaranteeing that they are within distance $\delta$ of the ground truth. Our proposed $\delta$-exact approach iteratively strengthens the approximated bounds through a synergistic combination of static analysis, approximation, and iterative refinement.  
Our empirical evaluation on 30 real-world probabilistic Datalog programs demonstrates that \textsc{Praline} can compute precise probability bounds, while scaling to large benchmarks with more than 200,000 relations. In contrast, the ablations of \toolname that are not $\delta$-exact infer significantly less accurate probability bounds, while exact inference does not scale. These experiments demonstrate that \toolname strikes an effective balance between precision and inference time.


\section{Data-Availability Statement}

%
%
An artifact supporting the results of this paper is available on Zenodo~\cite{wang2025probabilistic}.
Our tool depends on Datalog, SMT, and optimization solvers, which users will need to install separately. One requirement is a free academic license for Gurobi optimization solver, which can be easily obtained using an institutional email address.

\section{ACKNOWLEDGMENTS}
This work was conducted in a research group supported by NSF awards CCF-1762299, CCF-1918889, CNS-1908304, CCF-1901376, CNS-2120696, CCF- 2210831, and CCF-2319471, CCF-2422130, CCF-2403211 as well as a DARPA award under agreement HR00112590133.

\bibliographystyle{ACM-Reference-Format}
\bibliography{sample-base}

\newpage
\appendix
\section{Appendix}

\subsection{Encoding Input Correlations using \textsc{ProbLog}}
\label{appendix:evidence}

To the best of our knowledge, \textsc{ProbLog} primarily supports mutual independence as a form of input correlation. To encode more complex types of input correlations, we must creatively use other available syntax structures in \textsc{ProbLog} to implicitly represent them.
We have explored two methods to encode input correlations. However, both approaches rely on computations using a complete conditional probability table and can not capture the exact semantics as \textsc{Praline} does.

In our evaluation, we use the first method, specifically the \emph{evidence} feature, to encode input correlations in \textsc{ProbLog}.

\begin{figure}[ht]
\begin{lstlisting}[
  	basicstyle=  \footnotesize, %\tiny,or \small or \footnotesize
  	style      = Prolog-pygsty,
        %backgroundcolor=\color{white},
        escapechar=\%,
        frame=none,
        numbers=left, 
        numberstyle=\tiny\color{gray}, stepnumber=1, numbersep= 4pt,
	basicstyle=\footnotesize\ttfamily,
        xleftmargin=.3\textwidth
    ]
%\colorbox{lightestmain}{0.6::edge(1,4).}% 
%\colorbox{lightestmain}{0.6::edge(2,5).}%
%\colorbox{lightestmain}{0.6::edge(2,6).}%
%\colorbox{mygreen}{\textcolor{white}{0.8::edge(2,5) |   edge(1,4).}}%
%\colorbox{mygreen}{\textcolor{white}{0.83::edge(2,6)  | edge(1,4).}}%
%\colorbox{orange}{\textcolor{white}{r $\colonminus$ edge(1,4), edge(2,5), \textbackslash+edge(2,6).}}%
%query(r).%
\end{lstlisting}
    \caption{An example \textsc{Praline} program $P^*$.}
    \label{fig:motivationAppendix}
\end{figure}

\begin{figure}[ht]
\begin{lstlisting}[
  	basicstyle=  \footnotesize, %\tiny,or \small or \footnotesize
  	style      = Prolog-pygsty,
        %backgroundcolor=\color{white},
        escapechar=\%,
        frame=none,
        numbers=left, 
        numberstyle=\tiny\color{gray}, stepnumber=1, numbersep= 4pt,
	basicstyle=\footnotesize\ttfamily,
        xleftmargin=.1\textwidth
    ]
%\colorbox{lightestmain}{1/2::edge(2,5).}% 
%\colorbox{lightestmain}{1/2::edge(1,4).}% 
%\colorbox{lightestmain}{1/2::edge(2,6).}% 

%\colorbox{mygreen}{\textcolor{white}{1/2::c $\colonminus$ edge(2,5), edge(1,4), edge(2,6).}}%
%\colorbox{mygreen}{\textcolor{white}{1/8::c $\colonminus$ edge(2,5), edge(1,4), \textbackslash+edge(2,6).}}%
%\colorbox{mygreen}{\textcolor{white}{1/8::c $\colonminus$ edge(2,5), \textbackslash+edge(1,4), edge(2,6).}}%
%\colorbox{mygreen}{\textcolor{white}{1/32::c $\colonminus$ edge(2,5), \textbackslash+edge(1,4), \textbackslash+edge(2,6).}}%
%\colorbox{mygreen}{\textcolor{white}{19/128::c $\colonminus$ \textbackslash+edge(2,5), edge(1,4), edge(2,6).}}%
%\colorbox{mygreen}{\textcolor{white}{1/128::c $\colonminus$ \textbackslash+edge(2,5), edge(1,4), \textbackslash+edge(2,6).}}%
%\colorbox{mygreen}{\textcolor{white}{1/128::c $\colonminus$ \textbackslash+edge(2,5), \textbackslash+edge(1,4), edge(2,6).}}%
%\colorbox{mygreen}{\textcolor{white}{137/384::c $\colonminus$ \textbackslash+edge(2,5), \textbackslash+edge(1,4), \textbackslash+edge(2,6).}}%

%\colorbox{orange}{\textcolor{white}{e\_25\_14 $\colonminus$ edge(2,5), edge(1,4).}}%
%\colorbox{orange}{\textcolor{white}{e\_14\_26 $\colonminus$ edge(1,4), edge(2,6).}}%
%\colorbox{orange}{\textcolor{white}{r $\colonminus$ edge(1,4), edge(2,5), \textbackslash+edge(2,6).}}%

%evidence(c, true).%

%query(edge(2,5)).    \% \textsc{ProbLog} result: 0.6 -> encoding \colorbox{lightestmain}{0.6::edge(2,5).} %
%query(edge(1,4)).	  \% \textsc{ProbLog} result: 0.6 -> encoding \colorbox{lightestmain}{0.6::edge(1,4).} %
%query(edge(2,6)).    \% \textsc{ProbLog} result: 0.6 -> encoding \colorbox{lightestmain}{0.6::edge(2,6).} %
%query(e\_25\_14).    \% \textsc{ProbLog} result: 0.6*0.8 -> encoding \colorbox{mygreen}{\textcolor{white}{0.8::edge(2,5)|edge(1,4).}}%
%query(e\_14\_26).    \% \textsc{ProbLog} result: 0.6*0.83 -> encoding \colorbox{mygreen}{\textcolor{white}{0.83::edge(2,6)|edge(1,4).}}%
%query(r).			  \% \textsc{ProbLog} result: Pr(r) = \textcolor{red}{\textbf{0.096}} % 
\end{lstlisting}
    \caption{The first \textsc{Problog} program encoding  the same input fact and conditional probabilities as Figure~\ref{fig:motivationAppendix}.}
    \label{fig:problog1}
\end{figure}

\begin{figure}[ht]
\begin{lstlisting}[
  	basicstyle=  \footnotesize, %\tiny,or \small or \footnotesize
  	style      = Prolog-pygsty,
        %backgroundcolor=\color{white},
        escapechar=\%,
        frame=none,
        numbers=left, 
        numberstyle=\tiny\color{gray}, stepnumber=1, numbersep= 4pt,
	basicstyle=\footnotesize\ttfamily,
        xleftmargin=.1\textwidth
    ]
%\colorbox{lightestmain}{1/2::edge(2,5).}% 
%\colorbox{lightestmain}{1/2::edge(1,4).}% 
%\colorbox{lightestmain}{3/4::edge(2,6).}% 

%\colorbox{mygreen}{\textcolor{white}{5/8::c $\colonminus$ edge(2,5), edge(1,4), edge(2,6).}}%
%\colorbox{mygreen}{\textcolor{white}{1/2::c $\colonminus$ edge(2,5), edge(1,4), \textbackslash+edge(2,6).}}%
%\colorbox{mygreen}{\textcolor{white}{1/32::c $\colonminus$ edge(2,5), \textbackslash+edge(1,4), edge(2,6).}}%
%\colorbox{mygreen}{\textcolor{white}{1/2::c $\colonminus$ edge(2,5), \textbackslash+edge(1,4), \textbackslash+edge(2,6).}}%
%\colorbox{mygreen}{\textcolor{white}{377/1920::c $\colonminus$ \textbackslash+edge(2,5), edge(1,4), edge(2,6).}}%
%\colorbox{mygreen}{\textcolor{white}{3/640::c $\colonminus$ \textbackslash+edge(2,5), edge(1,4), \textbackslash+edge(2,6).}}%
%\colorbox{mygreen}{\textcolor{white}{263/1920::c $\colonminus$ \textbackslash+edge(2,5), \textbackslash+edge(1,4), edge(2,6).}}%
%\colorbox{mygreen}{\textcolor{white}{1871/1920::c $\colonminus$ \textbackslash+edge(2,5), \textbackslash+edge(1,4), \textbackslash+edge(2,6).}}%

%\colorbox{orange}{\textcolor{white}{e\_25\_14 $\colonminus$ edge(2,5), edge(1,4).}}%
%\colorbox{orange}{\textcolor{white}{e\_14\_26 $\colonminus$ edge(1,4), edge(2,6).}}%
%\colorbox{orange}{\textcolor{white}{r $\colonminus$ edge(1,4), edge(2,5), \textbackslash+edge(2,6).}}%

%evidence(c, true).%

%query(edge(2,5)).    \% \textsc{ProbLog} result: 0.6 -> encoding \colorbox{lightestmain}{0.6::edge(2,5).} %
%query(edge(1,4)).	  \% \textsc{ProbLog} result: 0.6 -> encoding \colorbox{lightestmain}{0.6::edge(1,4).} %
%query(edge(2,6)).    \% \textsc{ProbLog} result: 0.6 -> encoding \colorbox{lightestmain}{0.6::edge(2,6).} %
%query(e\_25\_14).    \% \textsc{ProbLog} result: 0.6*0.8 -> encoding \colorbox{mygreen}{\textcolor{white}{0.8::edge(2,5)|edge(1,4).}}%
%query(e\_14\_26).    \% \textsc{ProbLog} result: 0.6*0.83 -> encoding \colorbox{mygreen}{\textcolor{white}{0.83::edge(2,6)|edge(1,4).}}%
%query(r).			  \% \textsc{ProbLog} result: Pr(r) = \textcolor{red}{\textbf{0.10105263}} % 
\end{lstlisting}
    \caption{The second \textsc{Problog} program encoding  the same input fact and conditional probabilities as Figure~\ref{fig:motivationAppendix}.}
    \label{fig:problog2}
\end{figure}

\paragraph{\textbf{1. Using evidence features}} In \textsc{ProbLog}, we can utilize evidence features to encode the conditional probabilities among input facts. The idea is to introduce mutually-independent facts, and constraint them using the derived rules. For instance, given three independent facts $x,y,z$ (i.e., $P(x\wedge y) = P(x)P(y)$), two rules $\textcolor{blue}{p_1}::c \colonminus x,y,z, ~\textcolor{blue}{p_2}::c \colonminus \text{\textbackslash+}x,y,z$, and the statement \emph{evidence($c$)} (i.e., observing that $c$ is true), $x,y,z$ become dependent after observing $c$, i.e., $P(x\wedge y|c) \neq P(x|c) * P(y|c)$.
By reverse engineering the value of \textcolor{blue}{$p_1$} and \textcolor{blue}{$p_2$}, we can encode  $P(x|y,c)$ to any value as follows:

$
\begin{array}{ll}
    P(x|y,c) &=P(x,y|c)/P(y|c)  \\
     &=P(x,y,c)/P(y,c) \\
     &=\frac{P(c|x,y)P(x,y)}{P(c|y)P(y)} \\
     &=\frac{[P(c|x,y,z)*P(z) + P(c|x,y,\neg z)*(1-P(z))]P(x,y)}{[P(c|y,x,z)P(x,z) + P(c|y,\neg x,z)P(\neg x, z) + P(c|y,x,\neg z)P(x, \neg z)+ P(c|y,\neg x,\neg z)P(\neg x, \neg z)]P(y)} \\
     &=\frac{[\textcolor{blue}{p_1}*P(z)+0]*P(x)*P(y)}{[\textcolor{blue}{p_1}*P(x)P(z)+\textcolor{blue}{p_2}*P(\neg x)P(z)+0+0]P(y)}
\end{array}
$

$P(x|y,c)$ specifies the conditional probability between input facts $x$ and $y$, given $c$ is true.

\begin{example}
\label{example:pralineProblog}
    We first provide an example \textsc{Praline} program simplified from the motivation example of Figure~\ref{fig:probCond} in Section~\ref{motivation}. Given this \textsc{Praline} program in Figure~\ref{fig:motivationAppendix}, we provide two corresponding \textsc{ProbLog} programs that encode the same conditional probabilities as shown in Figures~\ref{fig:problog1} and \ref{fig:problog2}.
    For instance, in Figure~\ref{fig:problog1}, we denote \texttt{edge(2,5)} and \texttt{edge(1,4)} as $x$ and $y$ respectively. They are independent initially. We can correlate $x$ and $y$ by using the derived rules (Lines 5-12) and the observation that $c$ is true. Given $c$ is true, $x$ and $y$ are no longer independent. In this \textsc{ProbLog} encoding, we get $P(x,y |c) = 0.48 = 0.6*0.8 \neq P(x|c) * P(y|c) = 0.6*0.6$ (line 23), which encodes the input correlation \colorbox{mygreen}{
    \textcolor{white}{0.8::\texttt{edge(2,5)} $\colonminus$ \texttt{edge(1,4)}}
    }.
\end{example}

Given the \textsc{Praline} program $P^*$ as shown in Figure~\ref{fig:motivationAppendix}, we can generate infinitely many \textsc{ProbLog} programs that encode the same input fact probabilities and conditional probabilities from $P^*$ (\emph{Lines 1-6}). However, these \textsc{ProbLog} programs yield different results of \texttt{query(r)}, e.g., \textbf{0.096} ( Figure~\ref{fig:problog1}) and \textbf{0.101} (Figure~\ref{fig:problog2}).

In short, a single \textsc{ProbLog} program might not always capture the same semantics as a \textsc{Praline} program. \textsc{Praline} is equipped to handle scenarios where some conditional probabilities are unknown, allowing it to provide outputs as probability intervals as the \emph{exact} inference result. Conversely, \textsc{ProbLog} is designed to compute absolute probabilities as its \emph{exact} result, typically requiring complete information to do so.

Figures~\ref{fig:problog1} and \ref{fig:problog2} are two \textsc{ProbLog} examples that encode the same input fact and conditional probabilities as the \toolname program $P^*$ (in Figure~\ref{fig:motivationAppendix}), by correlating the independent input facts using derived rules, e.g., \texttt{edge(1,4), edge(2,5)} and \texttt{edge(2,6)} are independent input facts but they are dependent given event c is true (e.g., \texttt{evidence(c, true)}). We also want to highlight that to encode the same conditional probabilities using \textsc{ProbLo}g, users would need to reverse engineer all the rule probabilities, e.g., 19/128::c $\colonminus$ \textbackslash+\texttt{edge(2,5), edge(1,4), edge(2,6)} as shown in Line 9 of Figure~\ref{fig:problog1}.

\paragraph{\textbf{2. Using \( p :: I \colonminus S_I \)}}
Another approach involves using the syntax $p :: I \colonminus S_I$ to represent input correlations. In ProbLog, however, this syntax is not intended for articulating conditional probabilities among input facts (e.g., it is not designed to accommodate an input fact \( I \) as the rule head). Utilizing this syntax for input correlations might consequently yield unforeseen results.

To demonstrate how such unexpected outcomes occur, we present a new example illustrated in Figure~\ref{fig:compProbLog}. The figure highlights specified input correlations in green, and ProbLog computes the query result for the output fact \texttt{path(1,2)} as 0.824, shown in \emph{red}.

We analyze the derivation graph of \texttt{path(1,2)} to trace the source of this discrepancy. As detailed in Figure~\ref{fig:compProbLog}, ProbLog initially selects \texttt{edge(1,2)} with a probability of 0.6. If not, it assumes \texttt{edge(1,2)} is false with a probability of 0.4 and attempts to derive it using the input correlation rule (colored in green):
\[
\begin{array}{ll}
   P(\texttt{edge(1,2)}) &= P(\texttt{edge(1,2)} \wedge \texttt{edge(2,5)}) + P(\texttt{edge(1,2)} \wedge \neg \texttt{edge(2,5)}) \\
   &= P(\texttt{edge(1,2)} \wedge \texttt{edge(2,5)}) + 0 \\
   &= 0.8 \times P(\texttt{edge(2,5)}) = 0.8 \times 0.7
\end{array}
\]

This approach leads to an erroneous outcome, as ProbLog is not designed to handle input correlations using such rules and mistakenly processes \texttt{edge(1,2)} as both an input and output fact. This results in two conflicting calculations for \( P(\texttt{edge(1,2)}) \)=0.6 or \( 0.4 \times 0.8 \times 0.7 \), which are incorrectly aggregated. As a result, $P(\texttt{path(1,2)}) = P(\texttt{edge(1,2)}) = 0.6+0.4*0.8*0.7 = 0.824$.

\begin{figure}[ht]
  \hspace{5em}
  \begin{minipage}[c]{0.4\textwidth}
  \centering
     \setcounter{lstlisting}{1}
\begin{lstlisting}[
  	basicstyle=  \footnotesize, %\tiny,or \small or \footnotesize
  	style      = Prolog-pygsty,
	escapechar=\%,
	frame=none,
        numbers=left, numberstyle=\tiny\color{gray}, stepnumber=1, numbersep= 4pt,
	basicstyle=\footnotesize\ttfamily
    ]
0.6::edge%(%1,2%)%.
0.7::edge%(%2,5%)%.
%\colorbox{mygreen}{\textcolor{white}{0.8::edge(1,2) $\colonminus$ edge(2,5).}}%
1::path%(%X,Y%)% %$\colonminus$% edge%(%X,Y%)%.
query%(%path%(%1,2%)%)%.





\end{lstlisting}
  \end{minipage}%
  \hspace{-2em}
  \begin{minipage}[c]{0.56\textwidth}
    \centering
    \scalebox{0.9}{\begin{tikzpicture}[font=\small] 
 \tikzstyle{arrow1}=[->,>=stealth,black]
 \tikzstyle{arrow2}=[thick,->,>=stealth,darkmain]

 \tikzstyle{medRec}=[%
rectangle, draw, minimum width=1.5cm, minimum height=0.9cm, inner sep = 0cm, outer sep = 0cm, align=center, text width=1.3cm]

\node [rectangle,draw] (P12) at (0,0) {  \texttt{path(1,2)} };
\node [rectangle,draw,fill=lightestmain] (E12) at (0,-1.4) {  \texttt{edge(1,2)} };
\node [rectangle,draw,fill=lightestmain] (E25) at (2,-2.8) {  \texttt{edge(2,5)} };

\node [rectangle,draw] (E12old) at (2,-1.4) {  \texttt{edge(1,2)} };

\node [] (e4) at (0,-.7) { $e_4$ }; 
\node [] (e6) at (2,-1.95) { \textcolor{darkerGreen}{\textbf{$e_6$}} }; 
\node [] (e4old) at (1,-.7) { $e_4'$ }; 

\node [below=-0.2cm of e6] (probE6) {\textcolor{darkerGreen}{\textbf{0.8}}};
\node [left=0.02cm of E12] (probE12) {\textcolor{blue}{0.6}};
\node [right=0.05cm of E12old] (probE12old) {\textcolor{blue}{0.4}};
\node [left=0.05cm of E25] (probE25) {\textcolor{blue}{0.7}};
\node [right=0.05cm of P12] (probP12) {\textcolor{red}{\textbf{0.824}}};

  \path
 	(P12) [-] edge node {} (e4)	  
	(e4) [arrow1] edge node { } (E12)	
	(P12) [-] edge node {} (e4old)	  
	(e4old) [arrow1] edge node { } (E12old)
	;
	 \draw [-, darkerGreen, thick] (E12old) -- (e6);
	 \draw [arrow1, darkerGreen, thick] (probE6) -- (E25);
	
\

\node [] (calculation) at (1,-3.3) { \textbf{\textcolor{red}{0.824}} = \textcolor{blue}{0.6} + \textcolor{blue}{0.4} $*$ \textbf{\textcolor{darkerGreen}{0.8}} $*$ \textcolor{blue}{0.7} }; 
 \end{tikzpicture}}
  \end{minipage}
  \caption{Computation procedure of \texttt{path(1,2)} by \textsc{Problog}.}
    \label{fig:compProbLog}
\end{figure}
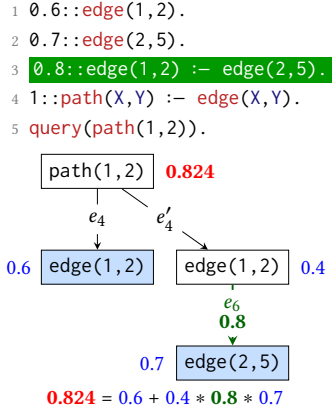

The unexpected result occurs because ProbLog is not typically designed to use rules such as $p:: R \colonminus S_R$ to directly express input correlations. These rules are primarily intended for specifying the probability of rules where the output fact serves as the rule head.

\subsection{Discussion of Syntax Constructs in  \textsc{Praline} and \textsc{Problog}}

In \textsc{Praline}, there are two types of rules: \( p :: R \colonminus S_R \) and \( p :: I ~|~ S_I \). In \textsc{Problog}, only the first type is available.

The first type defines how the output fact \( R \) is derived. If \( R \) can be derived by only one rule, then \( P(R) = P(R \wedge S_R) + P(R \wedge \neg S_R) = P(R \wedge S_R) = p \times P(S_R) \), since \( P(R \wedge \neg S_R) \) is zero by default.

If multiple rules infer \( R \), it is interpreted as follows:
\[
\begin{array}{lll}
    p_1 :: & R & \colonminus ~S_1 \\
    p_2 :: & R & \colonminus ~S_2 \\
    \dots & & \\
    p_n :: & R & \colonminus ~S_n 
\end{array}
\]
\[
P(R) = P(\underline{R \wedge S_1} \vee \underline{R \wedge S_2} \vee \dots \underline{R \wedge S_n}) + P(R \wedge \neg (S_1 \vee S_2 \dots S_n)) = P(\underline{R \wedge S_1} \vee \underline{R \wedge S_2} \vee \dots \underline{R \wedge S_n})
\]
where \( P(R \wedge \neg (S_1 \vee S_2 \dots S_n)) \) is zero. It is because if no rules are applicable to infer \( R \), i.e., $\neg (S_1 \vee S_2\vee \dots S_n)$, then \( R \) is false by default according to probabilistic Datalog semantics.

The second type, \( p :: I ~|~ S_I \), encodes statistical correlations among input facts \( I \), which are considered constraints, rather than the definition of $I$. It denotes \( P(I | S_I) = p \). Unlike the first type, \( P(I \wedge \neg S_I) \) is unknown and may not be zero. If \( P(I \wedge \neg S_I) \) is zero, we may get conflicting probabilities for \( P(I) \), such as \( P(I)_1 = p_1 \) from the rule \( p_1 :: I \) and \( P(I)_2 = P(I \wedge S_I) + P(I \wedge \neg S_I) = P(I \wedge S_I) = p \times P(S_I) \). To avoid conflicts, \( P(I \wedge \neg S_I) \) is not enforced as zero.

When user-provided correlation constraints are conflicting, such as $0.5::I_1$, $0.3::I_2$, $0.6::I_1 \mid I_2$, and $0.7::I_1 \mid \neg I_2$, which imply $P(I_1) = 0.5$ is inconsistent with $P(I_1) = P(I_1|I_2)P(I_2) + P(I_1|\neg I_2)(1-P(I_2)) = 0.6 \times 0.3 + 0.7 \times 0.7$, \textsc{Praline} will return "No solution" and produce no output.

The syntax structure \( p :: I \mid S_I \) enables users to represent conditional probabilities even when only partially known. For example, consider a scenario where a user is aware of \( P(I \mid S_I) = p \) but lacks knowledge about \( P(I \mid \neg S_I) \). Under such conditions, determining an exact probability may not be feasible. Our tool, \textsc{Praline}, is designed to handle such cases by inferring probabilities and providing them as intervals. This is not a reflection of imprecision in our tool, but rather a result of dealing unknown conditional dependencies. In contrast, traditional probabilistic Datalog tools like \textsc{ProbLog} calculate absolute probabilities and assume unspecified probabilities as zero, such as $P(I \mid \neg S_I) = 0$, instead of providing outputs as probability intervals.

\textsc{Praline} allows expressing conditional dependencies that are not precisely known, which is a common scenario in many real-world applications including program analysis tasks. 
In such cases, the probability of output facts cannot be determined precisely, but one can compute lower and upper bounds on the probability. In contrast, \textsc{ProbLog} cannot express such scenarios, as it always produces a single probability value. To capture the same semantics as \textsc{Praline}, one would need to write infinitely many \textsc{ProbLog} programs. Any \textsc{ProbLog} program inherently corresponds to a \textsc{Praline} program where all joint probabilities are known. Sec~\ref{appendix:evidence} shows a \textsc{Praline} program (a simplified version of Figure~\ref{fig:probCond}) and why it would take infinitely many \textsc{ProbLog} programs to encode the same semantics.

\subsection{Semantics of \toolname}
\label{appendix:semantics}

In this section, we formally define the semantics of \toolname. Since \toolname inherits its syntax from Probabilistic Datalog, we extend the possible-world semantics~\cite{fuhr2000probabilistic} of probabilistic Datalog to formalize \toolname.


\emph{Probabilistic Datalog}  
\begin{itemize}
    \item \emph{Rule:} A sentence of the form:  
    \[
    p ~::~ O \colonminus O_1, O_2, \dots, O_n.
    \]  
    Here, \( O \) is an atom, while \( O_1, O_2, \dots, O_n \) are literals. Each atom represents a predicate or relation, and each literal is either an atom or its negation. If the body of a rule is empty, the rule is considered an input fact declaration.
    
    \item \emph{Query:} A query is a conjunction of literals.
\end{itemize}

\emph{\toolname}  inherits the syntax of probabilistic Datalog but relaxes the mutual independence assumption among rule probabilities. In \toolname, rules that declare input facts are allowed to be dependent, whereas rules with non-empty bodies are still treated as mutually independent. Additionally, \toolname introduces new syntax features that allow users to flexibly specify \emph{partially known} or \emph{completely known} conditional dependencies.

\begin{itemize}
    \item \emph{Correlation classes:} \texttt{corr($I_1, I_2,\dots$)} or \textsf{$k~::~$Class($I$)} 
    \item \emph{Conditional dependencies:}  $p ~::~ I \mid S.$
\end{itemize}

The newly added syntax features from \toolname do not modify the probability expression of any derived fact but only affect its evaluation. We first describe how the probability expression is obtained from a \toolname  program according to the semantics of Probabilistic Datalog and then discuss how it is evaluated based on \toolname's semantics.

The probability expression is computed based on the probabilities of the ground facts and instantiated rules used in deriving the fact.

\begin{definition}
A Datalog program \( D \) is \emph{modularly stratified} if there exists an assignment of ordinal levels to ground atoms such that:  
(a) if a ground atom appears negatively in the body of a rule, then the ground atom in the head of that rule is assigned a strictly higher level;  
(b) if a ground atom appears positively in the body of a rule, then the ground atom in the head is assigned a level that is at least as high.
\end{definition}

For a Probabilistic Datalog program \( D \), let \( S_g \) denote the set of grounded rules, consisting of $S_d$ and $S_p$, i.e., $S_g = S_d \cup S_p$. We define \( S_d \) as the set of deterministic rules (\( p = 1 \)) and \( S_p \) as the set of probabilistic (indeterministic, $0 < p < 1$) rules.  
The set of all possible deterministic programs of \( D \) is given by:  
\[
S(D) = \{ S_d \cup x \mid x \in \text{PowerSet}(S_p) \}
\]

\begin{lemma}
A probabilistic Datalog program \( D \) is modularly stratified if every element in the set of its possible deterministic programs \( D' \in S(D) \) is modularly stratified.
\end{lemma}

\begin{lemma}
A modularly stratified Datalog program has a least fixed point.
\end{lemma}

The least fixed point of a Datalog program corresponds to its least Herbrand model, which captures all facts that can be derived from the program.

\begin{definition}
The Herbrand universe of a Datalog program \( D \) is the set of all ground terms that can be formed using the constants in \( D \). For a deterministic program \( D' \) (i.e., \( D' \in S(D) \)) with a complete, well-founded Herbrand model, let \( \omega(D') \) denote this model. Additionally, \( \omega(D') \) represents a possible world of \( D \).
\end{definition}

For each deterministic program $D'$, \( \omega(D') \) is defined as a set of all ground rules ($r$) used in $D'$, i.e., $\omega(D') = \{r\}$. An input fact can also be represented as a rule with a body that is always true.


\begin{definition}
Given a Probabilistic Datalog program \( D \), the set of possible worlds is defined as  
\[
W(D) = \{ \omega(D') \mid D' \in S(D) \}.
\]
\end{definition}

We also use the judgment $\omega \models D$ to denote that $\omega$ is a possible world consistent with the \toolname program $D$. 

\begin{definition}
Let \( \mu \) be a mapping that assigns each possible world \( \omega \) (\(\omega \in W(D)\)) to its full joint probability distribution over the ground terms/rules in \( \omega \).
\end{definition}

Given that \( \omega = \{r\} \), if all input facts and rules are assumed to be mutually independent, \( \mu \) maps each joint distribution of \( r \) to a unique value (\emph{one-to-one mapping}). Otherwise, without the independence assumption, \( \mu \) maps it to a set of values (\emph{one-to-many mapping}).

\begin{definition}
For a possible world \( \omega \), let \( \epsilon: \omega \to \texttt{EK} \) be a mapping that assigns each ground rule (\( r \in \omega \)) to an event key (\texttt{EK}). This mapping satisfies the following constraints:
\begin{itemize}
    \item \( \forall r \in \omega. \quad \mu(r) = 1 \leftrightarrow \epsilon(r) = \texttt{true} \).
    \item \( \forall r, r' \in \omega. \quad \epsilon(r) = \epsilon(r') \rightarrow (r = r' \vee (\mu(r) = 1 \wedge \mu(r') = 1)) \).
\end{itemize}
\end{definition}

If the probability of a rule \( r \) is constantly true (i.e., \( \mu(r) = 1 \)), it is not mapped to an \texttt{EK} variable. Instead, it is assigned a constant value. If two rules share the same \texttt{EK} variable, they are either identical (i.e., \( r = r' \)) or both have a probability of 1.

Given an event key variable $v$, the inverse mapping $\epsilon^{-1}(v)$ denotes the corresponding ground rule $r$.


\begin{definition}
For each output fact \( O \), we define a mapping \( \theta \) from \( O \) to a Boolean expression (\( \mathbb{B} \)) over event keys (\texttt{EK}). This mapping is defined as follows:

\begin{itemize}
    \item For an output fact \( O \),
    \[
    \theta(O) = \bigvee\nolimits_{r_o} \theta(r_o),
    \]
    where \( r_o \) is a rule whose head matches \( O \) and $r_o \in \omega,~\omega \in W(D)$.
    
    \item For a rule \( r \) of the form \( p :: O \colonminus O_1, O_2, \dots, O_n \),
    \[
    \theta(r) = \epsilon(r) \wedge \theta(O_1) \wedge \theta(O_2) \wedge \dots \wedge \theta(O_n).
    \]

    \item For a negated fact \( \neg O \),
    \[
    \theta(\neg O) = \neg \theta(O).
    \]
\end{itemize}
\end{definition}

The subtlety lies in the case of an input fact \( I \). If rule \( r \) is used to declare the input fact, i.e., \( p :: I \) is a special case of \( p :: I \colonminus O_1, O_2, \dots, O_n \) where \( O_1 \wedge O_2 \wedge \dots \wedge O_n = \texttt{true} \), then \( \theta(I) = \epsilon(r) \).

\begin{definition}
We define a mapping \( \mathcal{W} \) that assigns each Boolean expression \( B \) to the set of possible worlds in which \( B \) holds true. 

The mapping \( \mathcal{W} \) satisfies the following properties:

  \begin{itemize}
        \item $\mathcal{W}(\texttt{true}) = W(D)$
        \item $\mathcal{W}(\neg B) = W(D) \setminus \mathcal{W}(B)$
        \item $\mathcal{W}(B_1 \wedge B_2) = \mathcal{W}(B_1) \cap \mathcal{W}{B_2}$
        \item $\mathcal{W}(B_1 \vee B_2) = \mathcal{W}(B_1) \cup \mathcal{W}(B_2) $
        \item $\mathcal{W}(B) = \{\omega | r = \epsilon^{-1}(B),~ r \in \omega,~ \omega \models D \}$
    \end{itemize}

\end{definition}

For the last case, \( \mathcal{W}(B) \), the Boolean expression \( B \) represents an atomic formula corresponding to an input fact, and \( r \) is the rule that declares this input fact.

Specifically, for any output fact \( O \) in the \toolname program \( D \), let \( \theta(O) \) denote the Boolean expression over event keys representing \( O \). The set of possible worlds in which \( O \) holds is given by \( \mathcal{W}(\theta(O)) \).  

Given a possible world \( \omega \), if \( \omega \in \mathcal{W}(\theta(O)) \), we establish the judgment \( \omega \models O \), meaning that \( O \) can be inferred from \( \omega \).

\begin{definition}
    Given a full joint probability distribution \( \mu \), the probability expression of an output fact \( O \) is computed as follows:
    \[
    P_{\mu}(O) = \sum_{\omega \in \Omega} P_\mu (\omega),
    \]
    where \( \Omega = \{\omega \mid \omega \models O,~ (\omega, \mu) \models D\} \).
\end{definition}

The judgment \( (\omega, \mu) \models D \) indicates that the world \( \omega \) and the full joint probability distribution \( \mu \) are consistent with the input \toolname program \( D \).

For instance, consider a \toolname program where an input fact is specified as follows:  
\texttt{0.5 :: I1.}, denoted as \( r_1 \). If \( \mu(r_1) \neq 0.5 \), then the judgment \( (\omega, \mu) \models D \) is no longer valid.

In a classical probabilistic Datalog program, all rules are assumed to be mutually independent. Consequently, we have:

\[
P_\mu(\omega) = P_\mu\left(\bigwedge\nolimits_{r \in \omega} r\right) = \prod\nolimits_{r \in \omega} P_\mu(r) = \prod\nolimits_{r \in \omega} \mu(r).
\]

However, in a \toolname program, this independence assumption is relaxed. Specifically, rules that define input facts (i.e., rules with an empty body) are not required to be independent. As a result, in \toolname, we may have:

\[
P_\mu\left(\bigwedge\nolimits_{r \in \omega} r\right) \neq \prod\nolimits_{r \in \omega} P_\mu(r).
\]

To determine \( P_\mu(\omega) \), users must provide a complete conditional probability distribution \( \mu \), which uniquely determines \( P_\mu\left(\bigwedge\nolimits_{r \in \omega} r\right) \). However, specifying the full distribution may be impractical in many cases. To address this, \toolname allows users to provide \emph{partially specified} conditional probability information within the \toolname program \( D \). Consequently, multiple probability distributions \( \mu \) may satisfy \( D \), whereas in classical probabilistic Datalog, there is a unique \( \mu \) that satisfies the input program \( D \).

For example, consider the following input facts and their dependencies in a \toolname program:
\[
\texttt{0.6 :: I1}, \quad \texttt{0.7 :: I2}, \quad \texttt{0.6 :: I3}, \quad \texttt{0.8 :: I3 | I1, I2}.
\]
In this case, multiple probability distributions \( \mu \) satisfy these constraints. Consequently, the probability of an output fact \( P(O) \) is no longer a singular value but a set of values, defined as follows:
\[
P(O) = \{ P_\mu(O) \mid \exists \omega. (\omega, \mu) \models D \}.
\]

To obtain a valid set of probability distributions \( \mu \) consistent with the \toolname program \( D \), this can be formulated as a constraint-solving problem. Any complete model that satisfies the constraints specified by \( D \) represents a valid \( \mu \). 
The semantic encoding of these constraints are as follows:
\begin{align}
\llbracket p~::~I. \rrbracket ~&=\quad P(I) = p \\
\llbracket p~::~I_0~|~I_1,I_2,\dots I_n. \rrbracket ~&=\quad P(I_0\wedge I_1 \dots \wedge I_n) = p\times P(I_1 \wedge \dots \wedge I_n) \\
\llbracket k_1~::~\mathsf{Class}(I_1). ~k_2~::~\mathsf{Class}(I_2).~ \rrbracket ~&=\quad I_1 \indepSym I_2 \text{ if } k_1 \neq k_2; \\
~&~\quad \text{Otherwise}, I_1 \not\indepSym I_2 \\
\llbracket \mathsf{corr}(I_1,I_2) \rrbracket  ~&=\quad I_1 \not\indepSym I_2 \\
\llbracket p~::~O~\colonminus~O_1,O_2,...O_n. \rrbracket ~&=\quad P(r) = p \text{ if } r \text{ denotes the rule } p~::~O~\colonminus~O_1,O_2,...O_n. \\
~&=\quad \forall r,~I. ~\quad r \indepSym I. \\
 ~&=\quad \forall r_i,~r_j.~\quad ~ (i \neq j) \;\Rightarrow\; r_i \indepSym r_j
\end{align}

$\indepSym$ denotes mutual independence. 
The notation \( r \) refers only to rules with non-empty bodies, excluding input declarations.
If users do not explicitly declare the correlation class of an input fact $I$, we assign it a default class ID of $-1$, which satisfies the following property:
$
\forall I_i, I_j.~\textsf{Class}(I_i) = -1 \implies I_i \indepSym I_j
$.
 All the above probability computation $P(x)$ can be rewritten as $\mu(x)$.
The semantics above naturally extends to multiple correlation class declarations.

\begin{align}
\llbracket k_1 :: \mathsf{Class}(I_1).  \; \dots \; k_n :: \mathsf{Class}(I_n). \rrbracket 
&=
\begin{cases}
I_1 \indepSym I_2 \indepSym \dots I_n & \text{if } \forall i \neq j, \; k_i \neq k_j \\
\bigwedge\limits_{\substack{1 \leq i < j \leq n}} 
\begin{cases}
I_i \indepSym I_j & \text{if } k_i \neq k_j \\
I_i \not\indepSym I_j & \text{if } k_i = k_j\neq -1
\end{cases} & \text{otherwise}
\end{cases} 
\\
\llbracket \mathsf{corr}(I_1,I_2,\dots, I_n) \rrbracket  &= \bigwedge_{1 \leq i < j \leq n} I_i \not\indepSym I_j
\end{align}

\subsection{Optimized Satisfiability Check}
\label{appendix:SAT}

The iterative refinement from Section~\ref{sec:iterative} requires frequent satisfiability checks, most of which return unsatisfiable. To speed up this process, constraints are overapproximated by introducing joint probability variables over intermediate relations that are $k$ steps from the root node.

In the \textsc{BoundBounds} procedure, we iteratively check unsatisfiability until it reaches the SAT region. To optimize performance, we propose an overapproximated version $\phi'$ of the original set of constraints $\phi$, such that $\phi \implies \phi'$. Thus, $\neg(\phi' \wedge l \leq \textsf{Expr}(R) \leq u) \implies \neg(\phi \wedge l \leq \textsf{Expr}(R) \leq u)$. Hence, if we get the UNSAT result from the approximated version $\phi'$, it implies the unsatisfiability of the original version $\phi$.

To see why the approximated version improves performance and why $\phi \implies \phi'$, we provide an example to illustrate this.

\begin{figure}
    \centering
    \scalebox{0.75}{\begin{tikzpicture}[every node/.style={minimum size=8mm, inner sep=1pt}, sibling distance=20mm, level distance=20mm]

\tikzstyle{internal} = [rectangle, draw=black, fill=blue!30, thick]
\tikzstyle{input} = [rectangle, draw=black, fill=orange!30, thick]

\node[internal] (A) at (0, 0) {A};

\node[internal] (B) at (-3, -2) {B};
\node[internal] (C) at (-1, -2) {C};
\node[internal] (D) at (1, -2) {D};
\node[internal] (E) at (3, -2) {E};

\node[input] (I1) at (-4, -4) {I1};
\node[input] (I2) at (-3, -4) {I2};
\node[input] (I3) at (-2, -4) {I3};
\node[input] (I4) at (-1, -4) {I4};
\node[input] (I5) at (0, -4) {I5};
\node[input] (I6) at (1, -4) {I6};
\node[input] (I7) at (2, -4) {I7};
\node[input] (I8) at (3, -4) {I8};

\draw[thick] (A) -- ++(-0.2, -1) coordinate (mid1) -| (B);
\draw[thick] (mid1) -| (C);

\draw[thick] (A) -- ++(0.2, -1) coordinate (mid2) -| (D);
\draw[thick] (mid2) -| (E);

\draw[thick] (B) -- ++(0, -1) coordinate (mid3) -| (I1);
\draw[thick] (mid3) -| (I2);

\draw[thick] (C) -- ++(0, -1) coordinate (mid4) -| (I3);
\draw[thick] (mid4) -| (I4);

\draw[thick] (D) -- ++(0, -1) coordinate (mid5) -| (I5);
\draw[thick] (mid5) -| (I6);

\draw[thick] (E) -- ++(0, -1) coordinate (mid6) -| (I7);
\draw[thick] (mid6) -| (I8);

\draw [decorate,decoration={brace,amplitude=10pt,mirror,raise=5pt},thick] (-4.5,-4.5) -- (-0.5,-4.5) node[midway,yshift=-25pt]{$V_1$};

\draw [decorate,decoration={brace,amplitude=10pt,mirror,raise=5pt},thick] (-0.4,-4.5) -- (3.5,-4.5) node[midway,yshift=-25pt]{$V_2$};

\node at (-.6, -.8) {\textcolor{darkmain}{$e_1$}};
\node at (.6, -.8) {\textcolor{darkmain}{$e_2$}};
\node at (-3.5, -2.8) {\textcolor{darkmain}{$e_3$}};
\node at (-1.5, -2.8) {\textcolor{darkmain}{$e_4$}};
\node at (.5, -2.8) {\textcolor{darkmain}{$e_5$}};
\node at (2.5, -2.8) {\textcolor{darkmain}{$e_6$}};

\end{tikzpicture}}
    \caption{An example derivation graph.}
    \label{fig:exampleOptimize}
\end{figure}
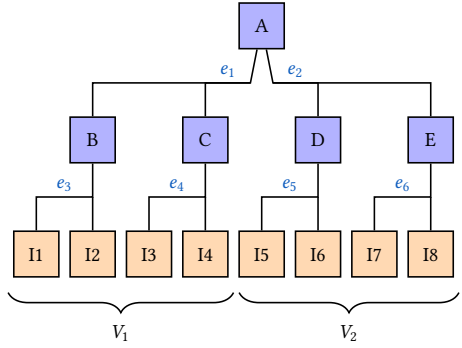

There is a derivation graph in Figure~\ref{fig:exampleOptimize}, focusing on computing the probability of the root node (relation $A$). $I_1$ till $I_8$ are input facts. There are two correlation classes $\{I_1, I_2, I_3, I_4\}$ and $\{I_5, I_6, I_7, I_8\}$. The original constraint $\phi$ is shown on the left of Figure~\ref{fig:constraint}. Here, $\phi_{\text{corr}}$ denotes the constraints from the correlation classes, and we omit the details of $\phi_{\text{corr}}$ here.

\begin{figure}
    \centering
    \[
\begin{array}{rl|rl}
     \phi: & \sum_{b \in \mathbb{B}^4} V_1[b] = 1 \quad (I_1,I2,I3,I4)  &\phi':& \sum_{b \in \mathbb{B}^2} V_1[b] = 1 \quad (B,C) \\
     & \sum_{b \in \mathbb{B}^4} V_2[b] = 1 \quad (I_5,I6,I7,I8)  & & \sum_{b \in \mathbb{B}^2} V_2[b] = 1 \quad (D,E) \\
     & \forall b \in \mathbb{B}^4.~V_1[b] \in [0,1],~V_2[b] \in [0,1] & &  
      \forall b \in \mathbb{B}^2.~V_1[b] \in [0,1],~V_2[b] \in [0,1] \\
      & \sum_{i,j,k} V_1[1ijk] = P(I_1)  & & \sum_{i} V_1[1i] \in [l_B, u_B] \\
      & \sum_{i,j,k} V_1[i1jk] = P(I_2)  & & \sum_{i} V_1[i1] \in [l_C, u_C] \\
      & \sum_{i,j,k} V_1[ij1k] = P(I_3)  & &  \\
      & \sum_{i,j,k} V_1[ijk1] = P(I_4)  & &  \\
    & \sum_{i,j,k} V_2[1ijk] = P(I_5)  & & \sum_{i} V_2[1i] \in [l_D, u_D] \\
    & \sum_{i,j,k} V_2[i1jk] = P(I_6)  & & \sum_{i} V_2[i1] \in [l_E, u_E] \\
    & \sum_{i,j,k} V_2[ij1k] = P(I_7)  & &  \\
    & \sum_{i,j,k} V_2[ijk1] = P(I_8)  & &  \\
    & \phi_{corr} & & \phi'_{corr}\\
\end{array}
\]
    \caption{Original constraints $\phi$ vs. overapproximated constraints $\phi'$.}
    \label{fig:constraint}
\end{figure}

The overapproximated version is shown on the right of Figure~\ref{fig:constraint}. The encoding starts from the \emph{intermediate relations} ($B, C, D, E$) instead of the input facts (See Figure~\ref{fig:exampleOptimize}). From the correlation analysis, we know that $B, C$ are dependent and $D, E$ are dependent. We also treat all unknown cases as if they are inside the same correlation class for soundness reasons.

In $\phi'$, there are also two correlation classes $V_1 = \{B, C\}$ and $V_2 = \{D, E\}$. For the constraint $\sum_{i} V_1[1i] \in [l_B, u_B]$, $\sum_{i} V_1[1i]$ represents the probability of relation $B$, i.e., $P(B)$, and $[l_B, u_B]$ is obtained from the correlation analysis. For example, assuming that $l^*_B, u^*_B$ denote the ground truth lower and upper bounds of $P(B)$ satisfying $\phi$. Since our correlation analysis is sound, we know that $l_B \leq l^*_B \leq u^*_B \leq u_B$. As a result, $\phi \implies l_B \leq P(B) \leq u_B$. A similar implication happens to $C, D, E$. Meanwhile, with the correlation analysis, we can get the correlation type between a pair of relations from $\{B, C, D, E\}$, and encode them as constraints $\phi'_{corr}$. Due to the soundness of correlation analysis, according to Theorem~\ref{theorem:corr}, we know that $\phi \implies \phi'_{\text{corr}}$.

To summarize, $\phi \implies \phi'$.

In terms of performance improvement, it is because the approximated version has fewer joint probability variables compared to the original version. We can see that the number of variables in $\phi'$ is significantly less than the number of variables in $\phi$. For instance, $\phi$ has 32 variables, while $\phi'$ only has 8 variables. It would 
significantly
reduce the number of variables when this derivation graph is larger, and we pick a suitable step $k$ to select \emph{intermediate relations}.

It is worth noting that $k$ is not a fixed value. The core idea behind our over-approximation is to partition the derivation graph $G$ into two disjoint sets, $S_s$ and $S_t$, such that $S_s \cup S_t = \mathsf{Nodes}(G)$ and $S_s \cap S_t = \emptyset$. We omit all edges that connect nodes from $S_s$ to nodes in $S_t$.

Assume that $S_s$ contains the root node (e.g., the output fact $A$ in Figure~\ref{fig:exampleOptimize}). Instead of expressing the probability of $A$ in terms of input facts, our over-approximated formulation encodes this probability based on the leaf nodes within $S_s$. Here, we define leaf nodes as those without outgoing edges.

For example, in Figure~\ref{fig:exampleOptimize}, one possible cut yields $S_s = \{A, B, C, D, E\}$ and $S_t = \{I1, I2, I3, I4, I5, I6, I7, I8\}$. Alternatively, a valid cut could be $S_s = \{A, B, C, D, E, I5, I6, I7, I8\}$ and $S_t = \{I1, I2, I3, I4\}$. We encode the cut selection as an ILP (Integer Linear Programming) problem, where the objective is to choose the leaf nodes of $S_s$ along the cut such that the number of incoming edges to these leaf nodes is maximized.


\subsection{Inferring Symbolic Probability Expressions}

\textsc{Theorem} 1.
Let $E_1$ and $E_2$  denote the probability expressions of events $A$ and $B$ respectively. Then, we have: (1) If  $\ominus E_1 \rightsquigarrow E$, then $E$ represents the probability of $\neg A$; (1) if \ $ E_1 \oplus E_2 \rightsquigarrow E$, then $E$ represents the probability of event $A \lor B$
(3) iff \  $ E_1 \otimes E_2 \rightsquigarrow E$,then $E$ represents the probability of event $A \land B$

In this subsection, we present soundness proofs for $\ominus$, $\otimes$ and $\oplus$ computations used in deriving symbolic probability expressions. We also present some Lemmas that are used to prove the soundness of both $\otimes$ and $\oplus$ computation.

\subsubsection{\textsc{Theorem~\ref{theorem:AndOr} ($\ominus$).} } Let $E_A$ denote the probability of event $A$, then we have:
\begin{center}
If \  $\ominus E_A \rightsquigarrow E$,then $E$ represents the probability of event $\neg A$.
\end{center}

\textsc{Proof}. Let $Pr(A) = \textsf{Expr}(A) = \exptemplate[\vec{\sigma_A}]$ where $\vec{\sigma}_A = \{\lambda_0, \lambda_1, \ldots, \lambda_i,\ldots\}$, as per Definition~\ref{def:temp-inst}. Hence, $E_A = Pr(A) = \sum_i \lambda_i \psi_i$.

Since $Pr(\neg A) = 1 - Pr(A)$, it follows that:
\[ Pr(\neg A) = 1 - E_A = 1 -  \sum_i \lambda_i \psi_i. \]

From Definition~\ref{def:exprTemplate}, we know $\sum_i \psi_i = 1$ as it spans the complete joint probability space.
This leads to:
\[ Pr(\neg A) = 1 - \sum_i \lambda_i \psi_i = \sum_i \psi_i - \sum_i \lambda_i \psi_i = \sum_i (1-\lambda_i) \psi_i. \]

By the $\ominus$ rule, $\ominus E_A \rightsquigarrow E$ yields $E = \sum_i (1-\lambda_i) \psi_i$. Therefore, we conclude:
\[ Pr(\neg A) = E. \]
\qed

\begin{lemma}
\label{lemma:term}
Given two events $A$ and $B$, if $Pr(A) = \psi_A$, $Pr(B) = \psi_B$ and $\psi_A \neq \psi_B$, then events $A$ and $B$ are mutually exclusive.    
\end{lemma}

\textsc{Proof}. 
Based on the definition of product terms (Definition~\ref{def:pterm}), we have 
$P(A) = \psi_A = \prod_{c \in \mathcal{C}} V_c[b]$, $P(B) = \psi_B = \prod_{c \in \mathcal{C}} V_c[b']$, and $b\neq b'$. Both $b$ and $b'$ represent bit-vectors.
Here, we use $V_c$ to denote \textsf{Rep}($c$) for the ease of presentation.

Given a boolean event $A$, we can rewrite $A = A_1 \wedge A_2 \wedge A_n$, where $n = |\mathcal{C}|,~P(A_i) = V_i[b]$. Based on the independence definition of distinct correlation classes, we know that $P(A) = P(A_1)P(A_2)\dots P(A_n)$. Similarly, $P(B) = P(B_1)P(B_2)\dots P(B_n)$. For the same correlation class $i$, $P(A_i) = V_i[b]$ and $P(B_i) = V_i[b']$. For instance, $b=001$ and $b'=100$, as for $b$, it represents the joint probability of the first two events being false and the third event being true, whereas $b'$ represents the joint probability of the first event being true while the last two events being false. Hence, $A_i$ and $B_i$ represent two mutually-exclusive events, i.e., $P(A_i B_i)=0$. The joint probability of $P(A\wedge B)$ is as follows:
\[
P(A \wedge B) = \prod_{1 \leq i \leq |\mathcal{C}|} P(A_iB_i) \quad \text{using definitions of correlation class}
\]

Hence, $P(A\wedge B) = 0$ as $P(A_iB_i) = 0$.
We can also get $P(A|B) = P(A \wedge B)/P(B) = 0$.
\qed


\subsubsection{\textsc{Theorem~\ref{theorem:AndOr} ($\otimes$).} } Let $E_A$ and $E_B$  denote the probability of events $A$ and $B$ respectively. Then, we have: 
\begin{center}
If \  $\vdash E_A \otimes E_B \rightsquigarrow E$,then $E$ represents the probability of event $A \land B$.
\end{center}

\textsc{Proof}. Given $E_A = P(A)$, $E_B = P(B)$, we will first compute $P(A\wedge B)$ and prove it is equivalent to $E$.

Assuming that $E_A = \sum \lambda_k^A\psi_k$, $E_B = \sum \lambda_k^B \psi_k$ and $ \psi_k = \prod_{c \in \mathcal{C}} V_c[b]$. 
Let $e_k$ be an event, where $P(e_k) = \psi_k$.
%
Let $e^{-}$ be the event which is a complement of the event $e_1 \vee \dots \vee e_k \dots \vee e_n$.

\[
\begin{array}{lll}
(1) P(A \wedge B) & = & P(A \wedge B \wedge e_1) + \dots  P(A \wedge B \wedge e_k)+ \dots  P(A \wedge B \wedge e_n) + P(A \wedge B \wedge e^-) \\
(2) & = & P(A \wedge B \wedge e_1) + \dots  P(A \wedge B \wedge e_k)+ \dots  P(A \wedge B \wedge e_n) \\
(3) & = & P(A \wedge B | e_1) P(e_1) + \dots P(A \wedge B | e_k)P(e_k) + \dots P(A \wedge B | e_n)P(e_n)  \\
(4) & = & P(A_1 \wedge B_1) P(e_1) + \dots P(A_n \wedge B_n) P(e_n)
\\
(5) & = & P(\bigvee_{i\in I} A_1^i \wedge \bigvee_{j\in J} B_1^j)P(e_1) + \dots
\\
(6) & = & P(\bigvee_{i,j \in I \times J} A_1^i \wedge B_1^j)P(e_1) + \dots
\\
(7) & = & (\sum_{i,j \in I \times J} P(A_1^i \wedge B_1^j) )P(e_1) + \dots
\end{array}
\]

\textsc{Step (1)}: Since each $e_i$ is mutually exclusive (according to Lemma~\ref{lemma:term}), we can get $P(e^-) + \sum_i P(e_i)  = 1$.
Hence, we can rewrite $P(A\wedge B)$ as $P(A \wedge B \wedge e_1) + \dots  P(A \wedge B \wedge e_k)+ \dots  P(A \wedge B \wedge e_n) + P(A \wedge B \wedge e^-)$.

\textsc{Transformation from (1) - (2)}: Given that $P(A) = \sum \lambda_k^AP(e_k)$ and each $e_k$ is mutually exclusive, we can infer that the boolean event $A = \vee e_k'$, where $e_k'$ is an event with probability $P(e_k') = \lambda_k^AP(e_k)$. Let's assume an event $c$ with probability $P(c) = \lambda_k^A$. As we know that $c$ is the coefficient which is independent of the term event $e_k$, we can get $e_k' = e_k \wedge c$. Hence $P(e_k'|e_k) = P(c \wedge e_k | e_k) = P(c)P(e_k)/P(e_k) = \lambda_k^A$, and $P(e_k \wedge e_k') = P(e_k')$.

We have the following:
\[
\begin{array}{lll}
     P(e^- \wedge (e_1' \vee e_2' \vee \dots e_n')) & = & P(\underline{e^- \wedge e_1'} \vee \underline{e^- \wedge e_2'} \vee \dots \underline{e^- \wedge e_n'})
     \\
     & = & P(\underline{e^- \wedge e_1' \wedge e_1} \vee \underline{e^- \wedge e_2' \wedge e_2} \vee \dots \underline{e^- \wedge e_n' \wedge e_n})
     \\
      & = & P(e^- \wedge e_1' \wedge e_1) + P(e^- \wedge e_2' \wedge e_2) + \dots P(e^- \wedge e_n' \wedge e_n) \\
      & = & 0
\end{array}
\]

From the above, we can get $e^-$ is also mutually exclusive with the union of $e_1' \vee e_2' \vee \dots e_n'$. As $A = \vee e_k'$, we can get $e^-$ is mutually exclusive with $A$. 
As a result, $P(A\wedge B \wedge e^-) = 0$.


\textsc{Transformation from (2) - (3)}: According to Bayes' theorem, $P(A\wedge B \wedge e_1)$ = $P(A \wedge B | e_1)P(e_1)$. This applies to $e_2 \dots, e_n$.

\textsc{Transformation from (3) - (4)}: 
For instance, we introduce events $A_1$ and $B_1$ such that $P(A_1) = \lambda_1^A$ and $P(B_1) = \lambda_1^B$. 
We have $(A_1\wedge B_1) \indepSym e_1$ (i.e., the product of coefficients $\lambda$ are independent of the product terms $\psi$.)
Since $A = (A_1 \wedge e_1) \vee \ldots (A_n \wedge e_n)$ and $B = (B_1 \wedge e_1) \vee \ldots (B_n \wedge e_n)$, we have $A \wedge B = (A_1 \wedge B_1 \wedge e_1) \vee \ldots (A_n \wedge B_n \wedge e_n)$. Thus $P(A\wedge B |e_1) = P(A_1 \wedge B_1 \wedge e_1 | e_1) = P(A_1\wedge B_1)$

\textsc{Transformation from (4) - (5)}: Event $A_1$ can be represented as a DNF formula: $A_1 = \bigvee_{i \in I} A_1^i$, where each  $A_1^i$ is a disjoint conjunctive clause. 
$A_1^i = \bigwedge_{x^+ \in X^+} {x^+}\bigwedge_{x^- \in X^-} {x^-}$, where $x^+$ represents the rule probability variable and $x^-$ represents the negated rule probability variable. We also use $P_{x^+}$ to represent the concrete probability value for variable $x^+$, similar to $P_{x^-}$.
The probability of each conjunctive clause $P(A_1^i) = \prod_{x^+ \in X^+}P_{x^+}\prod_{x^- \in X^-}(1-P_{x^-})$.

\textsc{Transformation from (5) - (6)}: We follow the distributive law of conjunction over disjunction operator.

\textsc{Transformation from (6) - (7)}: Each $A_1^i\wedge B_1^j$ is mutually disjoint from another one 
$A_1^{i'}\wedge B_1^{j'}$ if $i \neq i' \vee j \neq j'$. Due to this mutual disjoint relation, we can rewrite the probability of a DNF formula over disjoint conjunction as a summation of the conjunction probabilities.

When computing $P(A_1^i \wedge B_1^j)$, we have to consider two cases. Since $P(A_1^i) = \prod_{x^+ \in X^+}P_{x^+}\prod_{x^- \in X^-}(1-P_{x^-})$ and $P(B_1^j) = \prod_{y^+ \in Y^+}P_{y^+}\prod_{y^- \in Y^-}(1-P_{y^-})$, if the set $X^+ \cap Y^- \neq \emptyset$ or $X^- \cap Y^+ \neq \emptyset$, we know that one term in $A_1^i$ is a negated version of the term in $B_1^j$, so that their conjunction is 0. Otherwise, without the contradiction, we can compute the joint probability of $A_1^i$ and $B_1^j$. If there are no overlapping terms, e.g. $X^+ \cap Y^+ = \emptyset$, due to the mutual independence among rule probabilities, we can have $P(A_1^i \wedge B_1^j) = P(A_1^i)P(B_1^j)$. Otherwise, if there are overlapping terms and denote the conjunction of overlapping terms as S, let's assume that $A_1^i = a_1 \wedge \dots \wedge S \wedge a_n$ and $B_1^j = b_1 \wedge \dots \wedge S \wedge b_n$, then $A_1^i \wedge B_1^j = (a_1 \wedge \dots \wedge S \wedge a_n) \wedge (b_1 \wedge \dots \wedge S \wedge b_n) = a_1 \wedge \dots \wedge S \wedge a_n \wedge b_1 \wedge \dots \wedge b_n$ by removing the overlapping part $S$.

By following the above computation, 
we have the formula in $(\sum_{i,j \in I \times J} P(A_1^i \wedge B_1^j) )$ shown in Line (7) is evaluated the same as $\textsf{JointProb}(P(A_1), P(B_1))$. Hence, we can have the following:
\[
\begin{array}{ll}
   P(A\wedge B)  &= (\sum_{i,j \in I \times J} P(A_1^i \wedge B_1^j) ) \cdot P(e_1) \dots + \sum_{i,j \in I \times J} P(A_n^i \wedge B_n^j) ) P(e_n)  \\
     &= \textsf{JointProb}(P(A_1), P(B_1))  \cdot P(e_1) \dots + \textsf{JointProb}(P(A_n), P(B_n))  \cdot P(e_n) \\
     &= \sum \textsf{JointProb}((P(A_i), P(B_i))\cdot P(e_i) \\
     &= P(A) \otimes P(B)
\end{array}
\]



\qed





\subsubsection{\textsc{Theorem~\ref{theorem:AndOr} ($\oplus$) .} } Let $E_A$ and $E_B$  denote the probability of events $A$ and $B$ respectively. Then, we have: 
\begin{center}
If \ $\vdash E_A \oplus E_B \rightsquigarrow E$, then $E$ represents the probability of event $A \lor B$.
\end{center}

\textsc{Proof}. 
Here, we do not provide a formal proof for $\oplus$ as it is very similar to the proof of Theorem~\ref{theorem:AndOr} ($\otimes$), which mainly utilizes the Lemma~\ref{lemma:term}.

\subsection{Inferring Correlation Types}
\label{appendix:corr}

In this section, we give soundness proof for Theorem~\ref{thm:input} and Theorem~\ref{theorem:corr}.

\subsubsection{Soundness of Theorem~\ref{thm:input}}

 We prove the soundness of this theorem for each $\star \in \{+, -,\bot \}$:
\begin{itemize}[leftmargin=*]
    \item {\bf Case +}: There are four rules in Figure~\ref{fig:corr-input} to infer the positive correlation $I_1 \blacktriangleright^+ I_2$. 
    \begin{itemize}
        \item \textsc{Rule Id} states that $I \blacktriangleright^+ I$, because $P(I|I) = 1 > P(I)$. 
         \item \textsc{Rule Symm} states that $I_2 \blacktriangleright^+ I_1$ if $I_1 \blacktriangleright^+ I_2$. Given $I_1 \blacktriangleright^+ I_2$, we can infer that $P(I_1|I_2) > P(I_1)$. Thus, $P(I_2|I_1) = P(I_1|I_2)P(I_2)/P(I_1) > P(I_1)P(I_2)/P(I_1) = P(I_2)$, which implies $I_2 \blacktriangleright^+ I_1$.
         \item \textsc{Rule Semantic} states that $I_1 \blacktriangleright^+ I_2$ if $P(I_1\wedge I_2) > P(I_1)P(I_2)$. With this information, we can infer that $P(I_1|I_2) = P(I_1\wedge I_2)/P(I_2) > P(I_1)P(I_2)/P(I_2) = P(I_1)$, implying $I_1 \blacktriangleright^+ I_2$.
    \end{itemize} 
    \item {\bf Case -}: There are three rules in Figure~\ref{fig:corr-input} to infer the negative correlation $I_1 \blacktriangleright^- I_2$. 
    \begin{itemize} 
        \item \textsc{Rule Symm} states that $I_2 \blacktriangleright^- I_1$ if $I_1 \blacktriangleright^- I_2$. Given $I_1 \blacktriangleright^- I_2$, we can infer that $P(I_1|I_2) < P(I_1)$. Thus, $P(I_2|I_1) = P(I_1|I_2)P(I_2)/P(I_1) < P(I_1)P(I_2)/P(I_1) = P(I_2)$, which implies $I_2 \blacktriangleright^- I_1$.
         \item \textsc{Rule Semantic} states that $I_1 \blacktriangleright^- I_2$ if $P(I_1\wedge I_2) < P(I_1)P(I_2)$. With this information, we can infer that $P(I_1|I_2) = P(I_1\wedge I_2)/P(I_2) < P(I_1)P(I_2)/P(I_2) = P(I_1)$, implying $I_1 \blacktriangleright^- I_2$.
    \end{itemize}
    \item {\bf Case $\bot$}: There are three rules in Figure~\ref{fig:corr-input} to infer the independence relation $I_1 \blacktriangleright^\bot I_2$.
    \begin{itemize}
        \item \textsc{Rule Indep} states that $I_1 \blacktriangleright^\bot I_2$ if $\textsf{Class}(I_1) \neq \textsf{Class}(I_2)$. This is correct based on our assumption that elements from different correlation classes are independent.
        \item \textsc{Rule Symm} states that $I_2 \blacktriangleright^\bot I_1$ if $I_1 \blacktriangleright^\bot I_2$. Given $I_1 \blacktriangleright^\bot I_2$, we can infer that $P(I_1|I_2) = P(I_1)$. Thus, $P(I_2|I_1) = P(I_1|I_2)P(I_2)/P(I_1) = P(I_1)P(I_2)/P(I_1) = P(I_2)$, which implies $I_2 \blacktriangleright^\bot I_1$.
         \item \textsc{Rule Semantic} states that $I_1 \blacktriangleright^\bot I_2$ if $P(I_1\wedge I_2) = P(I_1)P(I_2)$. With this information, we can infer that $P(I_1|I_2) = P(I_1\wedge I_2)/P(I_2) = P(I_1)P(I_2)/P(I_2) = P(I_1)$, implying $I_1 \blacktriangleright^\bot I_2$.
    \end{itemize}
\end{itemize}

\subsubsection{Soundness of Theorem~\ref{theorem:corr}}

    In the rest of this section, we use the $\indepSym$ symbol to indicate statistical independence and prove a series of lemmas that are used to prove Theorem~\ref{theorem:corr}.

    \textbf{Proof of Theorem~\ref{theorem:corr}}. To establish the proof, we need to prove the correctness of three rules (Rule \textsc{Indep}, Rule \textsc{Pos} and Rule~\textsc{Neg}) in Figure~\ref{fig:corr}.

    In the following, we use Lemma~\ref{lemma:indep} to establish the correctness of Rule \textsc{Indep}, Lemma~\ref{thm:positive} to establish the correctness of Rule \textsc{Pos} and Lemma~\ref{thm:negative} for Rule \textsc{Neg}.

Lemma~\ref{thm:positive} establishes the correctness of Rule \textsc{Pos} in Figure~\ref{fig:corr}, used for inferring the positive correlation between a pair of expressions $E_1$ and $E_2$.
As $E_1$ and $E_2$ are arbitrary expressions in Lemma~\ref{thm:positive}, we can use DNF as a general format to represent both $E_1$ and $E_2$. In order to prove Lemma~\ref{thm:positive}, we need to use the following Lemmas~ \ref{lemma:monotonic}, \ref{lemma:factConjunct}, \ref{lemma:conjunct} and \ref{lemma:conjDisj} that are also used to prove the soundness of the rule \textsc{Pos} but with a restricted format of $E_1$ and $E_2$.

Similarly, to prove Lemma~\ref{thm:negative} for Rule \textsc{Neg} in Figure~\ref{fig:corr}, we use the following Lemmas: Lemma~\ref{lemma:monotonic}, Lemma~\ref{lemma:factConjunctNeg}, Lemma~\ref{lemma:conjunctNeg}, and Lemma~\ref{lemma:conjDisjNeg}. \\

    \begin{lemma}
    \label{lemma:monotonic}
        If $\frac{\partial f}{\partial \alpha} > 0$ and $\frac{\partial f}{\partial \beta} > 0$, function $f(\alpha,\beta)$ is strictly monotonically increasing, i.e., $f(\alpha_1, \beta_1) > f(\alpha_2, \beta_2)$, where $\alpha_1 > \alpha_2$ and $\beta_1 > \beta_2$.
    \end{lemma}

    \textsc{Proof} of Lemma~\ref{lemma:monotonic}.

We assume the following:
\[
f_\alpha(\alpha, \beta) = \frac{\partial f}{\partial \alpha} \quad
f_\beta(\alpha, \beta) = \frac{\partial f}{\partial \beta}
\]

For a function \( f \) of two variables, the Mean Value Theorem states that:
   \[
   f(\alpha_1, \beta_1) - f(\alpha_2, \beta_2) = f_\alpha(\chi, \eta) \cdot (\alpha_1 - \alpha_2) + f_\beta(\chi, \eta) \cdot (\beta_1 - \beta_2)
   \]
   
   where \((\chi, \eta)\) lies on the line segment joining \((\alpha_1, \beta_1)\) and \((\alpha_2, \beta_2)\).
   
Given that \( \alpha_1 >\alpha_2 \) and \( \beta_1 > \beta_2 \), and knowing that \( f_\alpha(\chi, \eta) > 0 \) and \( f_y(\chi, \eta) > 0 \), we have:
   \[
   f(\alpha_1, \beta_1) - f(\alpha_2, \beta_2) = f_\alpha(\chi, \eta) \cdot (\alpha_1 - \alpha_2) + f_\beta(\chi, \eta) \cdot (\beta_1 - \beta_2) > 0
   \]
   
   This is because each term in the sum is positive, given the positivity of the partial derivatives and the non-negativity of the differences \(\alpha_1 - \alpha_2 \) and \( \beta_1 - \beta_2 \).

Therefore, \( f(\alpha_1, \beta_1) > f(\alpha_2, \beta_2) \) if \( \alpha_1 > \alpha_2 \) and \( \beta_1 > \beta_2 \). This completes the proof that $f(\alpha, \beta)$ is strictly increasing in both \( \alpha \) and \( \beta \).
\qed
    
    \begin{lemma}
    \label{lemma:indep2}
        Given two different correlation classes $V_1$ and $V_2$, if we have two arbitrary boolean expressions $E_1 = f(x),~x \in V_1$, and $E_2 = g(y),~y \in V_2$, then $E_1 \indepSym E_2$.
    \end{lemma}

    \textsc{Proof}. We omit the proof here since this is our definition of correlation classes.
    
    \begin{lemma}
    \label{lemma:indep}
        Given a pair of $E_1$ and $E_2$, if $\forall~x,y \in \textsf{Dep}(E_1) \times \textsf{Dep}(E_2).~\textsf{Class}(x) \neq \textsf{Class}(y)$, then $E_1 \indepSym E_2$.
    \end{lemma}

    \textsc{Proof} of Lemma~\ref{lemma:indep}.


   Let $S_1 = \textsf{Dep}(E_1)$ and $S_2 = \textsf{Dep}(E_2)$.
%
     We define the following function:
    \begin{equation}
    \label{eq:class}
    \small
    \mathcal{V}(E) =\{V| V = \textsf{Class}(x), x\in \textsf{Dep}(E)\} 
    \end{equation}

    Based on the premises of rule \textsc{Indep}, we know that $\forall x, y \in S_1 \times S_2. ~ \textsf{Class}(x) \neq \textsf{Class}(y)$,
    we can get $\mathcal{V}(E_1) \cap \mathcal{V}(E_2) = \emptyset$.
    With this, we can assume that all the variables (input facts) appearing in $E_1$ belong to a single correlation class $V_1 = \bigcup_{i \in \mathcal{V}(E_1)} V_i$ and all the variables appearing in $E_2$ belong to a single correlation class $V_2 = \bigcup_{j \in \mathcal{V}(E_2)} V_j$.

    We can express $E_1$ and $E_2$ as follows: 
     $$E_1 = f(x_1,x_2,\dots x_n), x_i \in V_1,~ E_2 = g(y_1,y_2,\dots, y_n), y_i \in V_2$$


   Given that $V_1$ and $V_2$ are two different correlation classes, based on Lemma~\ref{lemma:indep2}, we know that any arbitrary boolean combinations of variables from $V_1$ are mutually independent from the boolean combinations of variables from $V_2$. Thus, $E_1 \indepSym E_2$.
   
   Note that, in general,  independence does not necessarily propagate to conjunctions of conditions, e.g., given $x_1 \indepSym x$ and $x_2 \indepSym x$, it does not imply that $x \indepSym (x_1\wedge x_2)$. However, this is a valid assumption in our domain. Since all the input facts $x,~y$ are atomic, it is natural to assume the mutual independence between arbitrary boolean formulas $f(x)$ and $g(y)$ if $x$ and $y$ are from different correlation classes.\\


    \qed

    \begin{lemma}
    \label{lemma:factConjunct}
    Given  an input fact $E_1 = I$ and  a general  DNF formula $E_2 = \bigvee \bigwedge_{j} I_j$,  if $\vdash \chi(E_1),~\vdash \chi(E_2),~ \vdash E_1 \rightharpoonup^{+} E_2$ and $\not \vdash  E_1 \rightharpoonup^{-} E_2$, then $P(E_1\wedge E_2) > P(E_1)P(E_2), i.e., E_1 \cpos E_2$.
    \end{lemma}

    \textsc{Proof} of Lemma~\ref{lemma:factConjunct}.  We prove this by induction and start with the base case.
    
    \textbf{\emph{Base cases}}: $E_1 =I,~E_2 = \bigwedge_i I_i$. 
    
    We can compute the expression $P(E_1 \wedge E_2)$ as follows:
        \[
        \small
        \begin{array}{lllr}
        (1)P(E_1 \wedge E_2) & = & P(I \wedge \bigwedge_i I_i) & \\
        (2)& = & P(I \wedge I_k \wedge \bigwedge_{i \neq k} I_i) & \exists I_k \in \textsf{Dep}(E_2).~I_k \blacktriangleright^+ I ~(\text{i.e.,} \vdash E_1 \rightharpoonup^{+} E_2) \\
        (3) & = & P(I \wedge I_k)\prod_{i\neq k}P(I_i) & \text{Definition of }\chi(E_2),~I_1 \indepSym I_2 \indepSym \dots (I_k, I) \indepSym \dots I_i \\
        (4) & = & P(I|I_k)P(I_k)\prod_{i\neq k}P(I_i) & \\
        (5) & = & P(I|I_k)\prod_{i} I_i &
        \end{array}
        \]
        
    The expression of $P(E_1)P(E_2)$ is as follows:
        \[
        \small
        \begin{array}{lllr}
        (1)P(E_1)P(E_2) & = & P(I)P(\bigwedge_i I_i) &  \\
        (2) & = & P(I)\prod_{i} I_i & \text{Definition of }\chi(E_2)
        \end{array}
        \]

    Based on the precondition $\vdash E_1 \rightharpoonup^{+} E_2$, we know that $\exists I_k \in \textsf{Dep}(E_2).~I_k \blacktriangleright^+ E_1$, which is $I_k  \blacktriangleright^+ I$. It implies that $P(I|I_k) > P(I)$, and thus $P(E_1 \wedge E_2) > P(E_1)P(E_2)$, i.e., $E_1 \cpos E_2$.
    

    \textbf{\emph{Inductive cases}}: $E_1 =I,~E_2 = E_2' \vee t$, where $E_2' =\bigvee \bigwedge_i I_i,~t= \bigwedge_i I_i$.

    Case 1: $E_1 \cpos E_2'$.
     Given $E_1 = I$, we have the following three subcases.
     \begin{itemize}
         \item Case 1.1: $E_1 \rightharpoonup^{+}  t$ violates the precondition $\chi(E_2)$.
         \item Case 1.2: $E_1 \rightharpoonup^{-}  t$ violates the precondition $\not \vdash  E_1 \rightharpoonup^{-} E_2$.
         \item  Case 1.3: $E_1 \indepSym t$ is valid.
    
    \textsc{Proof} of Case 1.3
    
    Given $E_1 \cpos E_2'$, and $E_1 = I$, we can get $\mathcal{V}(E_1) \subseteq \mathcal{V}(E_2')$, where $\mathcal{V}(E_1)$ is defined by Eq~\ref{eq:class}. 
    Given $\chi(E_2)$, we know that the correlation classes for each $x \in \textsf{Dep}(E_2)$ are disjoint. With $E_2 = E_2' \vee t$, we can get $\mathcal{V}(E_2') \cap \mathcal{V}(t) = \emptyset$. With $\mathcal{V}(E_1) \subseteq \mathcal{V}(E_2')$, we can get $\mathcal{V}(E_1) \cap \mathcal{V}(t) = \emptyset$, thus $E_1 \indepSym t$.

    The inductive case focuses on proving the following: given $E_1 \cpos E_2'$, and $E_1 \indepSym t$,
    it implies that $E_1 \cpos E_2' \vee t$.  


        \[
        \small
        \begin{array}{lllr}
        (1) P(E_1 \wedge E_2) & = & P(I \wedge E_2) & E_1 = I\\
        (2) & = & P((I \wedge E_2') \vee (I \wedge t)) & E_2 = E_2' \vee t \\
        (3) & = & P(I \wedge E_2') + P(I \wedge t) - P(I \wedge E_2'\wedge t)& \\
        (4) & = & P(I \wedge E_2') + P(I)P(t) - P(I\wedge E_2')P(t) & \text{definitions of } \chi(E): \{I, E_2'\} \indepSym t. \\
        (1) P(E_1)P(E_2) & = & P(I)P(E_2' \vee t) &  E_1 = I \\
        (2) & = & P(I)(P(E_2') + P(t) - P(E_2')P(t)) & \text{definitions of } \text{and } \chi(E) \\
        (3) & = & P(I)P(E_2') + P(I)P(t) - P(I)P(E_2')P(t) &
        \end{array}
        \]

        From above expressions, we can compute $P(E_1\wedge E_2) - P(E_1)P(E_2)$ as $\Delta$:
        \[
        \small
        \begin{array}{lllr}
        (1) \Delta & = & P(I \wedge E_2') - P(I \wedge E_2'\wedge t) - P(I)P(E_2') + - P(I)P(E_2')P(t) & \\
        (2) & = & P(I|E_2')P(E_2') - P(I|E_2')P(E_2')P(t) - P(I)P(E_2') + - P(I)P(E_2')P(t) & 
        \\
        (3) & = & (1-P(t))(P(I|E_2')P(E_2') - P(I)P(E_2')) &
        \end{array}
        \]
        As  $E_1 \cpos E_2'$ and $E_1 = I$, we can get $P(I|E_2')P(E_2') > P(I)P(E_2')$, and thus $\Delta > 0$, i.e., $P(E_1\wedge E_2) > P(E_1)P(E_2)$. Hence, we are able to prove $E_1 \cpos E_2' \vee t$.
    \end{itemize}

 Case 2: $E_1 \indepSym E_2'$.

    \begin{itemize}
        \item Case 2.1: $E_1  \indepSym  t$  violates the precondition $\vdash E_1 \rightharpoonup^{+} E_2$.
        \item Case 2.2: $E_1  \rightharpoonup^{-}  t$ violates the precondition $\not \vdash  E_1 
        \rightharpoonup^{-} E_2$.
        \item Case 2.3: $E_1  \rightharpoonup^{+}  t$ is valid.

        The proof of Case 2.3 is similar to Case 1.3, so we omit it here.
    \end{itemize}


    Other cases

    





    Case 3: $E_1 \cneg E_2'$ is invalid as it violates the precondition $\not \vdash  E_1 \rightharpoonup^{-} E_2$. 
    \qed

    \begin{lemma}
    \label{lemma:conjunct}
    Let $E_1$, $E_2$ be conjunctions over input facts, i.e.,  $E_1 = \bigwedge_{i} I_i$ and $E_2 = \bigwedge_{j} I_j$. If $\vdash \chi(E_1),~\vdash \chi(E_2),~ \vdash E_1 \rightharpoonup^{+} E_2$ and $\not \vdash  E_1 \rightharpoonup^{-} E_2$, then $P(E_1\wedge E_2) > P(E_1)P(E_2)$.
    \end{lemma}

    \textsc{Proof} of Lemma~\ref{lemma:conjunct}. We prove this by induction and start with the base cases.

    \textbf{\emph{Base cases}}:
    $E_1 = I_{ai} \wedge I_{bi}$ and $E_2 = I_{aj} \wedge I_{bj}$
    
    To compute $P(E_1 \wedge E_2)$, we first list all the possible dependency information among $\{I_{ai}, I_{bi}, I_{aj}, I_{bj}\}$. Given the precondition $\chi(E_1)$ and $\chi(E_2)$, we know that $I_{ai}\indepSym I_{bi}$ and $I_{aj} \indepSym I_{bj}$. Based on the precondition $\vdash E_1 \rightharpoonup^{+} E_2,~\not \vdash  E_1 \rightharpoonup^{-} E_2$, we can list all the cases satisfying these conditions:
    \begin{enumerate}
        \item $I_{ai} \blacktriangleright^+ I_{aj}, ~I_{bi} \blacktriangleright^+ I_{bj}$: In this case, these four input facts are grouped to two correlation classes: $\{I_{ai}, I_{aj}\}$, $\{I_{bi}, I_{bj}\}$, thus $P(E_1)P(E_2) = P(I_{ai}\wedge I_{bi})P(I_{aj} \wedge I_{bj}) = P(I_{ai})P(I_{bi})P(I_{aj})P(I_{bj})$.
        Meanwhile, we can rewrite the joint probability as follows:
        $P(E_1\wedge E_2) = P(I_{ai}\wedge I_{bi} \wedge I_{bi} \wedge I_{bj}) = P(I_{ai} \wedge I_{aj})P(I_{bi} \wedge I_{bj}) = P(I_{ai}|I_{aj})P(I_{aj})P(I_{bi}|I_{bj})P(I_{bj})$. Given $I_{ai} \blacktriangleright^+ I_{aj}, ~I_{bi} \blacktriangleright^+ I_{bj}$, we know that $P(I_{ai}|I_{aj}) > P(I_{ai})$ and $P(I_{bi}|I_{bj}) > P(I_{bi})$. 
        
        Thus, $P(E_1\wedge E_2) = P(I_{ai}|I_{aj})P(I_{aj})P(I_{bi}|I_{bj})P(I_{bj}) > P(I_{ai})P(I_{aj})P(I_{bi})P(I_{bj}) = P(E_1)P(E_2)$.
        \item $I_{ai} \blacktriangleright^+ I_{bj}, ~I_{bi} \blacktriangleright^+ I_{aj}$. The proof of this is similar to case (1).
        \item $I_{ai} \blacktriangleright^+ I_{aj}, ~I_{bi} \indepSym I_{bj}$:  In this case, these four input facts are grouped to three correlation classes: $\{I_{ai}, I_{aj}\}$, $\{I_{bi}\}, \{I_{bj}\}$, thus $P(E_1)P(E_2) = P(I_{ai}\wedge I_{bi})P(I_{aj} \wedge I_{bj}) = P(I_{ai})P(I_{bi})P(I_{aj})P(I_{bj})$. Similarly, $P(E_1\wedge E_2) = P(I_{ai}\wedge I_{bi} \wedge I_{bi} \wedge I_{bj}) = P(I_{ai} \wedge I_{aj})P(I_{bi} \wedge I_{bj}) = P(I_{ai}|I_{aj})P(I_{aj})P(I_{bi}|I_{bj})P(I_{bj})$. Given $I_{ai} \blacktriangleright^+ I_{aj}$ and $I_{bi} \indepSym I_{bj}$, we know that $P(I_{ai}|I_{aj}) > P(I_{ai})$, and $P(I_{bi}|I_{bj}) = P(I_{bi})$. As a result, $P(E_1 \wedge E_2) > P(E_1)P(E_2)$.
        \item $I_{ai} \blacktriangleright^+ I_{bj}, ~I_{bi} \indepSym I_{aj}$. 
        \item $I_{ai} \indepSym I_{bj}, ~I_{bi} \blacktriangleright^+  I_{aj}$. 
        \item $I_{ai} \indepSym I_{aj}, ~I_{bi} \blacktriangleright^+  I_{bj}$.
    \end{enumerate}

    The proof of cases (4)-(6) is simlar to case (3), so we omit the proof here.

    \textbf{\emph{Inductive cases}}. 
    \[
\begin{array}{lll|lll}
E_1 & = & E_1' \wedge I_1 \quad &E_1' & = &  \bigwedge_i I_i \\
E_2 & = & E_2' \wedge I_2 \quad &E_2' & = & \bigwedge_j I_j
\end{array}
\]

   Case 1: $E_1' \cpos E_2'$. 
    Based on the precondition $\chi(E_1)$ (resp. $\chi(E_2)$ ), we know that $E_1' \indepSym I_1$ (resp. $E_2' \indepSym I_2$).
    \begin{itemize}
        \item Case 1.1 $I_1 \rightharpoonup^{+} I_2,~ I_1 \indepSym E_2',~ I_2 \indepSym E_1'$.

        \textsc{Proof of Case 1.1 } $E_1' \cpos E_2',~ I_1 \rightharpoonup^{+} I_2,~ I_1 \indepSym E_2',~ I_2 \indepSym E_1'$
    
    \[
        \small
        \begin{array}{lllr}
        P(E_1 \wedge E_2) & = & P(E_1' \wedge I_1 \wedge E_2' \wedge I_2) & \\
        & = & P(E_1'\wedge E_2')P(I_1 \wedge I_2) & \text{definitions of }  \{E_1', E_2'\} \indepSym \{I_1,~I_2\} \\
        & = & P(E_1'|E_2')P(E_2')P(I_1|I_2)P(I_2) & \\
        P(E_1)P(E_2) & = & P(E_1' \wedge I_1)P(E_2' \wedge I_2) & \\
        & = & P(E_1')P(I_1)P(E_2')P(I_2) & \text{definitions of } \chi(E_1) \text{ and } \chi(E_2)
        \end{array}
    \]

    In case 1.1, given $E_1' \cpos E_2'$, we can infer that $P(E_1'|E_2') > P(E_1')$. Given $I_1 \rightharpoonup^{+} I_2$, as both $I_1$ and $I_2$ are input facts, we know that $I_1 \blacktriangleright^+ I_2$, and thus $P(I_1 |I_2) > P(I_1)$. As a result, $P(E_1 \wedge E_2) > P(E_1)P(E_2)$. \\
        
        \item Case 1.2 $I_1 \rightharpoonup^{+} I_2,~I_1 \indepSym E_2',~ I_2 \rightharpoonup^{+} E_1'$. It is invalid as it violates the precondition $\chi(E_1)$.
        \item Case 1.3 $I_1 \rightharpoonup^{+} I_2,~I_1 \rightharpoonup^{+} E_2',~ I_2 \indepSym E_1'$. It is invalid as it violates the preconditions $\chi(E_2)$.
        \item Case 1.4 $I_1 \rightharpoonup^{+} I_2,~I_1 \rightharpoonup^{+} E_2',~ I_2 \rightharpoonup^{+} E_1'$. It is invalid as it violates the precondition $\chi(E_1)$ and $\chi(E_2)$.
        \item Case 1.5 $I_1 \indepSym I_2,~ I_1 \indepSym E_2',~ I_2 \indepSym E_1'$

        \textsc{Proof of Case 1.5 } $E_1' \cpos E_2',~I_1 \indepSym I_2,~ I_1 \indepSym E_2',~ I_2 \indepSym E_1'$

        \[
        \small
        \begin{array}{lllr}
        P(E_1 \wedge E_2) & = & P(E_1' \wedge I_1 \wedge E_2' \wedge I_2) & \\
        & = & P(E_1' \wedge E_2')P(I_1)P(I_2) & \text{definitions of } I_1 \indepSym I_2 \indepSym \{E_1', E_2'\}  \\
        & = & P(E_1'|E_2')P(E_2')P(I_1)P(I_2) & \\
        P(E_1)P(E_2) & = & P(E_1' \wedge I_1)P(E_2' \wedge I_2) & \\
        & = & P(E_1')P(I_1)P(E_2')P(I_2) & \text{definitions of } \chi(E_1) \text{ and } \chi(E_2)
        \end{array}
    \]

        In case 1.5, given $E_1' \cpos E_2'$, we can infer that $P(E_1'|E_2') > P(E_1')$. Thus, $P(E_1 \wedge E_2) > P(E_1)P(E_2)$. \\
    
        \item Case 1.6 $I_1 \indepSym I_2,~ I_1 \rightharpoonup^{+} E_2',~ I_2 \indepSym E_1'$

        \textsc{Proof of Case 1.6 } $E_1' \cpos E_2',~I_1 \indepSym I_2,~ I_1 \rightharpoonup^{+} E_2',~ I_2 \indepSym E_1'$

        Given $E_1' \cpos E_2'$ and $I_1 \rightharpoonup^{+} E_2'$, we split $E_2'$ into $E_2'(1)$ and $E_2'(2)$, where $E_1' \cpos E_2'(1),~E_1' \indepSym E_2'(2)$ and $I_1 \rightharpoonup^{+} E_2'(2),~I_1 \indepSym E_2'(1)$. Meanwhile, given $E_2' = \bigwedge_j I_j$, $E_2'$ can be represented as $E_2' = E_2'(1) \wedge E_2'(2)$.

        \[
        \small
        \begin{array}{lll}
        P(E_1 \wedge E_2) & = & P(E_1' \wedge I_1 \wedge E_2' \wedge I_2)  \\
        & = & P(E_1'\wedge E_2'(1))P(I_1\wedge E_2'(2))P(I_2) \quad \text{definitions of } \{E_1', E_2'(1)\} \indepSym \{I_1, E_2'(2)\} \indepSym I_2  \\
        & = & P(E_2'(1)|E_1')P(E_1')P(E_2'(2)|I_1)P(I_1)P(I_2)  \\
        P(E_1)P(E_2) & = & P(E_1' \wedge I_1)P(E_2' \wedge I_2)  \\
        & = & P(E_1')P(I_1)P(E_2')P(I_2) \quad \text{definitions of } \chi(E_1) \text{ and } \chi(E_2)
        \end{array}
        \]

        In case 1.6, given $E_1' \cpos E_2'(1)$, we can get $P(E_2'(1)|E_1') > P(E_2'(1))$. Given the precondition $\not \vdash E_1 \rightharpoonup^{-} E_2 $, it implies that $\not \vdash I_1 \rightharpoonup^{-} E_2'$ since $E_1 = E_1'\wedge I_1$ and $E_2 = E_2' \wedge I_2$. As $E_2' = E_2'(1)\wedge E_2'(2)$, it implies that $\not \vdash I_1 \rightharpoonup^{-} E_2'(2)$.
        Together with $I_1 \rightharpoonup^{+} E_2'(2)$, from Lemma~\ref{lemma:factConjunct}, we know that $I_1 \cpos E_2'(2)$. As a result, $P(E_2'(2)|I_1) > P(E_2'(2))$.

        Thus, $P(E_1 \wedge E_2) = P(E_2'(1)|E_1')P(E_1')P(E_2'(2)|I_1)P(I_1)P(I_2) > $
        $P(E_2'(1))P(E_1')P(E_2'(2))P(I_1)P(I_2)$ 
        $ = P(E_1')P(E_2')P(I_1)P(I_2) = P(E_1)P(E_2)$. \\

        \item Case 1.7 $I_1 \indepSym I_2,~ I_1 \rightharpoonup^{+} E_2',~ I_2 \rightharpoonup^{+} E_1'$

         \textsc{Proof of Case 1.7 } $E_1' \cpos E_2',~I_1 \indepSym I_2,~I_1 \rightharpoonup^+ E_2',~I_2 \rightharpoonup^+ E_1'$

        In this case, we split $E_1'$ into $E_1'(1)$ and $E_1'(2)$, such that $E_1'(1) \cpos E_2,~E_1'(2) \rightharpoonup^+ I_2, E_1'(1) \indepSym I_2, E_1'(2) \indepSym E_2$. Similarly, we split $E_2'$ into $E_2'(1)$ and $E_2'(2)$, such that $E_2'(1) \cpos E_1,~E_2'(2) \rightharpoonup^+ I_1, E_2'(1) \indepSym I_1, E_2'(2) \indepSym E_1$.
    
        \[
        \small
        \begin{array}{lll}
        P(E_1 \wedge E_2) & = & P(E_1' \wedge I_1 \wedge E_2' \wedge I_2)  \\
        & = & P(E_1'(1 ) E_2'(1))P(I_1\wedge E_2'(2))P(I_2 \wedge E_1'(2)) \\
        & & \quad  \text{definitions of } \{E_1'(1), E_2'(1)\} \indepSym \{I_1, E_2'(2)\} \indepSym \{I_2, E_1'(1)\} \\
        & = & P(E_2'(1)|E_1'(1))P(E_1'(1))P(E_2'(2)|I_1)P(I_1)P(E_1'(2)|I_2)P(I_2) \\
        P(E_1)P(E_2) & = & P(E_1' \wedge I_1)P(E_2' \wedge I_2)  \\
        & = & P(E_1')P(I_1)P(E_2')P(I_2) \\
        & & \quad \text{definitions of } \chi(E_1) \text{ and } \chi(E_2)
        \end{array}
        \]

        Given $E_1'(2) \rightharpoonup^+ I_2$, and $\not \vdash E_1'(2) \rightharpoonup^- I_2$, according to Lemma~\ref{lemma:factConjunct}, we can get $E_1'(2) \cpos I_2$. Similarly, we can get $E_2'(2) \cpos I_1$. With this, we have $P(E_1'(2) |I_2) > P(E_1'(2))$ and $P(E_2'(2) |I_1) > P(E_2'(2))$. As a result, $P(E_1 \wedge E_2)$ $= P(E_2'(1)|E_1'(1))P(E_1'(1))P(E_2'(2)|I_1)P(I_1)P(E_1'(2)|I_2)P(I_2)$ $>P(E_2'(1))P(E_1'(1))P(E_2'(2))P(E_1'(2))P(I_1)P(I_2)$ $=P(E_1')P(E_2')P(I_1)P(I_2) = P(E_1)P(E_2)$. We then prove $E_1 \cpos E_2$.
        
        \item Case 1.8 $I_1 \indepSym I_2,~ I_1 \indepSym E_2',~ I_2 \rightharpoonup^{+} E_1'$
        
        This proof of case 1.8 is symmetric to proof of case 1.6, so we omit it here.

        \item We do not talk about $\rightharpoonup^{-}$ relation between a pair of expressions as it violates the precondition $\not\vdash  E_1 \rightharpoonup^{-} E_2$.
    \end{itemize}

    Case 2: $E_1' \indepSym E_2'$. We also know that $E_1' \indepSym I_1$ and $E_2' \indepSym I_2$.
    \begin{itemize}
        \item Case 2.1 $I_1 \rightharpoonup^{+} I_2,~I_1 \indepSym E_2',~I_2 \indepSym E_1'$

         This proof of case 2.1 is symmetric to proof of case 1.5, so we omit it here.
         
        \item Case 2.2 $I_1 \rightharpoonup^{+} I_2,~I_1 \indepSym E_2',~I_2 \rightharpoonup^{+} E_1'$.  It is invalid as it violates the preconditions $\chi(E_1)$.
        \item Case 2.3 $I_1 \rightharpoonup^{+} I_2,~I_1 \rightharpoonup^{+} E_2',~I_2 \indepSym E_1'$. It is invalid as it violates the preconditions $\chi(E_2)$.
        \item Case 2.4 $I_1 \rightharpoonup^{+} I_2,~I_1 \rightharpoonup^{+} E_2',~I_2  \rightharpoonup^{+} E_1'$. It is invalid as it violates the preconditions $\chi(E_1)$ and $\chi(E_2)$.
        \item Case 2.5 $I_1 \indepSym I_2,~I_1 \indepSym E_2',~I_2 \indepSym E_1'$. It is invalid as it violates the precondition $\vdash E_1 \rightharpoonup^{+} E_2$.
        \item Case 2.6 $I_1 \indepSym I_2,~I_1 \rightharpoonup^{+} E_2',~I_2 \indepSym E_1'$.
        
        This proof of case 2.6 is symmetric to proof of case 1.5, so we omit it here.
        \item Case 2.7 $I_1 \indepSym I_2,~I_1 \indepSym E_2',~I_2 \rightharpoonup^{+} E_1'$.
        
        This proof of case 2.7 is symmetric to proof of case 1.5, so we omit it here.
        \item Case 2.8 $I_1 \indepSym I_2,~I_1 \rightharpoonup^{+} E_2',~I_2 \rightharpoonup^{+} E_1'$.

        This proof of case 2.7 is symmetric to proof of case 1.1, so we omit it here.
        
    \end{itemize}

    Case 3: $E_1' \cneg E_2'$ is invalid as it violates the precondition $\vdash E_1 \rightharpoonup^{-} E_2$.\\
    \qed

    \begin{lemma}
    \label{lemma:conjDisj}
    Let $E_1$ be a conjunction over input facts (i.e.,  $E_1 = \bigwedge_{i} I_i$) and let  $E_2 = \bigvee \bigwedge_{j} I_j$ be a DNF formula over input facts. If $\vdash \chi(E_1),~\vdash \chi(E_2),~ \vdash E_1 \rightharpoonup^{+} E_2$ and $\not \vdash  E_1 \rightharpoonup^{-} E_2$, then $P(E_1\wedge E_2) > P(E_1)P(E_2)$.
    \end{lemma}

    \textsc{Proof} of Lemma~\ref{lemma:conjDisj}. We prove this by induction and start with the base cases.

    \textbf{Base case}: $E_1 = \bigwedge_i I_i,~E_2 = \bigwedge_j I_j$
    
    Given $\chi(E_1), \chi(E_2), \vdash E_1 \rightharpoonup^{+} E_2$, and $\not \vdash E_1 \rightharpoonup^{-} E_2$, according to Lemma~\ref{lemma:conjunct}, we can get $E_1 \cpos E_2$.

    \textbf{Inductive case}:
    \[
\begin{array}{lll|lll|lll}
E_1 & = & E_1'  \quad &E_1' & = &  \bigwedge_i I_i \quad & \\
E_2 & = & E_2' \vee C \quad & E_2' & = & \bigvee \bigwedge_j I_j \quad & C = \bigwedge_{j'} I_{j'}
\end{array}
\]

Case 1 $E_1' \cpos E_2'$
\begin{itemize}
    \item Case 1.1 $E_1' \rightharpoonup^+ C$

    Proof of Case 1.1 $E_1' \cpos E_2',~E_1' \rightharpoonup^+ C$ \\

    Given $\not \vdash E_1 \rightharpoonup^- E_2$, we can infer that $\not \vdash E_1' \rightharpoonup^- C$. Given $\chi(E_1)$ and $\chi(E_2)$, we can also infer that $\chi(E_1')$ and $\chi(C)$ hold.

    Given $E_1' \rightharpoonup^+ C$, $\not \vdash E_1' \rightharpoonup^- C$, $\chi(E_1')$ and $\chi(C)$, we get $E_1' \cpos C$ by applying Lemma~\ref{lemma:conjunct}.

    \[
    \small
    \begin{array}{lll}
    & & \textcolor{blue}{P(E_1)P(E_2)} \\
    & = & P(E_1')P(E_2')+P(E_1')P(C) - P(E_1')P(E_2')P(C) \\
    & & \textcolor{blue}{P(E_1\wedge E_2)} \\
    &=& P(\underline{E_1' \wedge E_2'} \vee \underline{E_1' \wedge C}) \\
    &=& P(E_1'\wedge E_2') + P(E_1'\wedge C) - P(E_1' \wedge E_2' \wedge C) \\
    &=& P(E_1'|E_2')P(E_2') + P(E_1'|C)P(C) - P(E_1'(1) \wedge E_2')P(E_1'(2)\wedge C) \quad \textsf{splitting } E_1': P(E_1')= P(E_1'(1))P(E_1'(2)) \\
    & = & P(E_1'|E_2')P(E_2') + P(E_1'|C)P(C) - P(E_1'(1) | E_2')P(E_1'(2)\ | C) P(E_2')P(C)
    \end{array}
    \]

    Given $E_1' \cpos E_2'$, we know that $P(E_1'|E_2') > P(E_1')$. Similarly, given $E_1' \cpos C$, we also have $P(E_1'|C) > P(E_1')$.
    
    We define a function $f(x,y)$ as follows ($x \in [0,1],~ y \in [0,1]$):
    \[
    f(\alpha,\beta) = \alpha*P(E_2') + \beta*P(C) - \alpha\beta
    P(E_2')P(C)/P(E_1')
    \]

    We then define the expression of the joint probability as follows:
    $P(E_1 \wedge E_2) = f(x_1, y_1)$ where $\alpha_1 = P(E_1'|E_2')$ and $\beta_1 = P(E_1'|C)$. Meanwhile $P(E_1)P(E_2) = f(x_2, y_2)$ where $\alpha_2 = P(E_1'),~\beta_2 = P(E_1')$. To summarize, $\alpha \in \{P(E_1'|E_2'), P(E_1')\},~ \alpha \geq P(E_1')$ and $\beta \geq P(E_1')$.

    The partial derivation $ f_\alpha(\alpha, \beta) = \frac{\partial f}{\partial \alpha} = P(E_2') - \beta P(E_2')P(C)/P(E_1') > 0$. Similarly, $f_\beta(\alpha,\beta) = \frac{\partial f}{\partial \beta} = P(C) - \alpha P(E_2')P(C)/P(E_1') > 0$. With both $f_\alpha(\alpha,\beta)$ and $f_\beta(\alpha,\beta)$ being positive, according to the Lemma~\ref{lemma:monotonic}, we can get the $f(\alpha_1, \beta_1) > f(\alpha_2, \beta_2)$ if $\alpha_1 > \alpha_2,~\beta_1 > \beta_2$.  Given $\alpha_1 > \alpha_2$ and $\beta_1 > \beta_2$, we can get $P(E_1 \wedge E_2) = f(\alpha_1,\beta_1) > f(\alpha_2,\beta_2) = P(E_1)P(E_2)$.  Therefore, we are able to prove $E_1 \cpos E_2$. 
    \\

    \item Case 1.2 $E_1' \indepSym C$

    Here we omit the proof of case 1.2 since it is similar to the proof of Case 1.1.
\end{itemize}

Case 2 $E_1' \indepSym E_2'$
\begin{itemize}
    \item Case 2.1 $E_1' \rightharpoonup^+ C$

    Here we omit the proof of case 2.1 since it is similar to the proof of Case 1.1.
    
    \item Case 12.2 $E_1' \indepSym C$

    Here we omit the proof of case 2.2 since it is similar to the proof of Case 1.1.
\end{itemize}

Case 3 $E_1' \cneg E_2'$: This case is invalid as it violates the precondition $\not \vdash E_1 \rightharpoonup^- E_2$. 

\qed

        \begin{lemma}[\textbf{Soundness of Rule \textsc{Pos}}]
    \label{thm:positive}
        The rule \textsc{Pos} in Figure~\ref{fig:corr} are sound. In particular, given a pair of DNF formulas $E_1 = \bigvee \bigwedge_{i} I_i, E_2 = \bigvee \bigwedge_{j} I_j$, if $\vdash \chi(E_1),~\vdash \chi(E_2),~ \vdash E_1 \rightharpoonup^{+} E_2$ and $\not \vdash  E_1 \rightharpoonup^{-} E_2$, then $P(E_1\wedge E_2) > P(E_1)P(E_2)$.
    \end{lemma}

     Proof of Lemma~\ref{thm:positive}.

    In Lemma~\ref{thm:positive}, both $E_1$ and $E_2$ are arbitrary boolean formulas so we use DNF format to represent both of them, i.e., $E_1 = \bigvee \bigwedge_i I_i$ and $E_2 = \bigvee \bigwedge_j I_j$. \\

    \textbf{Base case}: $E_1 = \bigwedge_i I_i,~E_2 = \bigwedge_j I_j$
    
    Given $\chi(E_1), \chi(E_2), \vdash E_1 \rightharpoonup^{+} E_2$, and $\not \vdash E_1 \rightharpoonup^{-} E_2$, according to Lemma~\ref{lemma:conjunct}, we can get $E_1 \cpos E_2$.

    \textbf{Inductive case}:
    \[
\begin{array}{lll|lll|lll}
E_1 & = & E_1' \vee C_1 \quad &E_1' & = &  \bigvee \bigwedge_i I_i & C_1 = \bigwedge_{i'} I_{i'}\\
E_2 & = & E_2' \vee C_2 \quad &E_2' & = & \bigvee \bigwedge_j I_j & C_2 = \bigwedge_{j'} I_{j'}
\end{array}
\]

Case 1: $E_1' \cpos E_2'$

    Based on the precondition $\chi(E_1)$ (resp. $\chi(E_2)$ ), we know that $E_1' \indepSym C_1$ (resp. $E_2' \indepSym C_2$). Following are the valid cases.
    \begin{itemize}
    \item Case 1.1 $C_1  \rightharpoonup^{+} C_2,~C_1 \indepSym E_2',~C_2 \indepSym E_1'$

     Proof of Case 1.1 $E_1' \cpos E_2',~C_1  \rightharpoonup^{+} C_2,~C_1 \indepSym E_2',~C_2 \indepSym E_1'$
        \[
        \small
        \begin{array}{lll}
        P(E_1 \wedge E_2) & = & P((E_1' \vee C_1) \wedge (E_2' \vee C_2))  \\
        & = & P(\underline{E_1'\wedge E_2'} \vee \underline{E_1'\wedge C_2} \vee \underline{C_1 \wedge E_2'} \vee \underline{C_1 \wedge C_2})  \\
        & = & P(E_1'\wedge E_2') + P(E_1'\wedge C_2) + P(C_1 \wedge E_2') + P(C_1 \wedge C_2)  \\
        &  & -P(E_1'E_2'C_2) -P(E_1'C_1E_2') - P(E_1'C_1C_2) - P(C_1E_2'C_2) + P(E_1'E_2'C_1C_2)  \\
        & = & P(E_1'|E_2')\textcolor{red}{P(E_2')} + \textcolor{blue}{P(E_1')P(C_2)} + \textcolor{cyan}{P(C_1)P(E_2')} + P(C_1|C_2)\textcolor{magenta}{P(C_2)}  \\
        & & -P(E_1'|E_2')\textcolor{brown}{P(E_2')P(C_2)} - P(E_1'|E_2')\textcolor{orange}{P(C_1)P(E_2')} - P(C_1|C_2)\textcolor{mygreen}{P(C_2)P(E_1')} - P(C_1|C_2)\textcolor{purple}{P(E_2')P(C_2)} \\
        & & P(E_1'|E_2')P(C_1|C_2)\textcolor{teal}{P(E_2')P(C_2)}  \quad \text{using } C_1 \indepSym E_2',~C_2 \indepSym E_1' \\
        P(E_1)P(E_2) & = & P(E_1' \vee C_1)P(E_2' \vee C_2)  \\
        & = & (P(E_1') + P(C_1) -P(E_1'C_1))(P(E_2')+P(C_2) - P(E_2'C_2))    \\
        & = & P(E_1')P(E_2') + P(E_1')P(C_2) + P(C_1)P(E_2') + P(C_1)P(C_2)  \\
        &  & - P(E_1')P(E_2'C_2) - P(E_1'C_1)P(E_2') - P(E_1'C_1)P(C_2) - P(C_1)P(E_2'C_2) + P(E_1'C_1)P(E_2'C_2)  \\
        & = & P(E_1')\textcolor{red}{P(E_2')} + \textcolor{blue}{P(E_1')P(C_2)} + \textcolor{cyan}{P(C_1)P(E_2')} + P(C_1)\textcolor{magenta}{P(C_2)}  \\
        &  & - P(E_1')\textcolor{brown}{P(E_2')P(C_2)} - P(E_1')\textcolor{orange}{P(C_1)P(E_2')} - P(C_1)\textcolor{mygreen}{P(C_2)P(E_1')} - P(C_1)\textcolor{purple}{P(E_2')P(C_2)} \\
        & & + P(E_1')P(C_1)\textcolor{teal}{P(E_2')P(C_2)} \quad \text{using } C_1 \indepSym E_2',~C_2 \indepSym E_1'  \\
        \end{array}
    \]

    As we can see from above, both $P(E_1 \wedge E_2)$ and $P(E_1)P(E_2)$ share some terms. Here, we use the same color to denote the same terms. The different parts between $P(E_1 \wedge E_2)$ and $P(E_1)P(E_2)$ are still colored in black. Based on this observation, we can define a function $f(\alpha,\beta)$ as follows: 
    \[
    \small
        \begin{array}{lll}
        f(\alpha,\beta) & = & \alpha\textcolor{red}{P(E_2')} + \textcolor{blue}{P(E_1')P(C_2)} + \textcolor{cyan}{P(C_1)P(E_2')} + \beta\textcolor{magenta}{P(C_2)}  \\
        &  &-\alpha\textcolor{brown}{P(E_2')P(C_2)} - \alpha\textcolor{orange}{P(C_1)P(E_2')} - \beta\textcolor{mygreen}{P(C_2)P(E_1')} - \beta\textcolor{purple}{P(E_2')P(C_2)} \\
        &  &\alpha\beta\textcolor{teal}{P(E_2')P(C_2)}
    \end{array}
    \]

    Based on the definition of $f(\alpha,\beta)$, we can infer that $P(E_1\wedge E_2) = f(\alpha_1, \beta_1),~\alpha_1 = P(E_1'|E_2'),~\beta_1=P(C_1|C_2)$ whereas $P(E_1)P(E_2) = f(\alpha_2, \beta_2),~\alpha_2 = P(E_1'),~\beta_2 = P(C_1)$. Hence $\alpha \in \{P(E_1'|E_2')$, $P(E_1)\},~ \beta \in \{P(C_1|C_2,~C_1)\}$. 

    Given $E_1' \cpos E_2'$, we know that $P(E_1'|E_2') > P(E_1')$, and thus $\alpha \geq P(E_1')$. Based on the precondition $\not \vdash E_1 \rightharpoonup^- E_2$, we can also imply that $\not \vdash C_1 \rightharpoonup^- C_2$ since $E_1 = E_1' \vee C_1,~E_2 = E_2' \vee C_2'$, and it does not introduce new negations here. In this case 1.1, we know that $C_1 \rightharpoonup^+ C_2$, and $\not \vdash C_1 \rightharpoonup^- C_2$, and both $C_1,~C_2$ is a conjunction of input facts, i.e., $C_1 = \bigwedge_{i'} I_{i'}$. By applying Lemma~\ref{lemma:conjunct}, we can infer that $C_1 \cpos C_2$, and thus $P(C_1|C_2) > P(C_1)$. As a result, $\beta \geq P(C_1)$.\\


   Compute the partial derivatives of \( f \) with respect to \( \alpha \) and \( \beta \):
   \[
   \begin{array}{lll}
        f_\alpha(\alpha, \beta) &=& \frac{\partial f}{\partial \alpha} = P(E_2') - P(E_2')P(C_2) -P(E_2')P(C_1) + \beta*P(E_2')P(C_2)  \\
        f_\beta(\alpha, \beta) &=&  \frac{\partial f}{\partial \beta} = P(C_2)-P(C_2)P(E_1') - P(E_2')P(C_2) + \alpha*P(E_2')P(C_2)
   \end{array}
   \]

   
   Prove that the partial derivatives \( f_\alpha(\alpha, \beta) \) and \( f_\beta(\alpha, \beta) \) are positive for all \( \alpha \) and \( \beta \):
    \[
    \small
    \begin{array}{lll}
     f_\alpha(\alpha, \beta)  &=& P(E_2') - P(E_2')P(C_2) -P(E_2')P(C_1) + y*P(E_2')P(C_2)  \\
    & &\geq P(E_2') - P(E_2')P(C_2) -P(E_2')P(C_1) + P(C_1)*P(E_2')P(C_2) \quad \text{ using } \beta \geq P(C_1) \\
    & & = (1-P(C_2))(P(E_2') - P(E_2')P(C_1)) \quad  \text{ using }  P(C_1),P(C_2) \in [0,1] \\
    & & > 0 
    \end{array}
    \]

    \[
    \small
    \begin{array}{lll}
    f_\beta(\alpha, \beta) &=& P(C_2)-P(C_2)P(E_1') - P(E_2')P(C_2) + \alpha*P(E_2')P(C_2)  \\
    & &\geq P(C_2)-P(C_2)P(E_1') - P(E_2')P(C_2) + P(E_1')*P(E_2')P(C_2) \quad \text{ using } \alpha \geq P(E_1') \\
    & & = (1-P(E_1'))(P(C_2) - P(C_2)P(E_2')) \quad \text{ using }  P(E_1'),P(E_2') \in [0,1] \\
    & & > 0 
    \end{array}
    \]

    
    As we can get from the above, both $\frac{\partial f}{\partial x\alpha}$ and $\frac{\partial f}{\partial \beta}$ are positive. According to Lemma~\ref{lemma:monotonic}, we know that $f(\alpha_1, \alpha_2) > f(\beta_1, \beta_2)$ if $\alpha_1 > \alpha_2$ and $\beta_1 > \beta_2$.

    As we know, $P(E_1\wedge E_2) = f(\alpha_1, \beta_1)$ while $P(E_1)P(E_2) = f(\alpha_2,\beta_2)$. Given $\alpha_1 = P(E_1'|E_2') > P(E_1') = \alpha_2$ and $\beta_1 = P(C_1|C_2) > P(C_1) = \beta_2$, we can get $f(\alpha_1, \beta_1) > f(\alpha_2,\beta_2)$, and thus $P(E_1 \wedge E_2) > P(E_1)P(E_2)$. \\
    \item Case 1.2 $C_1 \indepSym C_2,~ C_1 \rightharpoonup^{+} E_2,~C_2 \rightharpoonup^{+} E_1$

    Proof of Case 1.2 $E_1' \cpos E_2',~C_1  \indepSym C_2,~C_1 \rightharpoonup^+ E_2',~C_2 \rightharpoonup^+ E_1'$ \\

    Given $E_1' \cpos E_2',~C_1 \rightharpoonup^{+} E_2'$, and $C_2 \rightharpoonup^{+} E_1'$, we split $E_1'$ (resp. $E_2'$) into $E_1'(1)$ and $E_1'(2)$ (resp. $E_2'(1)$ and $E_2'(2)$), such that $E_1'(1) \cpos E_2'(1),~E_1'(1) \indepSym C_2,~ E_1'(2) \rightharpoonup^+ C_2,~E_2'(1) \indepSym C_1,~E_2'(2) \rightharpoonup^+ C_1$. There are two cases for the splitting, taking $E_1'$ for example, $E_1' = E_1'(1) \wedge E_1'(2)$ or  $E_1' = E_1'(1) \vee E_1'(2)$. We take the first form (i.e., $E_1' = E_1'(1) \wedge E_1'(2)$) to prove, and the proof of the second form is similar to that, so we focus on proving the first form and omit the proof of the second form here.

    After splitting, $E_1'(1) \cpos E_2'(1),~E_1'(2) \rightharpoonup^+ C_2,~E_2'(2) \rightharpoonup^+ C_1$. We also know that $\not \vdash E_1'(2) \rightharpoonup^- C_2$, according to Lemma~\ref{lemma:conjDisj}, we can get $E_1'(2) \cpos C_2$. Similarly $E_2'(2) \cpos C_1$.
    

From the proof of Case 1.1, we get the expression of $P(E_1 \wedge E_2)$ and $P(E_1)P(E_2)$ as follows:

        \[
        \small
        \begin{array}{lll}
        P(E_1 \wedge E_2) & = & P((E_1' \vee C_1) \wedge (E_2' \vee C_2))  \\
        & = & P(\underline{E_1'\wedge E_2'} \vee \underline{E_1'\wedge C_2} \vee \underline{C_1 \wedge E_2'} \vee \underline{C_1 \wedge C_2})  \\
        & = & P(E_1'\wedge E_2') +  \textcolor{blue}{P(E_1'\wedge C_2) + P(C_1 \wedge E_2')} + P(C_1 \wedge C_2)  \\
        &  & \textcolor{blue}{-P(E_1'E_2'C_2) -P(E_1'C_1E_2') - P(E_1'C_1C_2) - P(C_1E_2'C_2) + P(E_1'E_2'C_1C_2)}  \\
         P(E_1)P(E_2) & = & P(E_1' \vee C_1)P(E_2' \vee C_2)  \\
        & = & (P(E_1') + P(C_1) -P(E_1'C_1))(P(E_2')+P(C_2) - P(E_2'C_2))    \\
        & = & P(E_1')P(E_2') + \textcolor{blue}{P(E_1')P(C_2) + P(C_1)P(E_2')} + P(C_1)P(C_2) \\
        &  & \textcolor{blue}{- P(E_1')P(E_2')P(C_2) - P(E_1')P(C_1)P(E_2') - P(E_1')P(C_1)P(C_2) - P(C_1)P(E_2')P(C_2)} \\
        & & \textcolor{blue}{+ P(E_1')P(C_1)P(E_2')P(C_2)}  \\
        \end{array}
    \]

    We now define a new function $f(\alpha,\beta)$ to represent the blue colored formulas in $P(E_1 \wedge E_2)$ and $P(E_1)P(E_2)$.

    \[
    \small
        \begin{array}{lll}
    f(\alpha,\beta) & = & \alpha P(C_2) + \beta P(C_1) - \alpha\frac{P(E_1'(1)|E'_2)}{P(E_1'(1))}P(E'_2)P(C_2) -\beta\frac{P(E_2'(1)|E'_1)}{P(E_2'(1))}P(E'_1)P(C_1) \\
    & & - \alpha P(C_1)P(C_2) - yP(C_1)P(C_2) + \alpha\beta\frac{P(E_1'(1)E_2'(1))}{P(E_1'(1))P(E_2'(1))}P(I_1)P(I_2)
    \end{array}
    \]

    Joint probability $P(E_1 \wedge E_2)$ is equivalent to $P(E_1'\wedge E_2') + P(C_1 \wedge C_2) + \textcolor{blue}{f(\alpha_1,\beta_1)}$ while the formula $P(E_1)P(E_2)$ is defined as $P(E_1')P(E_2') + P(C_1)P(C_2) + \textcolor{blue}{f(\alpha_2,\beta_2)}$, where $\alpha_1,~\alpha_2,~\beta_1,~\beta_2$ are defined as follows:

    \[
    \small
    \begin{array}{lll|lll}
         \alpha_1 & = & P(E_1'|C_2) & \beta_1 & = & P(E_2'|C_1) \\
         \alpha_2 & = & P(E_1') & \beta_2 & = & P(E_2')
    \end{array}
    \]

    After splitting, we know that $E_1'(1) \cpos E_2'(1),~E_1'(1) \indepSym C_2,~ E_1'(2) \cpos C_2,~E_2'(1) \indepSym C_1,~E_2'(2) \cpos C_1$. Hence, we can infer that $P(E_1'\wedge C_2) = P(E_1'(1)\wedge E_1'(2) \wedge C_2)$ $= P(E_1'(1))P(E_1'(2)C_2) > P(E_1'(1))P(E_1'(2))P(C_2) = P(E_1')P(C_2)$. Thus, $E_1' \cpos C_2$. Similarly, we can get $E_2' \cpos C_1$. With this, we can get $\alpha_1 > \alpha_2$ and $\beta_1 > \beta_2$. In addition, $\alpha \geq P(E_1')$ and $\beta \geq P(E_2')$.

    Similarly, we compute the partial derivative of $f(\alpha,\beta)$ as follows:

    \[
    \small
    \begin{array}{lll}
       \frac{\partial f}{\partial \alpha} & = & P(C_2) - \frac{P(E'_1(1)|E'_2)}{P(E'_1(1))}P(E'_2)P(C_2) - P(C_1)P(C_2)
        + \textcolor{blue}{\beta}P(C_1)P(C_2)\frac{P(E'_1(1) E_2'(1))}{P(E_1'(1))P(E_2'(1))}\\
        & \geq & P(C_2) - \frac{P(E'_1(1)|E'_2)}{P(E'_1(1))}P(E'_2)P(C_2) - P(C_1)P(C_2)
        + \textcolor{blue}{P(E_2')}P(C_1)P(C_2)\frac{P(E'_1(1) E_2'(1))}{P(E_1'(1))P(E_2'(1))} \\
        & = & (1-P(C_1)P(C_2))(1-\textcolor{blue}{P(E'_2)\frac{P(E'_1(1) E'_2(1))}{P(E_1'(1))P(E_2'(1))}}) \\
        & = & (1-P(C_1)P(C_2))(1-\textcolor{blue}{P(E_2'(2))P(E_2'(1)|E_1'(1))}) \\
        & > & 0
    \end{array}
    \]

    Similarly, we can also prove that the partial derivative $\frac{\partial f}{\partial \beta}$ is positive. Given $\frac{\partial f}{\partial \alpha} > 0$ and $\frac{\partial f}{\partial \beta} > 0$, based on Lemma~\ref{lemma:monotonic}, we can get $f(\alpha_1, \beta_1) > f(\alpha_2, \beta_2)$ if $\alpha_1 > \alpha_2$ and $\beta_1 > \beta_2$.

    In addition, we also have $E'_1 \cpos E'_2$, $C_1 \indepSym C_2$, so $P(E_1'\wedge E_2') > P(E_1')P(E_2')$ and $P(C_1 \wedge C_2) = P(C_1)P(C_2)$.
    Thus we are able to prove $P(E_1 \wedge E_2) = P(E_1'\wedge E_2') + P(C_1 \wedge C_2) + \textcolor{blue}{f(\alpha_1,\beta_1)}$ $> P(E_1')P(E_2') + P(C_1)P(C_2) + \textcolor{blue}{f(\alpha_2,\beta_2)} = P(E_1)P(E_2)$. Thus, we can get $E_1 \cpos E_2$.\\

    \item Case 1.3 $C_1 \indepSym C_2,~ C_1 \rightharpoonup^{+} E_2,~C_2 \indepSym E_1$
    \item Case 1.4 $C_1 \indepSym C_2,~ C_1 \indepSym E_2,~C_2 \rightharpoonup^{+} E_1$
    \item Case 1.5 $C_1 \indepSym C_2,~ C_1 \indepSym E_2,~C_2 \indepSym E_1$
    \end{itemize}

Case 2: $E_1' \indepSym E_2'$
\begin{itemize}
    \item Case 2.1 $C_1 \rightharpoonup^+ C_2,~E_1' \indepSym C_2,~E_2' \indepSym C_1$
    \item Case 2.2 $C_1 \indepSym C_2,~E_1' \rightharpoonup^+ C_2,~E_2' \indepSym C_1$
    \item Case 2.3 $C_1 \indepSym C_2,~E_1' \indepSym C_2,~E_2' \rightharpoonup^+ C_1$
    \item Case 2.4 $C_1 \indepSym C_2,~E_1' \rightharpoonup^+ C_2,~E_2' \rightharpoonup^+ C_1$
\end{itemize}

Case 3: $E_1' \cneg E_2'~$
This case is invalid as it violates the precondition $\not \vdash E_1 \rightharpoonup^- E_2$. \\

Proof of other cases

Here we omit the proof of other valid cases (1.3, 1.4, 1.5, 2.1, 2.2, 2.3, and 2.4) since it is similar to the proof of Case 1.1 and Case 1.2.
\qed \\

\textbf{Assumptions}


For the above proof of Lemmas \ref{lemma:factConjunct}, \ref{lemma:conjunct}, \ref{lemma:conjDisj} and \ref{thm:positive},
we made two assumptions. First, we assume that input $I$ is a term, instead of the literal or input fact. For instance, $I \in \{a,~\neg a\}$ where $a \in \textsf{InputFacts}(D)$. Second, given $I_a \rightharpoonup^+ I_b$, we imply $P(I_aI_b) > P(I_aI_b)$.

Now we give a proof about the second assumption: if $I_a \rightharpoonup^+ I_b,~I_a \in \{a, \neg a\},~I_b \in \{b, \neg b\}$, $a \in \textsf{InputFacts}(D)$, and $b \in \textsf{InputFacts}(D)$,  then $P(I_aI_b) > P(I_aI_b)$.

\textsf{Proof}.

Based on the rules \textsc{May-1} and \textsc{May-2}, there 
are four cases for $I_a$ and $I_b$ that can infer $I_a \rightharpoonup^+ I_b$.

\begin{enumerate}
    \item Case 1: $I_a=a,~I_b=b,~a \blacktriangleright^+ b$.
    
    $P(I_a\wedge I_b) = P(a\wedge b) > P(a)P(b)$ according to $a \blacktriangleright^+ b$.
    \item Case 2: $I_a=a,~I_b=\neg b,~a \blacktriangleright^- b$.
    
    $P(I_a\wedge I_b) = P(a\wedge \neg b) = P(\neg b |a)P(a) = (1-P(b|a))P(a)$. Given $a \blacktriangleright^- b$, we have $P(b|a) < P(b)$, and thus $(1-P(b|a)) > (1-P(b))$. Hence $P(I_a\wedge I_b) = (1-P(b|a))P(a) > (1-P(b))P(a) = P(I_b)P(I_a)$
    \item Case 3: $I_a=\neg a,~I_b=b,~a \blacktriangleright^- b$.

    $P(I_a\wedge I_b) = P(\neg a\wedge b) = P(\neg a |b)P(b) = (1-P(a|b))P(b)$. Given $a \blacktriangleright^- b$, we have $P(a|b) < P(a)$, and thus $(1-P(a|b)) > (1-P(a))$. Hence $P(I_a\wedge I_b) = (1-P(a|b))P(b) > (1-P(a))P(b) = P(I_a)P(I_b)$
    
    \item Case 4: $I_a=\neg a,~I_b= \neg b,~a \blacktriangleright^+ b$.
    \[
    \begin{array}{lllr}
        P(I_a \wedge I_b) =  P(\neg a \wedge \neg b) = 1 - P(a \vee b) & = & 1 - P(a) - P(b) + P(a\wedge b) &\\
                          & > & 1 - P(a) - P(b) + P(a)P(b) & \textsf{ using } a \blacktriangleright^+ b \\
                          & = & (1-P(a))(1-P(b)) = P(I_a)P(I_b)
    \end{array}
    \]
    
\end{enumerate}
\qed

                \begin{lemma}
    \label{lemma:factConjunctNeg}
    Given  an input fact $E_1 = I$ and  a general  DNF formula $E_2 = \bigvee \bigwedge_{j} I_j$,  if $\vdash \chi(E_1),~\vdash \chi(E_2),~ \not \vdash E_1 \rightharpoonup^{+} E_2$ and $ \vdash  E_1 \rightharpoonup^{-} E_2$, then $P(E_1\wedge E_2) < P(E_1)P(E_2), i.e., E_1 \cneg E_2$.
    \end{lemma}

    \begin{lemma}
    \label{lemma:conjunctNeg}
    Let $E_1$, $E_2$ be conjunctions over input facts, i.e.,  $E_1 = \bigwedge_{i} I_i$ and $E_2 = \bigwedge_{j} I_j$. If $\vdash \chi(E_1),~\vdash \chi(E_2),~ \not \vdash E_1 \rightharpoonup^{+} E_2$ and $\vdash  E_1 \rightharpoonup^{-} E_2$, then $P(E_1\wedge E_2) < P(E_1)P(E_2)$.
    \end{lemma}

    \begin{lemma}
    \label{lemma:conjDisjNeg}
    Let $E_1$ be a conjunction over input facts (i.e.,  $E_1 = \bigwedge_{i} I_i$) and let  $E_2 = \bigvee \bigwedge_{j} I_j$ be a DNF formula over input facts. If $\vdash \chi(E_1),~\vdash \chi(E_2),~ \not \vdash E_1 \rightharpoonup^{+} E_2$ and $\vdash  E_1 \rightharpoonup^{-} E_2$, then $P(E_1\wedge E_2) < P(E_1)P(E_2)$.
    \end{lemma}

        \begin{lemma}[\textbf{Soundness of Rule \textsc{Neg}}]
    \label{thm:negative}
        The rule \textsc{Negative} in Figure~\ref{fig:corr} are sound. In particular, given a pair of relations $E_1 = \bigvee \bigwedge_{i} I_i, E_2 = \bigvee \bigwedge_{j} I_j$, if $\vdash \chi(E_1),~\vdash \chi(E_2),~ \not \vdash E_1 \rightharpoonup^{+} E_2$ and $\vdash  E_1 \rightharpoonup^{-} E_2$, then $P(E_1\wedge E_2) < P(E_1)P(E_2)$.
    \end{lemma}

Proof of Lemma~\ref{thm:negative}.

To prove negative correlation $E_1 \cneg E_2$, we also make two assumptions as we did for Lemma~\ref{thm:positive}. First, we assume that the input \( I \) is a term, instead of a literal or input fact. For instance, \( I \in \{a, \neg a\} \) where \( a \in \textsf{InputFacts}(D) \). Second, given \( I_a \rightharpoonup^- I_b \), we imply \( P(I_a I_b) < P(I_a I_b) \).

Now we give a proof about the second assumption: if $I_a \rightharpoonup^- I_b,~I_a \in \{a, \neg a\},~I_b \in \{b, \neg b\}$, $a \in \textsf{InputFacts}(D)$, and $b \in \textsf{InputFacts}(D)$,  then $P(I_aI_b) < P(I_aI_b)$.

\textsf{Proof}.

Based on the rules \textsc{May-1} and \textsc{May-2}, there 
are four cases for $I_a$ and $I_b$ that can infer $I_a \rightharpoonup^- I_b$.

\begin{enumerate}
    \item Case 1: $I_a=a,~I_b=b,~a \blacktriangleright^- b$.
    
    $P(I_a\wedge I_b) = P(a\wedge b) < P(a)P(b)$ according to $a \blacktriangleright^- b$.
    \item Case 2: $I_a=a,~I_b=\neg b,~a \blacktriangleright^+ b$.
    
    $P(I_a\wedge I_b) = P(a\wedge \neg b) = P(\neg b |a)P(a) = (1-P(b|a))P(a)$. Given $a \blacktriangleright^+ b$, we have $P(b|a) > P(b)$, and thus $(1-P(b|a)) < (1-P(b))$. Hence $P(I_a\wedge I_b) = (1-P(b|a))P(a) < (1-P(b))P(a) = P(I_b)P(I_a)$
    \item Case 3: $I_a=\neg a,~I_b=b,~a \blacktriangleright^+ b$.

    $P(I_a\wedge I_b) = P(\neg a\wedge b) = P(\neg a |b)P(b) = (1-P(a|b))P(b)$. Given $a \blacktriangleright^+ b$, we have $P(a|b) > P(a)$, and thus $(1-P(a|b)) < (1-P(a))$. Hence $P(I_a\wedge I_b) = (1-P(a|b))P(b) < (1-P(a))P(b) = P(I_a)P(I_b)$
    
    \item Case 4: $I_a=\neg a,~I_b= \neg b,~a \blacktriangleright^- b$.
    \[
    \begin{array}{lllr}
        P(I_a \wedge I_b) =  P(\neg a \wedge \neg b) = 1 - P(a \vee b) & = & 1 - P(a) - P(b) + P(a\wedge b) &\\
                          & < & 1 - P(a) - P(b) + P(a)P(b) & \textsf{ using } a \blacktriangleright^- b \\
                          & = & (1-P(a))(1-P(b)) = P(I_a)P(I_b)
    \end{array}
    \]
    
\end{enumerate}

To prove Lemma~\ref{thm:negative} for Rule \textsc{Neg} in Figure~\ref{fig:corr}, we use the following Lemmas: Lemma~\ref{lemma:monotonic},~Lemma~\ref{lemma:factConjunctNeg}, Lemma~\ref{lemma:conjunctNeg} and Lemma~\ref{lemma:conjDisjNeg}.
The proof of Lemma~\ref{thm:negative} ($E_1 \cneg E_2$) is symmetric to the proof of Lemma~\ref{thm:positive} ($E_1 \cpos E_2$). Similarly, the proofs of Lemma~\ref{lemma:factConjunctNeg}, Lemma~\ref{lemma:conjunctNeg}, Lemma~\ref{lemma:conjDisjNeg} are symmetric to the the proofs of Lemma~\ref{lemma:factConjunct}, Lemma~\ref{lemma:conjunct}, Lemma~\ref{lemma:conjDisj}.
Hence, we omit the detailed proof here.

\qed

\subsection{Computing approximate probability bounds}
\label{appendix:approx}

In this section, we give proof for Theorem~\ref{theorem:approx}. To prove Theorem~\ref{theorem:approx}, we first introduce the follow Lemmas which establish the soundness of the rules in Figure~\ref{fig:approx} and Table~\ref{table: CLCU}.

\begin{lemma}[Soundness of rules in Table~\ref{table: CLCU}]
\label{lemma:CLCU}
    Given Boolean expressions \( E_1 \) and \( E_2 \) with probabilities \( P(E_1) \in [l_1, u_1] \) and \( P(E_2) \in [l_2, u_2] \), and their correlation type \(\star\), applying the formulas in Table~\ref{table: CLCU} yields:
    
    $
    \textsf{CL}(l_1, l_2, \star) \leq P(E_1 \wedge E_2) \leq \textsf{CU}(u_1, u_2, \star)
    $
    and
    $
    \textsf{DL}(l_1, l_2, \star) \leq P(E_1 \vee E_2) \leq \textsf{DU}(u_1, u_2, \star).
    $
\end{lemma}

\textsc{Proof}.
We will prove the four cases (\textsf{CL, CU, DL, DU}) respectively.
\begin{itemize}[leftmargin=*]
    \item \textbf{\textsf{CL CU}} This case is used for computing the lower and upper bounds of conjunction $E_1 \wedge E_2$.
    \begin{enumerate}
        \item $\star = \bot$: In this case, $E_1$ and $E_2$ are independent, and thus $P(E_1 \wedge E_2) = P(E_1)P(E_2)$. Given $u_1 \geq P(E_1) \geq l_1$ and $u_2 \geq P(E_2) \geq l_2$, $u_1u_2 \geq P(E_1)P(E_2) \geq l_1l_2$. Thus, $P(E_1\wedge E_2) \in [l_1l_2,u_1u_2]$.
        \item $\star = \top$: In this case, the correlation between $E_1$ and $E_2$ is unknown. Given the conjunction $P(E_1 \wedge E_2)$ = $P(E_1)$*$P(E_2|E_1)$ = $P(E_2)$*$P(E_1|E_2)$, $P(E_1|E_2) \in [0,1]$ and $P(E_2|E_1) \in [0,1]$, in order to maximize $P(E_1 \wedge E_2)$, either $P(E_1|E_2)$ or $P(E_2|E_1)$ must be one. There are following three cases to compute the upper bound of $P(E_1 \wedge E_2)$:
        \begin{itemize}
            \item $P(E_1) = P(E_2)$: In this case, $P(E_1|E_2) = P(E_2|E_1) = 1$, the upper bound of $P(E_1 \wedge E_2)$ is $P(E_1)$.
            \item $P(E_1) < P(E_2)$: In this case, only $P(E_1|E_2)$ can be one, and $P(E_2|E_1) < 1$, the upper bound of $P(E_1 \wedge E_2)$ is $P(E_1)$.
            \item $P(E_1) > P(E_2)$: In this case, only $P(E_2|E_1)$ can be one, and $P(E_1|E_2) < 1$, the upper bound of $P(E_1 \wedge E_2)$ is $P(E_2)$.
        \end{itemize}
        To summarize the upper bound of $P(E_1 \wedge E_2)$ is \textsf{min}($P(E_1), P(E_2)$). As we focus on the upper bound, $P(E_1 \wedge E_2) < \textsf{min}(u1, u_2)$.

        As we know, $P(E_1 \vee E_2) = P(E_1) + P(E_2) - P(E_1 \wedge E_2) < 1$, we can get $P(E_1 \wedge E_2) > P(E_1) + P(E_2) - 1$. As we focus on computing the lower bound, $P(E_1 \wedge E_2) > l_1 + l_ 2 - 1$. Meanwhile, naturally, $P(E_1 \wedge E_2) \in [0,1]$, so the lower bound can be rewritten as \textsf{max}($0, l_1 + l_ 2 - 1$).

        As a result, $P(E_1 \wedge E_2) \in [\textsf{max}(0, l_1 + l_ 2 - 1), \textsf{min}(u_1, u_2)]$.
        \item $\star = +$:  In this case, $E_1 \cpos E_2$. According to the definition of $\cpos$, we know that $P(E_1 \wedge E_2) > P(E_1)P(E_2)$. Thus the lower bound of $P(E_1 \wedge E_2)$ is $P(E_1)P(E_2)$, which is $l_1l_2$. The upper bound in this case is same as the upper bound of the unknown ($\top$) case, which is $\textsf{min}(u_1, u_2)$. Hence, $P(E_1 \wedge E_2) \in [l_1l_2, \textsf{min}(u_1, u_2)]$.
        \item $\star = -$: In this case, $E_1 \cneg E_2$. According to the definition of $\cneg$, we know that $P(E_1 \wedge E_2) < P(E_1)P(E_2)$. Thus the upper bound of $P(E_1 \wedge E_2)$ is $P(E_1)P(E_2)$, which is $u_1u_2$. The lower bound in this case is same as the upper bound of the unknown ($\top$) case, which is \textsf{max}($0, l_1 + l_ 2 - 1$). Hence, $P(E_1 \wedge E_2) \in [\textsf{max}(0, l_1 + l_ 2 - 1), u_1u_2]$.
    \end{enumerate}
    \item \textbf{\textsf{DL DU}} This case is used for computing the lower and upper bounds of disjunction $E_1 \vee E_2$.
    \begin{enumerate}
        \item $\star = \bot$: In this case, $E_1$ and $E_2$ are independent, and thus $P(E_1 \vee E_2) = P(E_1) + P(E_2) - P(E_1)P(E_2)$. Given $u_1 \geq P(E_1) \geq l_1$ and $u_2 \geq P(E_2) \geq l_2$, $u_1u_2 \geq P(E_1)P(E_2) \geq l_1l_2$. Thus, $P(E_1\wedge E_2) \in [l_1+l_2-l_1l_2,u_1 + u_2 -u_1u_2]$.
        \item $\star = \top$: In this case, the correlation between $E_1$ and $E_2$ is unknown. Given the disjunction $P(E_1 \vee E_2)$ = $P(E_1) + P(E_2) - P(E_1\wedge E_2)$, in order to maximize $P(E_1 \vee E_2)$, $P(E_1 \wedge E_2)$ should be zero, i.e., $E_1$ and $E_2$ are mutually exclusive. Meanwhile, as $P(E_1 \vee E_2) \in [0,1]$. We refine the upper bound as $\text{min}(1, P(E_1)+P(E_2)$, which is $\text{min}(1, u_1 + u_2)$.

        To compute the lower bound of $P(E_1 \vee E_2)$, we want to maximize $P(E_1 \wedge E_2)$, which is $P(E_1|E_2)P(E_2)$ or $P(E_2|E_1)P(E_1)$. There are two cases as follows:
        \begin{itemize}
            \item $P(E_1) = P(E_2)$: in this case, the maximal value of $P(E_1|E_2)$ is 1 and $P(E_1 \wedge E_2) = P(E_1)$. Thus, $P(E_1 \vee E_2) = P(E_1) + P(E_2) - P(E_1) = P(E_2) = (E_1)$.
            \item $P(E_1) > P(E_2)$: in this case, the maximal value of $P(E_1|E_2)$ is one and $P(E_1 \wedge E_2) = P(E_2)$. Thus, $P(E_1 \vee E_2) = P(E_1) + P(E_2) - P(E_2) = P(E_1)$.
            \item $P(E_1) < P(E_2)$: in this case, the maximal value of $P(E_2|E_1)$ is one and $P(E_1 \wedge E_2) = P(E_1)$. Thus, $P(E_1 \vee E_2) = P(E_1) + P(E_2) - P(E_1) = P(E_2)$.
        \end{itemize}

         To summarize the lower bound of $P(E_1 \vee E_2)$ is \textsf{max}($P(E_1), P(E_2)$). As we focus on the lower bound, $P(E_1 \vee E_2) > \textsf{max}(l_1, l_2)$.

          As a result, $P(E_1 \vee E_2) \in [ \textsf{max}(l_1, l_2), \text{min}(1, u_1 + u_2)]$.
          
        \item $\star = +$: In this case, $E_1 \cpos E_2$. According to the definition of $\cpos$, we know that $P(E_1 \wedge E_2) > P(E_1)P(E_2)$. Thus $P(E_1 \vee E_2) = P(E_1) + P(E_2) - P(E_1 \wedge E_2) < P(E_1) + P(E_2) - P(E_1)P(E_2)$. The upper bound of $P(E_1 \vee E_2)$ is $u_1+u_2-u_1u_2$.
        The lower bound in this case is same as the lower bound of the unknown ($\top$) case, which is $\textsf{max}(l_1, l_2)$. Hence, $P(E_1 \vee E_2) \in [\textsf{max}(l_1, l_2), u_1+u_2-u_1u_2]$.
        \item $\star = -$: In this case, $E_1 \cneg E_2$. According to the definition of $\cneg$, we know that $P(E_1 \wedge E_2) < P(E_1)P(E_2)$. Thus $P(E_1 \vee E_2) = P(E_1) + P(E_2) - P(E_1 \wedge E_2) > P(E_1) + P(E_2) - P(E_1)P(E_2)$. Given a function $f(x,y) = x + y - xy$, it is evident that function $f$ is strictly increasing. Hence, $P(E_1 \vee E_2) > f(P(E_1), P(E_2)) \geq f(l_1, l_2)$.
        Thus, the lower bound of $f(x,y)$ is $l_1 + l_2 - l_1l_2$.
        
        The upper bound in this case is same as the upper bound of the unknown ($\top$) case, which is \textsf{min}($1, u_1+u_2$). Hence, $P(E_1 \vee E_2) \in [l_1 + l_2 - l_1l_2, \textsf{min}(1, u_1+u_2)]$.
    \end{enumerate}
\end{itemize}

\qed

\begin{lemma}[Soundness of rules in Figure~\ref{fig:approx}]
\label{lemma:approxRule}
    Given a Boolean expression \( E \), an environment mapping \( \Gamma \), a Datalog program \( D \), and a correlation environment \( \Sigma \), the rules in Figure~\ref{fig:approx} compute the lower and upper bounds \( l \) and \( u \) of \( P(E) \) such that \( l \leq P(E) \leq u \).
\end{lemma}

\textsc{Proof}. We prove that each rule is sound.
\begin{itemize}[leftmargin=*]
    \item \textsc{In}: Given $I \in \textsf{InputFacts}(D)$ and $p :: (I \ | \ \varnothing) \in \textsf{InputProbs}(D)$, we can get $P(I) = p$ and thus $P(I) \in [p,p]$.
    \item \textsc{Out}: $\Gamma$ is an environment mapping output relations $O$ to a pair of variables representing their upper and lower bounds, i.e., $\Gamma[O] = [l,u]$. Hence, $P(O) \in [l, u]$.
    \item \textsc{Neg}: Given $P(E) = [l,u]$, $P(\neg E) = 1 - P(E) \in [1-u,1-l]$.
    \item \textsc{Conjunct}: Given $P(E_1) \in  [l_1, u_1],~P(E_2) \in [l_2, u_2]$, and $\Sigma(E_1, E_2) = \star$, 
    according to Lemma~\ref{lemma:CLCU}, the probability of $E_1 \wedge E_2$ is $P(E_1 \wedge E_2) \in [\textsf{CL}(l_1,l_2,\star), \textsf{CU}(u_1,u_2,\star)]$.
    \item \textsc{Disjunct}: Given $(E_1, p_1)$ and $(E_2, p_2)$, $E_1 = \bigwedge_i I_i,~E_2 = \bigvee \bigwedge_j I_j$, this rule computes the probability of $(E_1,p_1) \vee (E_2, p_2)$.
    We prove this rule by listing two possible cases.

    \begin{enumerate}
    \item\textbf{Case 1}: $E_1 = \bigwedge_i I_i,~E_2 = \bigwedge_j I_j$.

    Given $P(E_1) \in [l_1,u_2]$, we use $e_1$ to denote the event $(E_1, p_1)$, and $P(e_1) = [p_1l_1, p_1u_1]$. Similarly, we use $e_2$ to denote the event $(E_2, p_2)$ and $P(e_2) \in [p_2l_2, p_2u_2]$. Given $\Sigma(E_1, E_2) = \star$, it's obvious that $\Sigma(e_1, e_2) = \star$. According to Lemma~\ref{lemma:CLCU}, the probability of $P(e_1 \vee e_2) \in [\textsf{DL}(p_1l_1, p_2l_2, \star), \textsf{DU}(p_1u_1, p_2u_2, \star)]$.

    \item \textbf{Case 2}: $E_1 = \bigwedge_i I_i$, $E_2 = \bigvee \bigwedge_j I_j$

    In the derivation graph, edge $E_1$ leads to one derivation $D_1$ with path probability $P(D_1|E_1) = p$. The probability of the union of two derivations is $P(D_i \vee D_j) = P((E_i, p_i) \vee (E_j, p_j))$, thus $P(D_i \vee D_j | (E_i, p_i) \vee (E_j, p_j)) = 1$. Therefore, the path probability of the union of multiple derivations is 1, making the path probability of $E_2$ equal to 1.

    Let $e_1$ denote $(E_1, p_1)$ and $e_2$ represent $(E_2, 1)$. Given $P(E_1) \in [l_1, u_1]$ and $P(E_2) \in [l_2, u_2]$, we have $P(e_1) = [p_1l_1, p_1u_1]$ and $P(e_2) \in [l_2, u_2]$. Since $\Sigma(E_1, E_2) = \star$, it follows that $\Sigma(e_1, e_2) = \star$. According to Lemma~\ref{lemma:CLCU}, the probability $P(e_1 \vee e_2)$ falls within $[\textsf{DL}(p_1l_1, l_2, \star), \textsf{DU}(p_1u_1, u_2, \star)]$.
    \end{enumerate}
    \qed

    
\end{itemize}

\begin{lemma}
\label{lemma:alg}
    Given a derivation graph $G$, a Datalog program $D$, and a correlation mapping $\Sigma$, Algorithm~\ref{alg:approxInterval} outputs a mapping $M$ that associates each output fact with its probability interval. For an output relation $R$, the mapping $M[R] = (l,u)$ ensures that the probability $P(R)$ lies within the interval $[l,u]$.
\end{lemma}

\textsc{Proof}.

The probability interval $(l, u)$ is directly computed by the \textsc{ApproxExpr} function. Since the \textsc{ApproxExpr} function applies the rules in Figure~\ref{fig:approx}, according to Lemma~\ref{lemma:approxRule}, we know that the computed interval $(l, u)$ is sound, i.e., $l \leq P(R) \leq u$, given the output relation $R$. Here, $R$ is also an internal node $n$ in $G$.

\qed

\noindent
\textbf{\textsc{Proof of Theorem~\ref{theorem:approx}}} \\
Based on Lemma~\ref{lemma:alg}, we know that Algorithm~\ref{alg:approxInterval} outputs a mapping $M$ that associates each output fact with its probability interval. For an output relation $R$, the mapping $M[R] = (l, u)$ ensures that the probability $P(R)$ lies within the interval $[l, u]$. That is, $l \leq P(R) \leq u$.

For the constrained optimization shown in Algorithm~\ref{alg:baseline:opt}, it also returns a mapping $M$. Let's denote $(l^*, u^*) = M(R)$. Since it is based on optimization, we know that $l^* = \textsf{min}(P(R))$ and $u^* = \textsf{max}(P(R))$.

Hence, we have $l \leq P(R)$, which implies $l \leq \textsf{min}(P(R)) = l^*$, and $u \geq P(R)$, which implies $u \geq \textsf{max}(P(R)) = u^*$. Thus, we have $l \leq l^* \leq u^* \leq u$.
\qed

\subsection{Iterative Refinement of Probability Bounds}

In this section, we provide the formal proof of Theorem~\ref{thm:delta}. To establish this proof, we need the following Lemmas: Lemma~\ref{lemma:makeSAT}, Lemma~\ref{lemma:BB}, and Lemma~\ref{lemma:binary}.

Given an output relation $R$, $l^*$ and $u^*$ denote the ground truth probability bounds for $R$, i.e., $l^* = \textsf{min}(P(R))$ and $u^* = \textsf{max}(P(R))$.

\begin{lemma}
    \label{lemma:makeSAT}
    Given the lower bound $l$, upper bound $u$, constraints $\phi$, expression $e$ and the boolean flag \textsf{low}, \textsc{MakeSat} procedure returns an interval $(l,u)$ such that $l \leq l^* \leq u$ if \textsf{low} and $l \leq u^* \leq u$ otherwise.
\end{lemma}

\noindent
\textbf{\textsc{Proof of Lemma~\ref{lemma:makeSAT}}}\\
We start the proof by listing all possible cases as follows:
\begin{itemize}[leftmargin=*]
    \item \textsf{low=true}: in this case, it focuses on the lower bound. It starts from $[l,l]$. As we know that $\textsf{min}(e)$ = $l^*$, and our lower bound is sound: $l \leq l^*$. Let's futher classify this case:
    \begin{enumerate}
        \item $l = l^*$: In this case, $\phi \wedge l \leq e \leq l$ is SAT, it does enter the loop in \emph{lines 1-2} and \textsc{MakeSat} directly return $(l,l)$. It is consistent with $l \leq l^* \leq l$.
        \item $l < l^*$: In this case, $\phi \wedge l \leq e \leq l$ is UNSAT due to the contradiction between $l < l^* = \textsf{min}(e)$ and $e \leq l$. Hence, it enters the loop in \emph{lines 1-2} to iteratively increasing $l$ until at $t$-iteration, $\phi \wedge l_t \leq e \leq \textsf{Incr}(l_t)$ turns SAT. In this case, \textsc{MakeSat} returns $(l_t, \textsf{Incr}(l_t))$.
        
        It also implies that at $t-1$ iteration, $\phi \wedge l_{t-1} \leq e \leq l_t$ is UNSAT. From the UNSAT case in $t-1$-th iteration, we know that either $l_t < \textsf{min}(e)$ or $l_{t-1} > \textsf{max}(e)$. It's obvious that $l_{t-1} > \textsf{max}(e)$ is an invalid case. Hence, we can infer $l_t < \textsf{min}(e)$. From the SAT case, we know that $[\textsf{min}(e), \textsf{max}(e)]$ overlaps with $[l_t, \textsf{Incr}(l_t)]$, thus we can get $l_t \leq \textsf{max}(e) \wedge \textsf{Incr}(l_t) \geq \textsf{min}(e)$. Hence, we have $l_t < \textsf{min}(e) \leq \textsf{Incr}(l_t)$, which also implies $l_t \leq \textsf{min}(e) \leq \textsf{Incr}(l_t)$. As $l^* = \textsf{min}(e)$, we have $l_t \leq l^* \leq \textsf{Incr}(l_t)$.       
    \end{enumerate}
     \item \textsf{low=false}: in this case, it focuses on the upper bound. It starts from $[u,u]$. As we know that $\textsf{max}(e)$ = $u^*$, and our lower bound is sound: $u \geq u^*$. Let's futher classify this case:
    \begin{enumerate}
        \item $u = u^*$: In this case, $\phi \wedge u \leq e \leq u$ is SAT, it does enter the loop in \emph{lines 1-2} and \textsc{MakeSat} directly return $(u,u)$. It is consistent with $u \leq u^* \leq u$.
        \item $u > u^*$: In this case, $\phi \wedge u \leq e \leq u$ is UNSAT due to the contradiction between $u > u^* = \textsf{max}(e)$ and $u \leq e$. Hence, it enters the loop in \emph{lines 1-2} to iteratively decreasing $u$ until at $t$-iteration, $\phi \wedge \textsf{Decr}(u_t) \leq e \leq u_t$ turns SAT. In this case, \textsc{MakeSat} returns $(\textsf{Decr}(u_t), u_t)$.
        
        It also implies that at $t-1$ iteration, $\phi \wedge u_t \leq e \leq u_{t-1}$ is UNSAT. From the UNSAT case in $t-1$-th iteration, we know that either $u_{t_1} < \textsf{min}(e)$ or $u_t > \textsf{max}(e)$. It's obvious that $u_{t_1} < \textsf{min}(e)$ is an invalid case. Hence, we can infer $u_t > \textsf{max}(e)$. From the SAT case, we know that $[\textsf{min}(e), \textsf{max}(e)]$ overlaps with $[\textsf{Decr}(u_t), u_t]$, thus we can get $\textsf{Decr}(u_t) \leq \textsf{max}(e) \wedge u_t \geq \textsf{min}(e)$. Hence, we have $\textsf{Decr}(u_t) < \textsf{max}(e) < u_t$, which also implies $\textsf{Decr}(u_t) < \textsf{max}(e) \leq u_t$. As $u^* = \textsf{max}(e)$, we have $\textsf{Decr}(u_t) \leq u^* \leq u_t$.       
    \end{enumerate}
\end{itemize}

\qed

\begin{lemma}
    \label{lemma:BB}
     \textsc{BoundBounds} Procedure in Algorithm~\ref{alg:epilson} outputs a mapping $B$ that maps $R$ to a quadruple $(l^-,l^+,u^-,u^+)$, such that $l^- \leq l^* \leq l^+$ and $u^-\leq u^* \leq u^+$.
\end{lemma}

\noindent
\textbf{\textsc{Proof of Lemma~\ref{lemma:BB}}}\\
We start the proof by listing all possible cases as follows:
\begin{itemize}[leftmargin=*]
    \item \emph{lower bound}: When the signal \textsf{low} is set to true, \textsc{MakeSat} function in \emph{Line 3} is invoked in Algorithm~\ref{alg:epilson} and returns $(l^-,l^+)$. According to Lemma~\ref{lemma:makeSAT}, we know that $l^- \leq l^* \leq l^+$.
    \item \emph{upper bound}: When the signal \textsf{low} is set to false, \textsc{MakeSat} function in \emph{Line 4} is invoked in Algorithm~\ref{alg:epilson} and returns $(u^-,u^+)$. According to Lemma~\ref{lemma:makeSAT}, we know that $u^- \leq u^* \leq u^+$.
\end{itemize}
\qed

\begin{lemma}
    \label{lemma:binary}
    \textsc{BinarySearch} procedure in Algorithm~\ref{alg:boundByDelta} takes 
    an output relation $R$, lower bound $l$, upper bound $u$, the boolean flag \textsf{low} and error bound $\delta$ as inputs, and outputs an interval $(L,U)$. If \textsf{low} is set to true, $L \leq l^* \leq U$ and $|U-L| < \delta$. Otherwise, if \textsf{low} is set to false, $L \leq u^* \leq U$ and $U-L < \delta$.
\end{lemma}

\noindent
\textbf{\textsc{Proof of Lemma~\ref{lemma:binary}}}\\
We start the proof by listing all possible cases as follows:
\begin{itemize}[leftmargin=*]
    \item \textsf{low = true}: 
    \begin{enumerate}
        \item $(u-l) < \delta$: \textsc{BinarySearch} procedure directly returns $(l,u)$. Here, the output $[L,U] = [l,u]$.
        \item $(u-l) \geq \delta$: it enters the loop in \emph{lines 1-6}. Given $mid = (l+u)/2$. If $\phi \wedge l \leq \textsf{Expr}(R) \leq mid$ is UNSAT, we know that intervals $[l, mid]$ and $[l^*, u^*]$ are disjoint, we have $l > u^* \vee mid < l^*$. It's obvious that $l > u^*$ is false. Given $mid < l^*$, $l^*$ falls into the interval $[mid, l]$. Next iteration focuses on splitting $[mid, l]$ until its distance is less than $\delta$.
        
        Otherwise, if $\phi \wedge l \leq \textsf{Expr}(R) \leq mid$ is SAT, it means that two intervals $[l, mid]$ and $[l^*, u^*]$ overlap. Hence, we can get $l \leq u^* \wedge mid \geq l^*$. Thus $l^*$ falls into the interval $[l, mid]$. Next iteration focuses on splitting $[l, mid]$ until its distance is less than $\delta$.
    \end{enumerate}

    Hence, after each iteration, the interval $[l,u]$ we get is guaranteed to contain $l^*$. When it terminates, the loop condition is falsified, and it returns the output interval $[L,U]$ where $U-L < \delta$.

    \item \textsf{low = false}:

    \begin{enumerate}
        \item $(u-l) < \delta$: \textsc{BinarySearch} procedure directly returns $(l,u)$. Here, the output $[L,U] = [l,u]$.
        \item $(u-l) \geq \delta$: it enters the loop in \emph{lines 1-6}. Given $mid = (l+u)/2$. If $\phi \wedge mid \leq \textsf{Expr}(R) \leq u$ is UNSAT, we know that intervals $[mid, u]$ and $[l^*, u^*]$ are disjoint, we have $mid > u^* \vee u < l^*$. It's obvious that $u < l^*$ is false. Given $mid > u^*$, $u^*$ falls into the interval $[l, mid]$. Next iteration focuses on splitting $[l,mid]$ until its distance is less than $\delta$.
        
        Otherwise, if $\phi \wedge mid \leq \textsf{Expr}(R) \leq u$ is SAT, it means that two intervals $[mid,u]$ and $[l^*, u^*]$ overlap. Thus, we can get $l \leq u^* \wedge mid \geq u^*$. Thus $u^*$ falls into the interval $[mid, u]$. Next iteration focuses on splitting $[mid, u]$ until its distance is less than $\delta$.
        \end{enumerate}

    Hence, after each iteration, the interval $[l,u]$ we get is guaranteed to contain $u^*$. When it terminates, the loop condition is falsified, and it returns the output interval $[L,U]$ where $U-L < \delta$.
\end{itemize}
\qed

\noindent
\textbf{\textsc{Proof of Theorem~\ref{thm:delta}}} \\
In \textsc{MakeDeltaPrecise} procedure, the last step invokes \textsc{BinaryProcedure}. For lower bound, it returns $(l^-,l^+)$. For upper bound, it returns $(u^-,u^+)$. According to Lemma~\ref{lemma:binary}, we know that $l^- \leq l^* \leq l^+$, $l^+-l^- < \delta$ and $u^- \leq u^* \leq u^+$, $u^+-u^- < \delta$. \textsc{MakeDeltaPrecise} procedure returns $(l,u)$ as its probability bounds, where $l = l^-, u = u^+$.

Given $l^* < l^+$, we have $l^*-l = l^*-l^- \leq l^+-l^- < \delta$. Hence, $l > l^* - \delta$. Given $l=l^- \leq l^*$, we can get $l^*-\delta \leq l \leq l^*$.

Given $u^* > u^-$, we have $u-u^* = u^+-u^* \leq u^+-u^- < \delta$. Given $u=u^+ \geq u^*$, we can get $u^* \leq u \leq u^*+\delta$.
\qed

\subsection{Correlation type of constantly true and false cases}

For cases that are constantly true or false, the correlation type—whether $+,-,\bot, \top$—does not affect the probability computation in Table~\ref{table: CLCU}. Regardless of the assigned correlation type, the computation remains sound and correct.

\begin{itemize}
    \item \textbf{$E_1$ = \texttt{true}, $E_2$ = \texttt{true}}. 
    $e_1 = Pr(E_1) = 1, e_2 = Pr(E_2) = 1$.
    \begin{itemize}
        \item $CL(e_1, e_2, +) = CL(e_1, e_2, -) = CL(e_1, e_2, \bot) = CL(e_1, e_2, \top) = 1$
        \item $CU(e_1, e_2, +) = CU(e_1, e_2, -) = CU(e_1, e_2, \bot) = CU(e_1, e_2, \top) = 1$
        \item  $DL(e1, e2, +) = DL(e1, e2, -) = DL(e1, e2, \bot) = DL(e1, e2, \top) = 1$
        \item $DU(e1, e2, +) = DU(e1, e2, -) = DU(e1, e2, \bot) = DU(e1, e2, \top) = 1$
    \end{itemize}
    As a result, $Pr(E_1 \wedge E_2) \in [1,1] = 1$ and $Pr(E_1 \vee E_2) \in [1,1] = 1$ for any correlation type. 
    \item \textbf{$E_1$ = \texttt{true}, $E_2$ = \texttt{false}}. 
    $e_1 = Pr(E_1) = 1, e_2 = Pr(E_2) = 0$.
    \begin{itemize}
        \item $CL(e_1, e_2, +) = CL(e_1, e_2, -) = CL(e_1, e_2, \bot) = CL(e_1, e_2, \top) = 0$
        \item $CU(e_1, e_2, +) = CU(e_1, e_2, -) = CU(e_1, e_2, \bot) = CU(e_1, e_2, \top) = 0$
        \item $DL(e_1, e_2, +) = DL(e_1, e_2, -) = DL(e_1, e_2, \bot) = DL(e_1, e_2, \top) = 1$
        \item $DU(e_1, e_2, +) = DU(e_1, e_2, -) = DU(e_1, e_2, \bot) = DU(e_1, e_2, \top) = 1$
    \end{itemize}
    As a result, $Pr(E_1 \wedge E_2) \in [0,0] = 0$ and $Pr(E_1 \vee E_2) \in [1,1] = 1$ for any correlation type.
    \item \textbf{$E_1$ = \texttt{false}, $E_2$ = \texttt{true}}. 
    $e_1 = Pr(E_1) = 0, e_2 = Pr(E_2) = 1$.
    \begin{itemize}
        \item Symmetric to Case ($E_1$ = \texttt{true}, $E_2$ = \texttt{false}) , so we will not provide a proof here.
    \end{itemize}
    \item \textbf{$E_1$ = \texttt{false}, $E_2$ = \texttt{false}}. 
    $e_1 = Pr(E_1) = 0, e_2 = Pr(E_2) = 0$.
    \begin{itemize}
        \item $CL(e_1, e_2, +) = CL(e_1, e_2, -) = CL(e_1, e_2, \bot) = CL(e_1, e_2, \top) = 0$
        \item $CU(e_1, e_2, +) = CU(e_1, e_2, -) = CU(e_1, e_2, \bot) = CU(e_1, e_2, \top) = 0$
        \item $DL(e_1, e_2, +) = DL(e_1, e_2, -) = DL(e_1, e_2, \bot) = DL(e_1, e_2, \top) = 0$
        \item $DU(e_1, e_2, +) = DU(e_1, e_2, -) = DU(e_1, e_2, \bot) = DU(e_1, e_2, \top) = 0$
    \end{itemize}
    As a result, $Pr(E_1 \wedge E_2) \in [0,0] = 0$ and $Pr(E_1 \vee E_2) \in [0,0] = 0$ for any correlation type.
\end{itemize}

In summary, the correlation type does not affect the computation of always-true and always-false cases. When assigning a positive correlation (+) to a pair of constantly-true and constantly-false, the computation remains correct.

\subsection{Implied Conditional Dependence from Overview}
\label{appendix:example}

In this subsection, we provide a proof  that input facts \texttt{edge(2,5)} and \texttt{edge(2,6)} from Figure~\ref{fig:probCond} must be dependent. 
Let $A, B, C$ denote  \texttt{edge(2,5)}, \texttt{edge(2,6)}, and \texttt{edge(1,4)} respectively. 
Then, we have $P(A \land C) = P(A | C) \times P(C) = 0.8 \times 0.6 = 0.48$. Since $P(A \land C) = P(A \land C \land B) + P(A \land C \land \neg B)$, we have $P(A \land B \land C) = 0.48 - x$ where $x = P(A \land C \land \neg B)$.  Next, consider $P(B \land C) = P(B | C) \times P(C) = 0.83 \times 0.6 = 0.498$. 
Next, observe that $P(C) = P(C \land B) + P(C \land \neg B)$; thus, $P(C \land \neg B) = 0.6 - 0.498 = 0.102$. Now, because $x = P(A \land C \land \neg B) \leq P(C \land \neg B) = 0.102$, we obtain $x \leq 0.102$.   Finally, note that $P(A \land B) \geq  P(A \land B \land C)   = 0.48-x \geq (0.48 - 0.102) = 0.378$. Thus, clearly, $P(A \land B) > P(A) \times P(B) = 0.36$, so $\texttt{edge}(2,5)$ and $\texttt{edge}(2,6)$ are positively correlated.

\end{document}